\documentclass[10pt,twocolumn,twoside]{IEEEtran} 
\usepackage{balance}
\usepackage{amsfonts}
\usepackage{amssymb}
\usepackage{amsmath,mathtools}
\usepackage{amsthm}
\usepackage{graphicx}
\usepackage{subfigure}
\usepackage{cite}
\usepackage{algpseudocode}
\usepackage{algorithm}
\usepackage{multicol}
\usepackage{blindtext}
\usepackage{morefloats}
\usepackage{slashbox}
\usepackage{multirow}
\usepackage{caption}

\scrollmode

\DeclareCaptionFormat{myformat}{#3}
\begin{document}
\pagenumbering{gobble}
\theoremstyle{definition}
\newtheorem{theorem}{Theorem}
\theoremstyle{definition}
\newtheorem{example}{Example}
\theoremstyle{definition}
\newtheorem{remark}{Remark}
\newcommand{\xB}{\mathbf{B}}
\newcommand{\xtE}{\text{E}}
\newcommand{\xe}{\mathbf{e}}
\newcommand{\xGF}{\text{GF}}
\newcommand{\xtr}{\text{tr}}
\newcommand{\xd}{\mathbf{d}}
\newcommand{\xH}{\mathbf{H}}
\newcommand{\xI}{\mathbf{I}}
\newcommand{\xtI}{\text{I}}
\newcommand{\xtlog}{\text{log}}
\newcommand{\xmkl}{_{k,l}}
\newcommand{\xmkj}{_{k,j}}
\newcommand{\xQ}{\mathbf{Q}}
\newcommand{\xP}{\mathbf{P}}
\newcommand{\xp}{\mathbf{p}}
\newcommand{\xR}{\mathbf{R}}
\newcommand{\xSF}{\text{SF}}
\newcommand{\xtSINR}{\text{SINR}}
\newcommand{\xu}{\mathbf{u}}
\newcommand{\xU}{\mathbf{U}}
\newcommand{\xv}{\mathbf{v}}
\newcommand{\xtv}{\text{v}}
\newcommand{\xV}{\mathbf{V}}
\newcommand{\xw}{\mathbf{w}}
\newcommand{\xy}{\mathbf{y}}
\newcommand{\xX}{\mathbf{X}}
\newcommand{\xx}{\mathbf{x}}
\newcommand{\xz}{\mathbf{z}}
\newcommand{\xCES}{\text{CES}}
\newcommand{\xDES}{\text{DES}}



\onecolumn

\title{Second-Order Statistics of MIMO Rayleigh Interference Channels: Theory, Applications, and Analysis}

\author{Ahmed O. D. Ali $^{\dagger}$, Cenk M. Yetis$^{\ddagger}$, Murat Torlak$^{\dagger}$
\\ $^{\dagger}$ Department of Electrical Engineering, University of Texas at Dallas, USA
\\ $^{\ddagger}$ Electrical and Electronics Engineering Department, Mevlana University, Turkey
\\ $\{\text{ahmed.ali,torlak}\}$@utdallas.edu, cenkmyetis@ieee.org
}

\maketitle


\begin{abstract}
While first-order channel statistics, such as bit-error rate (BER) and outage probability, play an important role in the design of wireless communication systems, they provide information only on the static behavior of fading channels. On the other hand, second-order statistics, such as level crossing rate (LCR) and average outage duration (AOD), capture the correlation properties of fading channels, hence, are used in system design notably in packet-based transmission systems. In this paper, exact \mbox{closed-form} expressions are derived for the LCR and AOD of the signal at a receiver where maximal-ratio combining (MRC) is deployed over flat Rayleigh fading channels in the presence of additive white Gaussian noise (AWGN) and co-channel interferers with unequal transmitting powers and unequal speeds. Moreover, in order to gain insight on the LCR behavior, a simplified approximate expression for the LCR is presented. As an application of LCR in system designs, the packet error rate (PER) is evaluated through finite state Markov chain (FSMC) model. Finally, as another application again by using the FSMC model, the optimum packet length to maximize the throughput of the system with stop-and-wait automatic repeat request (SW-ARQ) protocol is derived. Simulation results validating the presented expressions are provided.
\end{abstract}

\section{Introduction}
\label{sec:intro}

The first order statistics are commonly used as important performance metrics, however they lack information for the overall system design and performance since they only capture the static behavior of the channel. Outage probability, the probability that the received signal-to-interference plus noise ratio (SINR) is below a certain threshold, and bit error rate (BER), the rate at which errors occur during the transmission, are such static metrics. On the other hand, the second-order statistics reflect the correlation properties of the fading channel providing a dynamic representation of the system performance. The level crossing rate (LCR) and the average outage duration (AOD) are examples of second-order statistical metrics and have been used in many applications for designing and evaluating the performances of wireless communication systems. They have been used in finite-state Markov chain (FSMC) channel modeling \cite{515}, the analysis of handoff algorithms \cite{585}, Markov decision process (MDP) and partially observable Markov decision process (POMDP) formulations \cite{586,337,587}, and packet error rate (PER) evaluation \cite{505}. Moreover, since AOD determines the average length of error bursts, it plays an important role in choosing the packet length for uncoded and coded systems, designing interleaved and non-interleaved coding schemes \cite{567}, determining the buffer size and transmission rate for adaptive modulation schemes \cite{568}, and estimating the throughput of different communication protocols such as automatic repeat request (ARQ) \cite{506}. Since LCR and AOD are closely related metrics, it is imperative to obtain the exact closed-form expressions for both LCR and AOD.

In his pioneering work, Rice obtained the LCR expression for a Rician-distributed fading signal in a single-user additive white Gaussian noise (AWGN) channel \cite{569}. Many works have been conducted later on to find LCR and AOD expressions in noise-limited systems covering different combining schemes with correlated and uncorrelated branches \cite{570,571,572,573,574,576}. Fewer works addressed interference-limited systems due to its relative complexity and the hardship of finding the joint probability density function (PDF) of the signal-to-interference ratio (SIR) and its time-derivative, and its challenging integration steps. Examples of such works include \cite{575} which derived the LCR and AOD expressions for interference-limited selection combining (SC) system over Rayleigh distributed channel while the interferers' channels are Rician distributed. In \cite{520}, Yang and Alouini derived the LCR expression where the desired signal is received over i.i.d. diversity paths and are Nakagami, Rician or Rayleigh distributed and so are the interferers; however, background noise effect was ignored. Recently, Fukawa et al. \cite{505} derived an approximate LCR expression for a point-to-point system subject to frequency-selective Rayleigh fading with the receiver deploying maximal-ratio combining (MRC) and the interfering signal components are of unequal powers. Seldom attention was given to systems where noise and interference powers are comparable. In \cite{370} an exact closed-form expression was obtained for the LCR of a single-antenna receiver with multiple interferers traveling at different speeds with unequal powers under AWGN and Rayleigh fading channel. However, it cannot be applied to MRC and hence loses the diversity gain. In this paper, the exact closed-form expressions of LCR and AOD are derived for a more general setting where the receiver with multiple antennas uses MRC diversity combining in the presence of multiple co-channel interferers that have different powers and speeds while the effects of both AWGN and the Rayleigh flat fading channels are taken into consideration. To the best of our knowledge, this is the first time to report these exact closed-form LCR and AOD expressions for this setting. It is worth noting that the results in \cite{370} are the special cases of the expressions derived in this paper when the number of receive diversity branches is set to $1$ as shown in Section \ref{sec:MERL}. A summary of the most relevant works in the literature on the LCR derivation is listed in Table \ref{table_literature_summary}.

\begin{table}[!htb]
\centering
\caption{A summary of the scenarios solved w.r.t. LCR derivation.}
\label{table_literature_summary}
\begin{tabular}{c c c}
\hline
\textbf{Problem} & \textbf{Contribution} & \textbf{Research Papers} \\ [0.5ex]
\hline
Single user Rician fading channel & Exact closed-form solution (ECFS) & \cite{569} \\
\hline
Correlated Rayleigh-fading multiple-branch equal-gain & & \\[-1ex]
predetection diversity combiner & \raisebox{1.5ex}{Approximate closed-form solution (ACFS)} & \raisebox{1.5ex}{\cite{570}} \\[1ex]
\hline
Independent Rayleigh-fading with \\ two-branch selection combining (SC), equal-gain (EGC) combining & & \\[-1ex]
 and maximal-ratio combining (MRC) for a single-user AWGN channel & \raisebox{1.5ex}{ECFS} & \raisebox{1.5ex}{\cite{571}} \\[1ex]
\hline
\multirow{2}{*}{Independent Nakagami-fading multiple-branch SC, EGC and MRC} & ECFS for SC and MRC & \\[-1ex]
 &  Numerical integration solution for EGC & \raisebox{1.5ex}{\cite{572}} \\[1ex]
\hline
EGC and MRC over independent multiple-branch & ECFS for MRC with Rayleigh i.i.d. branches & \\[-1ex]
with different fading distributions &  Numerical integration solution for the other scenarios & \raisebox{1.5ex}{\cite{573}} \\[1ex]
\hline
Independent unequal-power multiple-branch & & \\[-1ex]
Rayleigh-fading MRC & \raisebox{1.5ex}{ECFS} & \raisebox{1.5ex}{\cite{574}} \\[1ex]
\hline
Independent unequal-power multiple-branch & & \\[-1ex]
Weibull-fading SC & \raisebox{1.5ex}{ECFS} & \raisebox{1.5ex}{\cite{576}} \\[1ex]
\hline
Interference-limited SC system with Rayleigh fading desired user & \multirow{3}{*}{ECFS} & \multirow{3}{*}{\cite{575}} \\
and i.i.d. Rician-distributed co-channel interferers & & \\
with equal average powers &  &  \\[1ex]
\hline
Interference-limited MRC system with i.i.d. Nakagami, & \multirow{3}{*}{ECFS} & \multirow{3}{*}{\cite{520}} \\
Rician or Rayleigh distributed desired signal branches & & \\
  and i.i.d. co-channel interferers &  & \\[1ex]
\hline
Interference-limited MRC point-to-point system with i.i.d. & \multirow{2}{*}{ECFS for equal average power case} & \multirow{3}{*}{\cite{505}} \\
 frequency-selective Rayleigh fading channel with & & \\
 equal and unequal average power interferers & ACFS for unequal average power case & \\[1ex]
\hline
Single antenna nodes with Rayleigh-fading channels with interferers & & \\[-1ex]
having unequal average powers and different speeds & \raisebox{1.5ex}{ECFS for the LCR of the SINR} & \raisebox{1.5ex}{\cite{370}} \\[1ex]
\hline
\end{tabular}
\end{table}

LCR and AOD take place in a variety of applications ranging from estimating the system PER by using the FSMC model of the system to optimizing the packet length of automatic repeat request (ARQ)-based systems. In \cite{577}, Yousefi’zadeh and Jafarkhani estimated the PER of a multi-input multi-output (MIMO) system in the presence of co-channel interferers under AWGN and fading channels by using FSMC model although strong approximations were made in the SINR expression, while Fukawa et al. \cite{505} used the FSMC model to evaluate the PER of the point-to-point system under frequency selective fading. Many works have been done in literature on automatic repeat request (ARQ)-based systems due to their practical importance in improving the data transmission efficiency \cite{593}. Different ARQ schemes, including stop and wait (SW) ARQ, go-back-N (GBN) ARQ, selective-repeat (SR) ARQ, and hybrid (H) ARQ, have been analyzed for different point-to-point AWGN and fading channels \cite{506},\cite{588,589,590,592,579}. In \cite{506}, the authors evaluate the throughput using FSMC modeling of the system under GBN and SR ARQ protocols. The system's throughput is analyzed for SW ARQ-based system under slow Rayleigh fading with adaptive packet length in \cite{588}, while it is analyzed for GBN ARQ-based system under Rician fading using Gilbert-Elliott channel (GEC) model in \cite{589}. The throughput is analyzed and maximized w.r.t. information rate for HARQ-based system over an orthogonal space-time block coding (OSTBC) MIMO block fading channel in \cite{592}. FSMC is used in \cite{590} to model the packet error process in an SR ARQ-based system under Rayleigh fading, and the delay statistics is investigated. While those works considered single-user systems for which ARQ was originally developed, recently more attention was given to \mbox{multi-user} (MU) systems deploying ARQ protocols such as \cite{594} which considers linear precoder optimization to maximize the throughput in a MU-MIMO HARQ-based system while \cite{595} deals with the user scheduling aspect to improve the throughput. \cite{596} considers minimizing the transmission power via resource allocation in LTE uplink where HARQ is utilized assuming block fading Rayleigh channel model. It is worth mentioning that these recent works, \cite{594,595,596}, do not consider the channel time-correlation aspect.

Since the throughput is an important metric in communications systems, many works tackled the throughput maximization problem in different settings under different parameters, e.g. \cite{599} maximizes the throughput of an abstract system deploying one-bit ARQ suffering from delay through \mbox{multi-user} scheduling where POMDP framework is adopted. An important parameter affecting the throughput, specially in ARQ systems, is the packet length. Packet length and rate adaptation is tackled in \cite{597} to maximize the throughput in wireless LANs assuming Nakagami-$m$ fading while no retransmissions are allowed, hence, allowing constrained packets loss. \cite{598} proposed an algorithm to maximize the throughput in Bluetooth piconets through selecting the optimum packet length given a finite selection set. The authors in \cite{600} maximize the throughput of a point-to-point system utilizing truncated ARQ scheme under flat Rayleigh fading, where the optimum packet length and modulation size are chosen out of a given finite set based on an iterative suboptimum algorithm. In \cite{579}, again a point-to-point uncoded transmission system utilizing SW-ARQ with infinite retransmissions is studied under slow Rayleigh flat fading and an expression for the optimum data packet length maximizing the throughput was derived though the results are restricted to systems where the BER has exponential form. Again, we note that these works do not consider the channel time-correlation in the analysis.

In this paper, the exact LCR and AOD expressions are derived, in contrast to e.g., \cite{505}, under the effects of both AWGN and fading channels, in contrast SIR or signal-to-noise ratio (SNR) focused works e.g., \cite{575,573}, with a receiver employing MRC, in contrast to \cite{370}. Further, to easily acquire the LCR behavior with lesser computational complexity under different conditions, an approximate LCR expression is derived. By developing an FSMC model that makes use of the exact \mbox{closed-form} LCR expression and captures the channel \mbox{time-correlation} via incorporating the LCR in the model, the PER of the system, and the optimum data packet length for throughput maximization for a \mbox{multi-user} \mbox{SW-ARQ} system are derived. The main contributions of this work are summarized as follows.
\begin{itemize}

 \item Exact closed-form expressions for the LCR and AOD of the received SINR for an MRC receiver are derived in the presence of AWGN and multiple co-channel interferers with unequal powers, and moving with unequal speeds subject to Rayleigh flat fading channels.

 \item Approximate closed-form expressions of LCR and AOD are derived. These simplified expressions can help system designers streamline the design process without explicitly evaluating the LCR with every change of system parameters. Moreover, the computation complexity of the LCR is significantly reduced, especially for high number of interferers and receive antennas.

 \item The system is modelled by FSMC via the exact LCR expression and closed-form PER is derived.

 \item The formula for optimum packet length maximizing the throughput of the SW-ARQ system is derived.

\end{itemize}

Finding an exact closed-form solution for the LCR is a difficult process in general due to involved derivations including the joint PDF of the SINR and its time derivative along with difficult integrals that are challenging to be transformed to easier integrals. The difficulties arise from the following aspects included in this paper.
\begin{itemize}
\item	Both noise and interference are considered. As well known, noise-limited systems are rather easier since only SNR is tackled and hence the interferers' channels in the SINR denominator are absent. Again as well known, although interference-limited systems are more difficult than noise-limited systems, it is still an easier problem than the one incorporating both noise and interference. This is evident from the abundance of works considering either noise or interference and scarcity of works considering both.
\item	The interferers are assumed to be having different transmitting powers and speeds that lead to mathematical difficulties in obtaining the joint PDFs and in solving the integrations.
\item	An integral emanate corresponding to each interferer. Since a solution for any number of interferers is expected, an analytical solution for arbitrary number of integrations is not trivial contrary to the approaches that use numerical solutions for integrations.
\item	The receiver has multiple antennas and deploys MRC which constitute complicated integrations.

\end{itemize}

The theoretical expressions are validated by comparisons with the simulation results under varying system parameters. Some interesting findings in this work are

\begin{itemize}

 \item Given a fixed ratio of the SINR threshold to the average SINR, i.e. normalized threshold $\gamma_{\text{th}} / \gamma_{\text{avg}}$, the LCR value of the received SINR is independent of the value of the desired signal's power relative to the interferers', Section \ref{section:effect_pD_pI}.




 \item The exact LCR expression is very accurate in terms of matching the simulation results, Section \ref{sec:LCR_num_results}.

 \item The PER obtained from the LCR expression via the FSMC modeling is very accurate in terms of matching the simulation results, Section \ref{sec:FSMC}.

\end{itemize}

The paper is organized as follows. The system model is presented in Section \ref{sec:system_model}. The joint PDF of the SINR and its time derivative are derived in Section \ref{sec:joint_pdf_SINR_time_der} and used to obtain the LCR in Section \ref{sec:LCR_derivation}. The LCR expressions of simplified systems are derived as special cases in Section \ref{sec:special_cases} and are compared with previously reported expressions in the literature. The simplified approximate LCR expression is derived in Section \ref{sec:approx}. Section \ref{sec:LCR_num_results} presents the validation of the derived expressions via numerical results along with the investigation of the system parameters' effects on the LCR. The AOD derivation is given in Section \ref{sec:AOD_deriv} and verified via numerical examples. As an application of the new exact LCR expression, the system is modeled as an FSMC model and the PER is evaluated in Section \ref{sec:FSMC}. Another application of the LCR expression is presented in Section \ref{sec:ARQ} where the throughput of the system under SW-ARQ protocol is maximized by obtaining the optimal packet length as a function of the system parameters which includes the LCR. Finally, conclusions are drawn in Section \ref{sec:conclusion}.

\section{System Model}
\label{sec:system_model}

We consider a mobile radio system with $N+1$ transmitter nodes and a receiver node. Each transmitter and the receiver have a single antenna and $L$ antennas, respectively. $N$ transmitter nodes are interfering users, thus there is a single desired transmitter. All channel paths in the system are subject to flat Rayleigh fading with unit average power. The time autocorrelation of each channel path is given by $J_0(2\pi f_{\scriptscriptstyle \text{D}} \tau)$, where $J_0(.)$, $f_{\scriptscriptstyle \text{D}}$, and $\tau$ are the Bessel function of the zeroth order, the maximum Doppler frequency associated with this path and the time difference between the two correlated samples, respectively \cite{56}. The desired transmitter power and the interfering transmitter power are $p_{\scriptscriptstyle D}$ and $p_n$,  $\forall n \in \mathcal{N} \triangleq \{ 1, \dots, N \}$, respectively. The receiver performs MRC over the received signal. The received noise $\mathbf{z}$ is an additive white Gaussian noise (AWGN) with zero mean and covariance matrix $E\{\mathbf{z} \mathbf{z}^H \} = N_o \mathbf{I}$. This model can be generalized to any system with $L$ i.i.d. diversity paths, where again the receiver performs MRC over the $L$ diversity paths. The diversity can be also achieved in time or frequency, not necessarily in space.

Let $\mathbf{h}_{\scriptscriptstyle \text{D}}(t)$ and $\mathbf{h}_n(t)$ denote the $L \times 1$ channel vectors between the receiver and the desired user and $n^{th}$ interferer, respectively. The received signal before applying MRC, $\mathbf{y}(t)$, is given by
\begin{equation}
\mathbf{y}(t) = \mathbf{h}{\scriptscriptstyle \text{D}}(t) s_{\scriptscriptstyle \text{D}}(t) + \sum\limits_{n=1}^N \mathbf{h}_n(t) s_n(t) + \mathbf{z}(t)
\end{equation}
where $s_{\scriptscriptstyle \text{D}}(t)$ and $s_n(t)$ are the transmitted symbols at time instant $t$ from the desired user and the $n^{th}$ interferer, respectively, with average powers $E\{ |s_{\scriptscriptstyle \text{D}}|^2 \} = p_{\scriptscriptstyle \text{D}}$ and $E\{ |s_n|^2 \} = p_n$. Let $A_{\scriptscriptstyle \text{D},l}(t)$, $\forall l \in \mathcal{L} \triangleq \{ 1, \dots, L \}$, denote the Rayleigh distributed random envelope of the received signal of the desired user over the $l^{th}$ diversity path. The output signal-to-interference-plus-noise ratio (SINR) $\mathit{\Gamma}(t)$ of the system is given by
\begin{equation}
\label{eq:SINR_def}
\mathit{\Gamma}(t) = \frac{p_{\scriptscriptstyle \text{D}} \| \mathbf{h}_{\scriptscriptstyle \text{D}}(t) \|^2} {N_o + \sum\limits_{n=1}^N p_n |\mathbf{w}(t)^H \mathbf{h}_n(t)|^2} = \frac{p_{\scriptscriptstyle \text{D}} \sum\limits_{l=1}^L A_{\scriptscriptstyle \text{D},l}^2(t)} {N_o + \sum\limits_{n=1}^N p_n \ A_{\scriptscriptstyle \text{I},n}^2(t)} = \frac{Y}{Z},
\end{equation}
where $Y$ and $Z$ denote the numerator and denominator of the SINR, respectively, $\mathbf{w}(t) = \mathbf{h}_{\scriptscriptstyle D}(t) / \|\mathbf{h}_{\scriptscriptstyle D}(t)\|$ is the receive filter, and $A_{\scriptscriptstyle \text{I},n}(t) = |\mathbf{w}(t)^H \mathbf{h}_n(t)|$. Then, $A_{\scriptscriptstyle \text{D},l}^2(t)$ and $A_{\scriptscriptstyle \text{I},n}^2(t)$ are standard exponential random processes with unit means \cite{537}.

Given an SINR threshold $\gamma_{\text{th}}$ for the desired user, the level crossing rate (LCR) is defined as \cite{56}
\begin{equation}
\label{eq:LCR_def}
\text{LCR}(\gamma_{\text{th}}) = \int\limits_{\dot{\gamma}=0}^{\infty} \dot{\gamma} f_{\mathit{\Gamma}, \dot{\mathit{\Gamma}}} (\gamma_{\text{th}}, \dot{\gamma}) d\dot{\gamma},
\end{equation}
where the time index $t$ is dropped to simplify the notation, and hence, $\dot{\mathit{\Gamma}}$ is the time derivative of the SINR $\mathit{\Gamma}$, and $f_{\mathit{\Gamma}, \dot{\mathit{\Gamma}}} (.,.)$ is the joint probability density function (PDF) of $\dot{\mathit{\Gamma}}$ and $\mathit{\Gamma}$.

The average duration for which the SINR remains below a threshold, namely the average outage duration (AOD), is defined as \cite{56}
\begin{equation}
\label{eq:AOD_def}
\text{AOD} \left( \gamma_{\text{th}} \right) = \frac{F_{\mathit{\Gamma}}\left(\gamma_{\text{th}}\right)}
{\text{LCR}\left(\gamma_{\text{th}}\right)},
\end{equation}
where $F_{\mathit{\Gamma}}\left(\gamma_{\text{th}}\right) = \text{Prob}(\mathit{\Gamma} \leq \gamma_{\text{th}})$ is the cumulative distribution function (CDF) of $\mathit{\Gamma}$.

In Sections \ref{sec:joint_pdf_SINR_time_der} and \ref{sec:LCR_derivation}, the LCR for the system presented above is derived. For the reader's convenience, the main derivation steps are summarized as follows.
\begin{algorithm}
  \caption{Main LCR Derivation Steps}
  \label{derivation_summary}
  \begin{algorithmic}
    \State 1: The transformations of random variables (RVs) in \eqref{eq:variable_trans_1}-\eqref{eq:variable_trans_2} are introduced to find the joint PDF of the SINR and its time derivative.
    \State 2: Baye's rule is used on the conditional PDFs to find the joint PDF of the SINR, its time derivative and the envelopes of the interferers' paths, given in \eqref{eq:joint_pdf1}.
    \State 3: Integration is performed first over the SINR's time derivative in the LCR integration as shown in \eqref{eq:LCR1} to get \eqref{eq:LCR4}.
    \State 4: The $N$-multiple integral in \eqref{eq:Ia_def} is reduced into a double integral through regarding this integral as an expectation of a function of exponential RVs as in \eqref{eq:Ia_def2}.
    \State 5: The transformations of RVs \eqref{eq:Q1}-\eqref{eq:Qn} are introduced, then their joint PDF is obtained as in \eqref{eq:joint_pdf_allQ} to solve \eqref{eq:Ia_def2}. Using these transformations, the expectation is rewritten in terms of $Q_1$ and $Q_2$ only as in \eqref{eq:Ia1}.
    \State 6: The joint characteristic function of $Q_1$ and $Q_2$ is obtained as in \eqref{eq:cc_fn2}, then their joint PDF is obtained as in \eqref{eq:f_Q1Q2}.
    \State 7: The joint PDF is substituted in the integral \eqref{eq:Ia5} and after a series of binomial expansions and integrations, the final exact closed-form LCR expression in \eqref{eq:LCR_final} is achieved.
  \end{algorithmic}
\end{algorithm}

\section{Joint PDF of The SINR and Its Time Derivative}
\label{sec:joint_pdf_SINR_time_der}

Following the SINR definition in \eqref{eq:SINR_def}, the time derivative of SINR $\dot{\mathit{\Gamma}}$ is given by \small
\begin{equation}
\dot{\mathit{\Gamma}} = \frac{2 p_{\scriptscriptstyle D} } {Z} \sum\limits_{l=1}^L A_{\scriptscriptstyle \text{D},l} \dot{A}_{\text{D},l} - \frac{2 \mathit{\Gamma} } {Z} \sum\limits_{n=1}^N p_n A_{\scriptscriptstyle \text{I},n} \dot{A}_{\text{I},n},
\end{equation}
\normalsize where $\dot{A}_{\scriptscriptstyle \text{D},l}$, and $\dot{A}_{\scriptscriptstyle \text{I},n}$ are i.i.d Gaussian RVs with zero mean. The variances of the latter are $\sigma_n^2 = \pi^2 f_{\scriptscriptstyle \text{I},n}^2$ \cite{370}, where $f_{\scriptscriptstyle \text{I},n} = \frac{\nu_n}{\lambda}$ is the maximum Doppler shift associated with the $n^{th}$ interferer, $\nu_n$ and $\lambda$ are the relative corresponding mobile speed of $n^{th}$ interferer and the carrier wavelength, respectively. Note that for the desired user, $\{\dot{A}_{\scriptscriptstyle \text{D},l}\}$ is assumed to have the same variance $\sigma_{\scriptscriptstyle \text{D},l}^2 = \sigma_{\scriptscriptstyle D}^2=\pi^2 f_{\scriptscriptstyle \text{D}}^2, \forall \ l \in \mathcal{L}$, where $f_{\scriptscriptstyle \text{D}} = \frac{\nu_{\scriptscriptstyle \text{D}}}{\lambda}$ is the maximum Doppler shift associated with the desired transmitter (Tx), and $\nu_{\scriptscriptstyle \text{D}}$ is its relative mobile speed. This is due to the fact that these variances correspond to the same Tx, and $\sigma_{\scriptscriptstyle \text{D},l}^2$ depends only on the relative speed which is the same for all the
diversity paths between the same Tx and receiver (Rx). Then, $\dot{\mathit{\Gamma}}$ is Gaussian (conditioned on $A_{\scriptscriptstyle \text{D},l}^2$, and $A_{\scriptscriptstyle \text{I},n}^2, \forall \ l,n$) with zero mean and variance $\sigma_{\dot{\gamma}}^2$, and its conditional PDF is given as \cite{370}
\begin{equation}
\label{eq:cond_alpha_PDF_Jafarkhani1}
f_{\dot{\mathit{\Gamma}}|A_{\scriptscriptstyle \text{D},l}^2, A_{\scriptscriptstyle \text{I},n}^2} (\dot{\gamma}|\alpha_{\scriptscriptstyle \text{D},l}^2, \alpha_{\scriptscriptstyle \text{I},n}^2) = \frac{1}{\sqrt{2 \pi
\sigma_{\dot{\gamma}}^2}} e^{-\frac{\dot{\gamma}^2}{2 \sigma_{\dot{\gamma}}^2}}, \ \ \ \forall \ l   ,n
\end{equation}
where the variance $\sigma_{\dot{\gamma}}^2$ is given by

\begin{equation}
\label{eq:variance_Jafarkhani}
\sigma_{\dot{\gamma}}^2 = \frac{4 p_{\scriptscriptstyle D}^2} {z^2} \sum\limits_{l=1}^L \alpha_{\scriptscriptstyle \text{D},l}^2 \sigma_{\scriptscriptstyle \text{D}}^2 + \frac{4 \gamma^2} {z^2} \sum\limits_{n=1}^N p_n^2 \alpha_{\scriptscriptstyle \text{I},n}^2 \sigma_n^2 = \frac{4 p_{\scriptscriptstyle D} \sigma_{\scriptscriptstyle \text{D}}^2 \gamma} {z} + \frac{4 \gamma^2} {z^2} \sum\limits_{n=1}^N p_n^2 \alpha_{\scriptscriptstyle \text{I},n}^2 \sigma_n^2.
\end{equation}

It is noticed from \eqref{eq:variance_Jafarkhani} that $\sigma_{\dot{\gamma}}^2$ is independent of $\alpha_{\scriptscriptstyle \text{D},l}^2$ and the knowledge of $\gamma$ and $\alpha_{\scriptscriptstyle \text{I},n}^2$ is sufficient. Hence, \eqref{eq:cond_alpha_PDF_Jafarkhani1} is rewritten as
\begin{equation}
\label{eq:cond_alpha_PDF_Jafarkhani}
f_{\dot{\mathit{\Gamma}}|\mathit{\Gamma}, A_{\scriptscriptstyle \text{I},n}^2} (\dot{\gamma}|\gamma, \alpha_{\scriptscriptstyle \text{I},n}^2) = \frac{1}{\sqrt{2 \pi
\sigma_{\dot{\gamma}}^2}} e^{-\frac{\dot{\gamma}^2}{2 \sigma_{\dot{\gamma}}^2}}, \ \ \ \forall \ l   ,n
\end{equation}
and $\sigma_{\dot{\gamma}}^2$ is still given by \eqref{eq:variance_Jafarkhani}. To obtain $f_{\scriptscriptstyle \mathit{\Gamma}, \dot{\mathit{\Gamma}}} \left( \gamma, \dot{\gamma} \right)$ in (\ref{eq:LCR_def}), the following variable transformations can be introduced
\begin{align}
\label{eq:variable_trans_1}
\dot{\mathit{\Gamma}} \ \ &= \ \dot{\mathit{\Gamma}} \\
\label{eq:mu_Jafarkhani}
\tilde{U}_n &= \ A_{\scriptscriptstyle \text{I},n}^2, \ \ \ \ \ \ \ \ \ \  \ \forall n \in \mathcal{N} \\
\label{eq:variable_trans_2}
\mathit{\Gamma} \ \ &= \ \frac{Y} {N_o + \sum_{n=1}^N p_n A_{\scriptscriptstyle \text{I},n}^2} = \frac{Y} {N_o + \sum_{n=1}^N p_n \tilde{U}_n}.
\end{align}

The joint PDF can be derived by using the Jacobian approach as follows
\begin{equation}
\label{eq:big_joint_Jafarkhani}
f_{\mathit{\Gamma}, \dot{\Gamma}, \tilde{U}_1,\dots,\tilde{U}_N} \left( \dot{\gamma}, \gamma, \tilde{u}_1,\dots,\tilde{u}_{\scriptscriptstyle N} \right) = f_{\dot{\Gamma} | \Gamma, \tilde{U}_1,\dots,\tilde{U}_{\scriptscriptstyle N}} \left( \dot{\gamma} | \gamma, \tilde{u}_1,\dots,\tilde{u}_{\scriptscriptstyle N} \right) f_{\Gamma | \tilde{U}_1,\dots,\tilde{U}_{\scriptscriptstyle N}} \left( \dot{\gamma}, \gamma, \tilde{u}_1,\dots,\tilde{u}_{\scriptscriptstyle N} \right) f_{\tilde{U}_1,\dots,\tilde{U}_{\scriptscriptstyle N}} \left(\tilde{u}_1,\dots,\tilde{u}_n\right),
\end{equation}
where
\begin{align}
\label{eq:conditional_pdfs}
&f_{\mathit{\Gamma} | \tilde{U}_1,\dots,\tilde{U}_{\scriptscriptstyle N}} \left( \dot{\gamma}, \gamma, \tilde{u}_1,\dots,\tilde{u}_{\scriptscriptstyle N} \right) = \frac{1}{\begin{vmatrix}\frac{ \partial \mathit{\Gamma} }{ \partial Y }\end{vmatrix}} f_{Y} \left(\gamma \left(N_o + \sum\limits_{n=1}^N p_n \tilde{u}_n \right)\right), \nonumber\\
&f_{\tilde{U}_1,\dots,\tilde{U}_n} \left( \tilde{u}_1,\dots,\tilde{u}_{\scriptscriptstyle N} \right) = \frac{1} {\begin{vmatrix} \frac{\partial \left( \tilde{U}_1,\dots,\tilde{U}_{\scriptscriptstyle N} \right) } {\partial \left( A_{\scriptscriptstyle I,1}^2, \dots, A_{\scriptscriptstyle I,N}^2 \right) }\end{vmatrix}} f_{A_{\scriptscriptstyle I,1}^2,\dots,A_{\scriptscriptstyle I,N}^2} \left( \alpha_{\scriptscriptstyle I,1}^2,\dots,\alpha_{\scriptscriptstyle I,N}^2 \right) = \frac{1} {\begin{vmatrix}\frac{ \partial \left( \tilde{U}_1,\dots,\tilde{U}_{\scriptscriptstyle N} \right) } {\partial \left( A_{\scriptscriptstyle I,1}^2, \dots, A_{\scriptscriptstyle I,N}^2 \right) }\end{vmatrix}} \prod\limits_{n=1}^N f_{A_{\scriptscriptstyle \text{I},n}^2} \left( \alpha_{\scriptscriptstyle \text{I},n}^2 \right), \nonumber\\
&f_{\dot{\mathit{\Gamma}} | \Gamma, \tilde{U}_1,\dots,\tilde{U}_{\scriptscriptstyle N}} \left( \dot{\gamma} | \gamma, \tilde{u}_1,\dots,\tilde{u}_{\scriptscriptstyle N} \right) = \frac{1} {\begin{vmatrix}\frac{\partial \dot{\Gamma}} {\partial \dot{\Gamma}}\end{vmatrix}} f_{\dot{\Gamma} | Y, \tilde{U}_1, \dots, \tilde{U}_{\scriptscriptstyle N}} \left( \dot{\gamma} | \gamma \left(N_o + \sum\limits_{n=1}^N p_n \tilde{u}_n \right), \tilde{u}_1, \dots, \tilde{u}_{\scriptscriptstyle N}  \right).
\end{align}
The Jacobians used above are given by
\begin{equation}
\begin{vmatrix}
\frac{\partial \dot{\mathit{\Gamma}}} {\partial \dot{\mathit{\Gamma}}} \end{vmatrix} = 1, \  \ \ \ \ \ \ \ \ \ \ \  \ \ \ \ \ \ \
\begin{vmatrix}\frac{\partial \mathit{\Gamma}}{\partial Y } \end{vmatrix} = \frac{1} {N_o + \sum\limits_{n=1}^N p_n \tilde{U}_n}, \  \ \ \ \ \ \ \ \ \ \ \  \ \ \ \ \ \ \
\begin{vmatrix}
\frac{\partial \left( \tilde{U}_1,\dots,\tilde{U}_{\scriptscriptstyle N}\right)} {\partial \left( A_{\scriptscriptstyle I,1}^2,\dots,A_{\scriptscriptstyle I,N}^2\right)} \end{vmatrix}
= \begin{vmatrix}
  \begin{array}{cccc}
    \frac{\partial \tilde{U}_1}{\partial A_{\scriptscriptstyle I,1}^2} & \frac{\partial \tilde{U}_1}{\partial A_{\scriptscriptstyle I,2}^2} & \dots & \frac{\partial \tilde{U}_1}{\partial A_{\scriptscriptstyle I,N}^2} \\
    \frac{\partial \tilde{U}_2}{\partial A_{\scriptscriptstyle I,1}^2} & \frac{\partial \tilde{U}_2}{\partial A_{\scriptscriptstyle I,2}^2} & \dots & \frac{\partial \tilde{U}_2}{\partial A_{\scriptscriptstyle I,N}^2} \\
    \vdots & \vdots & \dots & \vdots \\
    \frac{\partial \tilde{U}_{\scriptscriptstyle N}}{\partial A_{\scriptscriptstyle I,1}^2} & \frac{\partial \tilde{U}_{\scriptscriptstyle N}}{\partial A_{\scriptscriptstyle I,2}^2} & \dots & \frac{\partial \tilde{U}_{\scriptscriptstyle N}}{\partial A_{\scriptscriptstyle I,N}^2}
  \end{array}
\end{vmatrix} = |\mathbf{I}| = 1.
\end{equation}

Since the distribution of the sum of i.i.d. standard exponential RVs is known to be an Erlang distribution, then the pdf of $X = \sum\limits_{l=1}^L A_{\scriptscriptstyle \text{D},l}^2$ is given as
\begin{equation}
\label{eq:pdf_X}
f_X(x) = \frac{x^{L-1} e^{-x}} {\Gamma_{\text{s}}(L)},
\end{equation}
where $\Gamma_{\text{s}}(L)$ is the Gamma function. Then the pdf of $Y = p_{\scriptscriptstyle D} X$ defined in \eqref{eq:SINR_def} as the numerator of the SINR is given as
\begin{equation}
\label{eq:pdf_Y}
f_Y(y) = \frac{1}{\begin{vmatrix} \frac{\partial Y} {\partial X} \end{vmatrix}} f_X\left( y/p_{\scriptscriptstyle D} \right) = \frac{y^{L-1} e^{-\frac{y}{p_{\scriptscriptstyle D}}}} {p_{\scriptscriptstyle D}^L \Gamma_{\text{s}}(L)}.
\end{equation}

Given the fact that $A_{\scriptscriptstyle \text{I},n}^2, \forall n \in \mathcal{N}$ are i.i.d. standard exponential RVs, and substituting \eqref{eq:cond_alpha_PDF_Jafarkhani} and \eqref{eq:pdf_Y} into \eqref{eq:conditional_pdfs}, we get
\begin{equation}
\label{eq:joint_pdf1}
f_{\dot{\mathit{\Gamma}}, \Gamma, \tilde{U}_1,\dots,\tilde{U}_{\scriptscriptstyle N}} \left( \dot{\gamma}, \gamma, \tilde{u}_1,\dots,\tilde{u}_{\scriptscriptstyle N} \right) = \frac{ N_o + \sum\limits_{n=1}^N p_n \tilde{u}_n } {\sqrt{2 \pi \sigma_{\dot{\gamma}}^2}} e^{-\frac{\dot{\gamma}^2} {2\sigma_{\dot{\gamma}}^2}} \frac{\gamma^{L-1} \left( N_o + \sum\limits_{n=1}^N p_n \tilde{u}_n \right)^{L-1}} {p_{\scriptscriptstyle D}^L \Gamma_s(L)} e^{-\frac{\gamma \left(N_o + \sum\limits_{n=1}^N p_n \tilde{u}_n \right)}{p_{\scriptscriptstyle D}} } e^{- \sum\limits_{n=1}\tilde{u}_n}.
\end{equation}

\section{LCR Derivation}
\label{sec:LCR_derivation}

By substituting (\ref{eq:joint_pdf1}) in \eqref{eq:LCR_def}, the level crossing rate given a certain SINR threshold $\gamma_{\text{th}}$ can be obtained as
\begin{align}
\label{eq:LCR1}
\text{LCR}(\gamma_{\text{th}}) &= \frac{\gamma_{\text{th}}^{L-1} e^{-\frac{\gamma_{\text{th}} N_o}{p_{\scriptscriptstyle D}}}} {\sqrt{2 \pi} p_{\scriptscriptstyle D}^L \Gamma_{\text{s}}(L)} \int\limits_{\tilde{u}_{\scriptscriptstyle N}=0}^{\infty} \dots \int\limits_{\tilde{u}_1=0}^{\infty} \left( N_o + \sum\limits_{n=1}^N p_n \tilde{u}_n \right)^L e^{-\sum\limits_{n=1}^N \left( 1 + \frac{\gamma_{\text{th}} p_n} {p_{\scriptscriptstyle D}} \right) \tilde{u}_n}       \int\limits_{\dot{\gamma}=0}^{\infty} \frac{\dot{\gamma}} {\sqrt{\sigma_{\dot{\gamma}}^2}} e^{-\frac{\dot{\gamma}^2} {2 \sigma_{\dot{\gamma}}^2}} \ d\tilde{u}_1 \dots d\tilde{u}_{\scriptscriptstyle N} \\
\label{eq:LCR2}
&= \frac{\gamma_{\text{th}}^{L-1} e^{-\frac{\gamma_{\text{th}} N_o}{p_{\scriptscriptstyle D}}}} {\sqrt{2 \pi} p_{\scriptscriptstyle D}^L \Gamma_{\text{s}}(L)} \int\limits_{\tilde{u}_{\scriptscriptstyle N}=0}^{\infty} \dots \int\limits_{\tilde{u}_1=0}^{\infty} \left( N_o + \sum\limits_{n=1}^N p_n \tilde{u}_n \right)^L e^{-\sum\limits_{n=1}^N \left( 1 + \frac{\gamma_{\text{th}} p_n} {p_{\scriptscriptstyle D}} \right) \tilde{u}_n} \sqrt{\sigma_{\dot{\gamma}}^2} \ d\tilde{u}_1 \dots d\tilde{u}_{\scriptscriptstyle N} \\
\label{eq:LCR3}
&= \sqrt{\frac{2 \sigma_{\scriptscriptstyle \text{D}}^2} {\pi}} \left( \frac{\gamma_{\text{th}} N_o}{p_{\scriptscriptstyle D}} \right)^{L-\frac{1}{2}} \frac{e^{-\frac{\gamma_{\text{th}} N_o}{p_{\scriptscriptstyle D}}}} {\Gamma_{\text{s}}(L)} \int\limits_{\tilde{u}_{\scriptscriptstyle N}=0}^{\infty} \dots \int\limits_{\tilde{u}_1=0}^{\infty} \sqrt{1 + \sum\limits_{n=1}^N \left( \frac{p_n} {N_o} + \frac{\gamma_{\text{th}} p_n^2 \sigma_n^2} {N_o p_{\scriptscriptstyle D} \sigma_{\scriptscriptstyle \text{D}}^2} \right) \tilde{u}_n } \nonumber\\
 & \times \left( 1 + \sum\limits_{n=1}^N \frac{p_n}{N_o} \tilde{u}_n \right)^{L-1} e^{-\sum\limits_{n=1}^N \left( 1 + \frac{\gamma_{\text{th}} p_n} {p_{\scriptscriptstyle D}} \right) \tilde{u}_n} \ d\tilde{u}_1 \dots d\tilde{u}_{\scriptscriptstyle N} \\
 \label{eq:LCR4}
 &= \frac{ \sqrt{2 \sigma_{\scriptscriptstyle \text{D}}^2 } \left( \frac{\gamma_{\text{th}} N_o}{p_{\scriptscriptstyle D}} \right)^{L-\frac{1}{2}} e^{-\frac{\gamma_{\text{th}} N_o}{p_{\scriptscriptstyle D}}} }   {\sqrt{\pi} \Gamma_{\text{s}}(L) \prod\limits_{n=1}^N \left(1 + \frac{\gamma_{\text{th}} p_n} {p_{\scriptscriptstyle D}}\right)} \ I_a,
\end{align}
where \eqref{eq:LCR3} follows by substituting by \eqref{eq:variance_Jafarkhani} in \eqref{eq:LCR2} and doing simple algebraic manipulation. By introducing the variable transformation $u_n = (1 + \gamma_{\text{th}} p_n / p_{\scriptscriptstyle \text{D}}) \tilde{u}_n$, \eqref{eq:LCR4} follows directly where the integral $I_a$ is defined as
\begin{equation}
\label{eq:Ia_def}
I_a = \int\limits_{u_{\scriptscriptstyle N}=0}^{\infty} \dots \int\limits_{u_1=0}^{\infty} \left( 1 + \sum\limits_{n=1}^N a_n u_n \right)^{L-1} \sqrt{1 + \sum\limits_{n=1}^N b_n u_n } e^{-\sum_{n=1}^N u_n} \ du_1 \dots du_{\scriptscriptstyle N}
\end{equation}
and the constants $a_n$ and $b_n$ are given by
\begin{align}
\label{eq:anbneps}
a_n &= \frac{p_n / N_o} {1 + \frac{\gamma_{\text{th}} p_n} {p_{\scriptscriptstyle D}} }, \nonumber\\
b_n &= a_n \left( 1 + \varepsilon_n  \right), \text{ and} \nonumber\\
\varepsilon_n &= \frac{\gamma_{\text{th}} p_n \sigma_n^2} {p_{\scriptscriptstyle D} \sigma_{\scriptscriptstyle \text{D}}^2}.
\end{align}
Note that from the above definitions, it can be seen that $a_n, b_n$ and $\varepsilon_n$ are all positive $\forall n \in \mathcal{N}$.

Now to proceed forward in solving the integral $I_a$, new i.i.d. standard exponential RVs denoted by $V_n, \ \forall n \in \mathcal{N}$ are introduced. Hence, their joint PDF is given by $f_{V_1,\dots,V_{\scriptscriptstyle N}}(v_1,\dots,v_{\scriptscriptstyle N}) = e^{-\sum_{n=1}^N v_n}$. Define $g(V_1,\dots,V_{\scriptscriptstyle N})$ as a function of these RVs as
\begin{equation}
g(V_1,\dots,V_{\scriptscriptstyle N}) = \left( 1 + \sum\limits_{n=1}^N a_n V_n \right)^{L-1} \sqrt{1 + \sum\limits_{n=1}^N b_n V_n }.
\end{equation}
Then the expectation of this function is given by
\begin{align}
E\left\{ g\left( V_1,\dots,V_{\scriptscriptstyle N} \right) \right\} &= \int\limits_{v_{\scriptscriptstyle N}=0}^{\infty} \dots \int\limits_{v_1=0}^{\infty} g\left( v_1,\dots,v_{\scriptscriptstyle N} \right) f_{V_1,\dots,V_{\scriptscriptstyle N}}(v_1,\dots,v_{\scriptscriptstyle N})  \ dv_1 \dots dv_{\scriptscriptstyle N} \nonumber\\
\label{eq:g_expectation}
&= \int\limits_{v_{\scriptscriptstyle N}=0}^{\infty} \dots \int\limits_{v_1=0}^{\infty} \left( 1 + \sum\limits_{n=1}^N a_n v_n \right)^{L-1} \sqrt{1 + \sum\limits_{n=1}^N b_n v_n } e^{-\sum_{n=1}^N v_n} \ dv_1 \dots dv_{\scriptscriptstyle N},
\end{align}
where we notice that the integration in \eqref{eq:g_expectation} is exactly the same as $I_a$ in \eqref{eq:Ia_def}. Hence, only for the sake of solving $I_a$, we shall regard the integration variables $U_n, \ \forall n \in \mathcal{N}$ as i.i.d. standard exponential RVs and thus $I_a$ can be rewritten as
\begin{equation}
\label{eq:Ia_def2}
I_a = E\left\{ \left( 1 + \sum\limits_{n=1}^N a_n U_n \right)^{L-1} \sqrt{1 + \sum\limits_{n=1}^N b_n U_n } \right\}.
\end{equation}



In order to find the above expectation, the following RV transformations are introduced
\begin{align}
\label{eq:Q1}
Q_1 &= \sum\limits_{n=1}^N b_n U_n, \\
\label{eq:Q2}
Q_2 &= \sum\limits_{n=2}^N \left(a_n - \frac{a_1 b_n}{b_1} \right) U_n,\text{ and}  \\
\label{eq:Qn}
Q_n &= a_n U_n, \ \ \ \ \ n = 3,\dots,N.
\end{align}

After finding the domains of the new variables as derived in Appendix \ref{sec:LCR_deriv_appendix}, the joint PDF $f_{Q_1,\dots,Q_{\scriptscriptstyle N}}(q_1,\dots,q_{\scriptscriptstyle N})$ is given as
\begin{equation}
\label{eq:joint_pdf_allQ}
f_{Q_1,\dots,Q_{\scriptscriptstyle N}}(q_1,\dots,q_{\scriptscriptstyle N}) = \frac{1}{\begin{vmatrix} J \end{vmatrix}} f_{U_1,\dots,U_{\scriptscriptstyle N}}(u_1,\dots,u_{\scriptscriptstyle N}) = \frac{1}{\begin{vmatrix} J \end{vmatrix}} e^{-\sum\limits_{n=1}^N u_n} = \frac{1}{\begin{vmatrix} J \end{vmatrix}} e^{-\sum\limits_{n=1}^N \alpha_n Q_n},
\end{equation}
where $|J|$ is the Jacobian. $|J|$ and $\alpha_n$ are given as
\begin{equation}
\label{eq:jacob}
|J| = \begin{vmatrix} \frac{\partial \left( Q_1,\dots,Q_{\scriptscriptstyle N} \right)} {\partial \left( U_1,\dots,U_{\scriptscriptstyle N} \right)} \end{vmatrix} = \left( a_2 b_1 - a_1 b_2 \right) \prod\limits_{n=3}^N a_n, \text{and} \\
\end{equation}
\begin{equation}
\label{eq:alpha}
\alpha_n = \begin{cases}
\frac{1} {b_1}, &n = 1 \\
\frac{b_1 - b_2} {a_2 b_1 - a_1 b_2}, &n = 2 \\
\frac{\left( b_2 - b_1 \right) \left( a_n b_1 - a_1 b_n \right)} {a_n b_1 \left( a_2 b_1 - a_1 b_2 \right)} - \frac{b_n} {a_n b_1} + \frac{1}{a_n},  &n = 3,\dots,N.
\end{cases}
\end{equation}

After tedious algebraic manipulation, detailed in Appendix \ref{sec:fQ1Q2_deriv_appendix}, the joint characteristic function $\varphi_{\scriptscriptstyle Q_1,Q_2}(s_1,s_2)$ is written as
\begin{equation}
\label{eq:cc_fn2}
\varphi_{\scriptscriptstyle Q_1,Q_2}(s_1,s_2) = \Lambda \sum\limits_{n=1}^N \delta_n \sum\limits_{\substack{t=1\\t\neq n}}^N \frac{\psi_{t,n}}{\left( s_1 + f_n s_2 + g_n \right) \left(s_2 + \lambda_{t,n}\right)},
\end{equation}
where $s_1$, and $s_2$ are the Laplace variables and $f_n, g_n, \delta_n, \lambda_{t,n}, \Lambda$ and $\psi_{t,n}$ are defined in Appendix \ref{sec:fQ1Q2_deriv_appendix} by equations \eqref{eq:fn}, \eqref{eq:gn}, \eqref{eq:delta}, \eqref{eq:lambda}, \eqref{eq:psi} and \eqref{eq:Lambda} and further simplified in Appendix \ref{sec:simplification_appendix} by equations \eqref{eq:fn_simple}-\eqref{eq:psi_simp}.

In order to find the joint PDF $f_{Q_1,Q_2}(q_1,q_2)$, double inverse Laplace transform over $s_1$ and $s_2$ is performed, resulting in the closed-form expression
\begin{equation}
\label{eq:f_Q1Q2}
f_{Q_1,Q_2}(q_1,q_2) = \Lambda \sum\limits_{n=1}^N \delta_n \sum\limits_{\substack{t=1 \\ t \neq n}} \psi_{t,n} e^{-g_n q_1} e^{-\lambda_{t,n} \left( q_2 - f_n q_1\right)},
\end{equation}
where this expression is valid under condition $q_2 \geq f_n q_1, \forall n \in \mathcal{N}$, with emphasis that $f_1 = 0$. These conditions result from the inverse Laplace transformation and will affect the range upon which $Q_2$ is integrated as will be seen later.

The joint PDF of $Q_1$ and $Q_2$ is then used to solve the integration in (\ref{eq:Ia_def}) as follows. Using the RVs transformation \eqref{eq:Q1}-\eqref{eq:Qn}, \eqref{eq:Ia_def2} can be rewritten as
\begin{align}
\label{eq:Ia1}
&I_a = E\left\{ \sqrt{1+q_1} \left( 1 + \frac{a_1}{b_1}q_1 + q_2 \right)^{L-1} \right\} \\
\label{eq:Ia5}
&= \sum\limits_{k=0}^{L-1} \binom{L-1}{k} \sum\limits_{m=0}^{L-1-k} \binom{L-1-k}{m} \left( \frac{a_1}{b_1}\right)^{L-1-m} \left( \frac{b_1}{a_1} - 1 \right)^{L-1-k-m} \underbrace{\int\limits_{q_1} \int\limits_{q_2} \left( 1 + q_1 \right)^{k+\frac{1}{2}} q_2^m f_{Q_1,Q_2}(q_1,q_2) dq_2 \ dq_1}_{I_b},
\end{align}
where \eqref{eq:Ia5} follows from a series of binomial expansions of the bracketed term in \eqref{eq:Ia1}.

Substituting \eqref{eq:f_Q1Q2} in \eqref{eq:Ia5}, $I_b$ can be rewritten as
\begin{equation}
\label{eq:Ib}
I_b = \Lambda \left[ \delta_1 \sum\limits_{t=2}^N \psi_{t,1} I_1 + \sum\limits_{n=2}^N \delta_n \sum\limits_{\substack{t=1 \\ t \neq n}}^N \psi_{t,n} I_2 \right],
\end{equation}
where
\begin{align}
\label{eq:I1_LCR}
I_1 &= \int\limits_{q_1=0}^{\infty} \left( 1 + q_1 \right)^{k+\frac{1}{2}} e^{-g_1 q_1} \ dq_1 \ \int\limits_{q_2=0}^{\infty} q_2^m e^{-\lambda_{t,1} q_2} \ dq_2 = \frac{m! \ e^{g_1} \Gamma_{\text{inc}}\left(k+\frac{3}{2}, \ g_1\right)} {g_1^{k+\frac{3}{2}} \ \lambda_{t,1}^{m+1}}, \\
I_2 &= \int\limits_{q_1=0}^{\infty} \left( \left( 1 + q_1 \right)^{k+\frac{1}{2}} e^{-\left( g_n - \lambda_{t,n} f_n \right) q_1} \ \int\limits_{q_2 = f_n q_1}^{\infty} q_2^m e^{-\lambda_{t,n} q_2} dq_2 \right) dq_1 \nonumber\\
\label{eq:I2_LCR}
&= \frac{m! e^{g_n} }{\lambda_{t,n}^{m+1}} \sum\limits_{r = 0}^m \frac{\left(f_n \lambda_{t,n}\right)^r} {r!} \sum\limits_{w=0}^r (-1)^{r-w} \binom{r}{w} \frac{\Gamma_{\text{inc}}\left( k+w+\frac{3}{2}, \ g_n \right)} {g_n^{k+w+\frac{3}{2}}},
\end{align}
under condition that $\lambda_{t,1} > 0$ and $g_n > 0$ which are discussed in detail in Appendix \ref{sec:notes}. $\Gamma_{\text{inc}}(a,x)=\int_{x}^{\infty} e^{-t} t^{y-1} dt$ is the upper incomplete Gamma function \cite[Sec. 8.350, Eq. 2]{372}. We note that the integration limits are derived based on the conditions imposed by the double inverse Laplace transform as detailed in Appendix \ref{sec:fQ1Q2_deriv_appendix} along with the detailed derivation of the above integrations.

Finally, using \eqref{eq:LCR4}, \eqref{eq:Ia5}, \eqref{eq:Ib}, \eqref{eq:I1_LCR}, and \eqref{eq:I2_LCR}, the closed-form expression of the LCR can be written as
\begin{align}
\label{eq:LCR_final}
\text{LCR} \left(\gamma_{\text{th}}\right) &= \frac{ \sqrt{2 \sigma_{\scriptscriptstyle \text{D}}^2 } \left( \frac{\gamma_{\text{th}} N_o}{p_{\scriptscriptstyle D}} \right)^{L-\frac{1}{2}} e^{-\frac{\gamma_{\text{th}} N_o}{p_{\scriptscriptstyle D}}} }   {\sqrt{\pi} \prod\limits_{n=1}^N \left(1 + \frac{\gamma_{\text{th}} p_n} {p_{\scriptscriptstyle D}}\right)} \ \Lambda \sum\limits_{k=0}^{L-1} \sum\limits_{m=0}^{L-1-k} \Xi_{k,m} \Bigg[ \sum\limits_{t=2}^N \delta_1 \psi_{t,1} \frac{e^{g_1} \Gamma_{\text{inc}}\left(k+\frac{3}{2}, g_1 \right) } {g_1^{k+\frac{3}{2}} \ \lambda_{t,1}^{m+1} } \nonumber\\
 & +  \sum\limits_{n=2}^N \sum\limits_{\substack{t=1 \\ t \neq n}}^N \sum\limits_{r=0}^m \sum\limits_{w=0}^r \delta_n \psi_{t,n} e^{g_n} \frac{(-1)^{r-w} \binom{r}{w} f_n^r \Gamma_{\text{inc}}\left(k+w+\frac{3}{2}, \ g_n \right)} {r! \ \lambda_{t,n}^{m-r+1} g_n^{k+w+\frac{3}{2}}}  \Bigg],
\end{align}
where $\Xi_{k,m}$ is defined as
\begin{align}
\label{eq:Xi_def}
\Xi_{k,m} &= \binom{L-1}{k} \ \binom{L-1-k}{m} \left( \frac{a_1}{b_1} \right)^{L-1-m} \left( \frac{b_1}{a_1} - 1 \right)^{L-1-k-m} \frac{m!}{\Gamma_{\text{s}}(L)} = \frac{a_1^k \left( b_1 - a_1 \right)^{L-1-m-k}} {k! \ \left( L-1-k-m \right)! b_1^{L-1-m} } \nonumber\\
&= \frac{1} {k! (L-1-k-m)!} \frac{1} { \varepsilon_1^k \left( 1+\frac{1}{\varepsilon_1} \right)^{L-1-m}},
\end{align}
since $\Gamma_{\text{s}}(L)=(L-1)!$.

\section{Special Cases}
\label{sec:special_cases}
\subsection{Interferers with Equal Powers and Equal Speeds}
\label{sec:Equal-Power Equal-Speed Interferers}

In case all the interferers have equal transmitting powers and move with the same speed, i.e. $ p_n=p_{\scriptscriptstyle \text{I}} $ and $ f_{n} = f_{\scriptscriptstyle \text{I}}, \forall n \in \mathcal{N}$, the LCR can be significantly simplified as follows
\begin{equation}
\label{eq:LCR_equal_pow_equal_speed}
\text{LCR}( \gamma_{\text{th}} ) = \frac{ \sqrt{2 \sigma_{\scriptscriptstyle \text{D}}^2 } \left( \frac{\gamma_{\text{th}} N_o}{p_{\scriptscriptstyle D}} \right)^{L-\frac{1}{2}} e^{-\frac{\gamma_{\text{th}} N_o}{p_{\scriptscriptstyle D}}} e^{\frac{1}{b}}}   {\sqrt{\pi} (1 + \varepsilon)^{L-1} \left(1 + \frac{\gamma_{\text{th}} p_{\scriptscriptstyle \text{I}}} {   p_{\scriptscriptstyle \text{D}}}\right)^N} \sum\limits_{l=0}^{L-1} \sum\limits_{m=0}^{N-1} \frac{ \left( -1 \right)^{N-m-1} \varepsilon^{L-l-1} } {l! \ m! \ \left(L-l-1\right)! \ \left(N-m-1\right)! }  \frac{\Gamma_{\text{inc}} \left( m+l+\frac{3}{2}, \frac{1}{b} \right)} {b^{N-m-l-\frac{3}{2}}}.
\end{equation}
We note that $a_n = a, \ b_n = b,$ and $\varepsilon_n = \varepsilon$, where they are still defined as in \eqref{eq:anbneps} but with replacing $p_n$ and $f_n$ by $p_{\scriptscriptstyle \text{I}}$ and $f_{\scriptscriptstyle \text{I}}$, respectively. The detailed derivation of \eqref{eq:LCR_equal_pow_equal_speed} is given in Appendix \ref{sec:eq_pow_eq_speed_LCR_deriv_appendix}.

\subsection{Interferers with Equal Powers and Equal Speeds in an Interference-Limited System}
\label{sec:int_limited_eqpow_eqspeed}

In an interference-limited system, the LCR expression in \eqref{eq:LCR_equal_pow_equal_speed} can be further simplified to
\begin{equation}
\label{eq:int_lim_eqpow_eqspeed}
\text{LCR} (\gamma_{\text{th}}) = \sqrt{2 \pi} \frac{ \Gamma_{\text{s}}(N+L-\frac{1}{2}) } { \Gamma_{\text{s}}(N) \Gamma_{\text{s}}(L) } \left( f_{\scriptscriptstyle \text{D}}^2 + \frac{f_{\scriptscriptstyle \text{I}}^2} {\Omega_{\text{th}}} \right)^{\frac{1}{2}} \frac{ \Omega_{\text{th}}^N } { \left( 1 + \Omega_{\text{th}} \right)^{N+L-\frac{1}{2}}},
\end{equation}
where $\Omega_{\text{th}} \triangleq p_{\scriptscriptstyle \text{D}} / \gamma_{\text{th}} p_{\scriptscriptstyle \text{I}}$. The detailed derivation of \eqref{eq:int_lim_eqpow_eqspeed} from \eqref{eq:LCR_final} is given in Appendix \ref{sec:int_lim_eq_pow_eq_speed_LCR_deriv_appendix}. It can be seen that \eqref{eq:int_lim_eqpow_eqspeed} is in agreement with (18) in \cite{520} for the Rayleigh fading scenario.

Due to the simplicity of the LCR expression of this case, we derive next an expression for the $\gamma_{\text{max}}$, which is defined as the $\gamma_{\text{th}}$ at which the maximum LCR value occurs. This would give us insight on the behavior of the LCR w.r.t. the speeds of the desired user and the interferers, i.e. their associated Doppler frequencies. Taking the derivative of \eqref{eq:int_lim_eqpow_eqspeed} w.r.t. $\gamma_{\text{th}}$, then equating it to zero results in a quadratic equation in $\gamma_{\text{th}}$ whose solution is
\begin{equation}
\label{eq:gamma_max}
\gamma_{\text{max}} = \frac{p_{\scriptscriptstyle \text{D}}} {\left( 2N-1 \right) p_{\scriptscriptstyle \text{I}}} \left( L - N \frac{ f_{\scriptscriptstyle \text{D}}^2 } {f_{\scriptscriptstyle \text{I}}^2}  + \sqrt{ \left( L - N \frac{ f_{\scriptscriptstyle \text{D}}^2 } {f_{\scriptscriptstyle \text{I}}^2} \right)^2 + \left( 2L-1 \right) \left( 2N-1 \right) \frac{ f_{\scriptscriptstyle \text{D}}^2 } {f_{\scriptscriptstyle \text{I}}^2} } \right)
\end{equation}
where the negative root was ignored since $\gamma_{\text{th}}$ has to be positive since it is an SINR value. From \eqref{eq:gamma_max} we see that as $f_{\scriptscriptstyle \text{D}}$ increases, $\gamma_{\text{max}}$ decreases, while it increases as $f_{\scriptscriptstyle \text{I}}$ increases. Moreover, the dependence on the Doppler frequencies is always in the form of the ratio $f_{\scriptscriptstyle \text{D}} / f_{\scriptscriptstyle \text{I}}$ which suggests that if the desired user's and interferers' speeds (or equivalently Doppler frequencies) increase by the same ratio, $\gamma_{\text{max}}$ does not change. This has been confirmed through simulations but the figures are not presented in this paper due to space limitation. While \eqref{eq:gamma_max} is valid for the special case of interferers with equal powers and equal speeds in interference-limited systems, nevertheless it gives us insight on its behavior in more general cases as will be shown in sections \ref{sec:desired_speed} and \ref{sec:interferer_speed}.

\subsection{Single-Antenna Receiver}
\label{sec:MERL}

For a receiver with a single antenna, $L=1$, \eqref{eq:LCR_final} simplifies to
\begin{align}
\label{eq:MERL_ours1}
\text{LCR} (\gamma_{\text{th}}) &= \sqrt{ \frac{ 2 \sigma_{\scriptscriptstyle \text{D}}^2 \gamma_{\text{th}} N_o } { \pi p_{\scriptscriptstyle \text{D}} } } \frac{ e^{-\frac{\gamma_{\text{th}} N_o} {p_{\scriptscriptstyle \text{D}}} } \Lambda } { \prod\limits_{n=1}^N \left( 1 + \frac{ \gamma_{\text{th}} p_n } { p_{\scriptscriptstyle \text{D}} } \right) } \sum\limits_{n=1}^N \delta_n \sum\limits_{\substack{t=1 \\ t \neq n}} \psi_{t,n} \frac{ e^{g_n} \Gamma_{\text{inc}}\left( \frac{3}{2}, g_n \right) } { g_n^{\frac{3}{2}} \lambda_{t,n} } \\
\label{eq:MERL_ours2}
&= \sqrt{ \frac{ 2 \sigma_{\scriptscriptstyle \text{D}}^2 \gamma_{\text{th}} N_o } { \pi p_{\scriptscriptstyle \text{D}} } } \frac{ e^{-\frac{\gamma_{\text{th}} N_o} {p_{\scriptscriptstyle \text{D}}} } } { \prod\limits_{n=1}^N \left( 1 + \frac{ \gamma_{\text{th}} p_n } { p_{\scriptscriptstyle \text{D}} } \right) } \sum\limits_{n=1}^N \delta_n^{\text{MERL}} \frac{e^{g_n}} {\sqrt{g_n}} \Gamma_{\text{inc}}\left(\frac{3}{2}, g_n\right).
\end{align}
The first equality \eqref{eq:MERL_ours1} follows from direct substitution of $L=1$ in \eqref{eq:LCR_final} and the second equality \eqref{eq:MERL_ours2} follows from some algebraic manipulation as detailed in Appendix \ref{sec:appendix_MERL}. Equation \eqref{eq:MERL_ours2} agrees with the LCR expression reported in \cite{370}.

\section{Approximated LCR Expression}
\label{sec:approx}

For the sake of simplifying the LCR expression in \eqref{eq:LCR_final} and getting insight on the behavior of the LCR w.r.t. different parameters easily, an approximate LCR is derived in this section. In \cite{505}, Fukawa et al. derived LCR expressions for a lower bound of the SIR of a point-to-point system subject to multipath with MRC for two different cases. The first case is referred to as equal average power (EAP) where the components of the desired signal and interference signals have equal average powers, i.e., $p_{\scriptscriptstyle \text{D},l}=p_{\scriptscriptstyle D}, \forall l \in \mathcal{L}$ and $p_n=p_{\scriptscriptstyle \text{I}}, \forall n \in \mathcal{N}$. The second case is referred to as unequal average power (UAP) where the components of the desired signal and interference signals have unequal average powers, i.e., $p_{\scriptscriptstyle \text{D},l}\neq p_{\scriptscriptstyle \text{D},j} \text{ for }l\neq j$ and $p_n\neq p_m \text{ for } n\neq m$. The LCR expression is exact for the EAP case while it is an approximation for the UAP case, however for both cases LCR expressions are derived for a lower bound of the SIR as mentioned earlier. Next the approach in \cite{505} is utilized to find the approximate LCR of our system. The SINR $\mathit{\Gamma}$ in \eqref{eq:SINR_def} can be rewritten as
\begin{equation}
\label{eq:SINR_def2}
\mathit{\Gamma} = \frac{p_{\scriptscriptstyle D} \sum\limits_{l=1}^L A_{\scriptscriptstyle \text{D},l}^2} {N_o + \sum\limits_{n=1}^N p_n \ A_{\scriptscriptstyle \text{I},n}^2} = \frac{R_{\scriptscriptstyle \text{D}}^2}{R_{\scriptscriptstyle \text{I}}^2},
\end{equation}
where $R_{\scriptscriptstyle \text{D}}^2$ and $R_{\scriptscriptstyle \text{I}}^2$ denote the numerator and denominator of the SINR, respectively. Further, as shown in \cite{505}, the LCR in \eqref{eq:LCR_def} can be rewritten as
\begin{align}
\label{eq:LCR_def2}
\text{LCR}&(\gamma_{\text{th}}) = \int\limits_{\dot{r}_{\scriptscriptstyle \text{D}}=-\infty}^{\infty} \int\limits_{\dot{r}_{\scriptscriptstyle \text{I}}=-\infty}^{\dot{r}_{\scriptscriptstyle \text{D}} / \sqrt{\gamma_{\text{th}}}} \int\limits_{r_{\scriptscriptstyle \text{I}}=\sqrt{N_o}}^{\infty} \left( \dot{r}_{\scriptscriptstyle \text{D}} - \sqrt{\gamma_{\text{th}}} \dot{r}_{\scriptscriptstyle \text{I}} \right) \nonumber\\
&\times f_{R_{\scriptscriptstyle \text{D}},\dot{R}_{\scriptscriptstyle \text{D}},R_{\scriptscriptstyle \text{I}},\dot{R}_{\scriptscriptstyle \text{I}}} \left( \sqrt{\gamma_{\text{th}}} r_{\scriptscriptstyle \text{I}}, \dot{r}_{\scriptscriptstyle \text{D}}, r_{\scriptscriptstyle \text{I}}, \dot{r}_{\scriptscriptstyle \text{I}} \right) \ dr_{\scriptscriptstyle \text{I}} \ d\dot{r}_{\scriptscriptstyle \text{I}} \ d\dot{r}_{\scriptscriptstyle \text{D}},
\end{align}
where $f_{R_{\scriptscriptstyle \text{D}},\dot{R}_{\scriptscriptstyle \text{D}},R_{\scriptscriptstyle \text{I}},\dot{R}_{\scriptscriptstyle \text{I}}}(r_{\scriptscriptstyle \text{D}},\dot{r}_{\scriptscriptstyle \text{D}},r_{\scriptscriptstyle \text{I}},\dot{r}_{\scriptscriptstyle \text{I}})$ is the joint PDF of the envelopes of the total desired and total interference signals and their time derivatives. It is given by
\begin{equation}
\label{eq:joint_all}
f_{R_{\scriptscriptstyle \text{D}},\dot{R}_{\scriptscriptstyle \text{D}},R_{\scriptscriptstyle \text{I}},\dot{R}_{\scriptscriptstyle \text{I}}}(r_{\scriptscriptstyle \text{D}},\dot{r}_{\scriptscriptstyle \text{D}},r_{\scriptscriptstyle \text{I}},\dot{r}_{\scriptscriptstyle \text{I}}) = f_{R_{\scriptscriptstyle \text{D}},\dot{R}_{\scriptscriptstyle \text{D}}}(r_{\scriptscriptstyle \text{D}},\dot{r}_{\scriptscriptstyle \text{D}}) f_{R_{\scriptscriptstyle \text{I}},\dot{R}_{\scriptscriptstyle \text{I}}}(r_{\scriptscriptstyle \text{I}},\dot{r}_{\scriptscriptstyle \text{I}}) \approx f_{R_{\scriptscriptstyle \text{D}}}(r_{\scriptscriptstyle \text{D}}) f_{\dot{R}_{\scriptscriptstyle \text{D}}}(\dot{r}_{\scriptscriptstyle \text{D}}) f_{R_{\scriptscriptstyle \text{I}}}(r_{\scriptscriptstyle \text{I}}) f_{\dot{R}_{\scriptscriptstyle \text{I}}}(\dot{r}_{\scriptscriptstyle \text{I}}),
\end{equation}
where the first equality follows from the fact that the desired and interfering signals are independent, while the second equality follows from the independence of each signal and its time derivative as explained in Appendix \ref{sec:deriv_of_rdot_pdf_appendix} along with the reason of using the approximation sign. The marginal PDFs in the above equation can be obtained as
\begin{align}
\label{eq:rD_pdf}
f_{R_{\scriptscriptstyle \text{D}}}(r_{\scriptscriptstyle \text{D}}) &= \frac{2}{(L-1)!} \frac{ r_{\scriptscriptstyle \text{D}}^{2L-1} } {p_{\scriptscriptstyle D}^L} e^{-\frac{r_{\scriptscriptstyle \text{D}}^2} {p_{\scriptscriptstyle D}}}, \\
\label{eq:rD_dot_pdf}
f_{\dot{R}_{\scriptscriptstyle \text{D}}}(\dot{r}_{\scriptscriptstyle \text{D}}) &= \frac{1} {\sqrt{2 \pi p_{\scriptscriptstyle D} \sigma_{\scriptscriptstyle \text{D}}^2 }} e^{-\frac{\dot{r}_{\scriptscriptstyle \text{D}}^2} {2 p_{\scriptscriptstyle D} \sigma_{\scriptscriptstyle \text{D}}^2}}, \\
\label{eq:rI_pdf}
f_{R_{\scriptscriptstyle \text{I}}}(r_{\scriptscriptstyle \text{I}}) &= \sum\limits_{n=1}^N \frac{2 \mu_n }{p_n} r_{\scriptscriptstyle \text{I}} e^{-\frac{r_{\scriptscriptstyle \text{I}}^2} {p_n}}, \ \ \ \ \ \ \ r_{\scriptscriptstyle \text{I}} \geq \sqrt{N_o}, \text{ and} \\
\label{eq:rdot_pdf}
f_{\dot{R}_{\scriptscriptstyle \text{I}}}(\dot{r}_{\scriptscriptstyle \text{I}}) &= \frac{1} {\sqrt{2 \pi \dot{\sigma}_{\scriptscriptstyle \text{I}}^2}} e^{-\frac{\dot{r}_{\scriptscriptstyle \text{I}}^2}{2 \dot{\sigma}_{\scriptscriptstyle \text{I}}^2}},
\end{align}
where
\begin{equation}
\label{eq:mu}
\mu_n = \prod\limits_{\substack{k=1\\ k\neq n}}^N \frac{p_n}{p_n-p_k}, \text{ and} \ \ \ \ \ \ \ \ \ \dot{\sigma}_{\scriptscriptstyle \text{I}}^2 = \frac{\pi^2 \sum\limits_{n=1}^N p_n^2 f_{\scriptscriptstyle \text{I},n}^2} { N_o + \sum\limits_{n=1}^N p_n }.
\end{equation}
The PDFs (\ref{eq:rD_pdf}-\ref{eq:rD_dot_pdf}) can be obtained by properly adapting the variables in equations (23) and (24) of the EAP case obtained in \cite{505} to match the notations in this paper. Note that an important difference between the system in this paper and in \cite{505} is that here multiple interferers possess multiple maximum Doppler frequencies due to different speeds while in \cite{505} there is one maximum Doppler frequency only since all components belong to the same interfering user. Another important difference is that here a more general case is considered by obtaining the LCR of the SINR while in \cite{505} the LCR of the SIR is obtained, i.e. the system under consideration in \cite{505} is an interference-limited system. The details on the derivation of (\ref{eq:rD_pdf}-\ref{eq:rdot_pdf}) can be found in Appendix \ref{sec:deriv_of_rdot_pdf_appendix}.

By substituting \eqref{eq:joint_all} into \eqref{eq:LCR_def2}, it is found that $\text{LCR}(\gamma_{\text{th}}) = AB$, where
\begin{align}
\label{eq:AB_def}
A &= \int\limits_{\dot{r}_{\scriptscriptstyle \text{D}}=-\infty}^{\infty} f_{\dot{R}_{\scriptscriptstyle \text{D}}}(\dot{r}_{\scriptscriptstyle \text{D}}) \int\limits_{\dot{r}_{\scriptscriptstyle \text{I}}=-\infty}^{\dot{r}_{\scriptscriptstyle \text{D}} / \sqrt{\gamma_{\text{th}}}} \left( \dot{r}_{\scriptscriptstyle \text{D}} - \sqrt{\gamma_{\text{th}}} \dot{r}_{\scriptscriptstyle \text{I}} \right) f_{\dot{R}_{\scriptscriptstyle \text{I}}}(\dot{r}_{\scriptscriptstyle \text{I}}) \ d\dot{r}_{\scriptscriptstyle \text{I}} \ d\dot{r}_{\scriptscriptstyle \text{D}}, \text{ and} \nonumber\\
B & = \int\limits_{r_{\scriptscriptstyle \text{I}}=\sqrt{N_o}}^{\infty} f_{R_{\scriptscriptstyle \text{D}}}(\sqrt{\gamma_{\text{th}}} r_{\scriptscriptstyle \text{I}}) f_{R_{\scriptscriptstyle \text{I}}}(r_{\scriptscriptstyle \text{I}}) \ dr_{\scriptscriptstyle \text{I}}.
\end{align}
$A$ can be rewritten as $A=I_3 - I_4$ where
\begin{align}
\label{eq:I3}
I_3 &= \int\limits_{\dot{r}_{\scriptscriptstyle \text{D}}=-\infty}^{\infty} \dot{r}_{\scriptscriptstyle \text{D}} f_{\dot{R}_{\scriptscriptstyle \text{D}}}(\dot{r}_{\scriptscriptstyle \text{D}}) \int\limits_{\dot{r}_{\scriptscriptstyle \text{I}}=-\infty}^{\dot{r}_{\scriptscriptstyle \text{D}} / \sqrt{\gamma_{\text{th}}}} f_{\dot{R}_{\scriptscriptstyle \text{I}}}(\dot{r}_{\scriptscriptstyle \text{I}}) \ d\dot{r}_{\scriptscriptstyle \text{I}} \ d\dot{r}_{\scriptscriptstyle \text{D}} = \frac{\pi^2 f_{\scriptscriptstyle \text{D}}^2 p_{\scriptscriptstyle D}} {\sqrt{2 \pi^3 f_{\scriptscriptstyle \text{D}}^2 p_{\scriptscriptstyle D} + 2 \pi \gamma_{\text{th}} \dot{\sigma}_{\scriptscriptstyle \text{I}}^2}}, \text{ and} \\
\label{eq:I4}
I_4 &= \sqrt{\gamma_{\text{th}}} \int\limits_{\dot{r}_{\scriptscriptstyle \text{D}}=-\infty}^{\infty} f_{\dot{R}_{\scriptscriptstyle \text{D}}}(\dot{r}_{\scriptscriptstyle \text{D}}) \int\limits_{\dot{r}_{\scriptscriptstyle \text{I}}=-\infty}^{\dot{r}_{\scriptscriptstyle \text{D}} / \sqrt{\gamma_{\text{th}}}} \dot{r}_{\scriptscriptstyle \text{I}} f_{\dot{R}_{\scriptscriptstyle \text{I}}}(\dot{r}_{\scriptscriptstyle \text{I}}) \ d\dot{r}_{\scriptscriptstyle \text{I}} \ d\dot{r}_{\scriptscriptstyle \text{D}} = \frac{ -\gamma_{\text{th}} \dot{\sigma}_{\scriptscriptstyle \text{I}}^2 } { \sqrt{2 \pi^3 f_{\scriptscriptstyle \text{D}}^2 p_{\scriptscriptstyle D} + 2 \pi\gamma_{\text{th}} \dot{\sigma}_{\scriptscriptstyle \text{I}}^2} }.
\end{align}
The above derivation follows similar steps as in \cite[Appendix B]{505} and is not included here for the sake of brevity. Subtracting \eqref{eq:I4} from \eqref{eq:I3}, $A$ is given as
\begin{equation}
\label{eq:A_def}
A = \sqrt{\frac{\pi^2 f_{\scriptscriptstyle \text{D}}^2 p_{\scriptscriptstyle D} + \gamma_{\text{th}} \dot{\sigma}_{\scriptscriptstyle \text{I}}^2} {2 \pi} }.
\end{equation}
However, for the derivation of $B$, a different approach from \cite{505} is needed as shown next. By directly substituting $f_{R_{\text{D}}}(r_{\scriptscriptstyle \text{D}})$ \eqref{eq:rD_pdf} and $f_{R_{\text{I}}}(r_{\scriptscriptstyle \text{I}})$ \eqref{eq:rI_pdf} in \eqref{eq:AB_def}, respectively, $B$ can be obtained as
\begin{align}
\label{eq:B1}
B &= \frac{2 \ \gamma_{\text{th}}^{L-0.5}} {(L-1)! \ p_{\scriptscriptstyle D}^L} \sum\limits_{n=1}^N \frac{2 \mu_n e^{\frac{N_o}{p_n}} }{p_n} \int\limits_{r_{\scriptscriptstyle \text{I}}=\sqrt{N_o}}^{\infty} r_{\scriptscriptstyle \text{I}}^{2L} \ e^{-\frac{\gamma_{\text{th}} p_n + p_{\scriptscriptstyle D}} {p_{\scriptscriptstyle D} p_n} r_{\scriptscriptstyle \text{I}}^2} \ dr_{\scriptscriptstyle \text{I}} \\
\label{eq:B_def}
&= \frac{2 \sqrt{p_{\scriptscriptstyle D}} \gamma_{\text{th}}^{L-\frac{1}{2}} } {(L-1)!} \sum\limits_{n=1}^N \frac{ \mu_n p_n^{L-\frac{1}{2}} e^{\frac{N_o}{p_n}} \Gamma_{\text{inc}}\left(L+\frac{1}{2}, \frac{ \gamma_{\text{th}} N_o } {p_{\scriptscriptstyle \text{D}}} + \frac{N_o}{p_n} \right) } {\left( \gamma_{\text{th}} p_n + p_{\scriptscriptstyle D} \right)^{L+\frac{1}{2}}},
\end{align}
where \cite[Sec. 3.381, Eq. 9]{372} is used to solve the integral in \eqref{eq:B1}.

Hence an approximate closed-form expression for the LCR of the SINR given in \eqref{eq:SINR_def} is obtained by the multiplication of \eqref{eq:A_def} and \eqref{eq:B_def}. Thus, denoting the approximate LCR given an SINR threshold,$\gamma_{\text{th}}$ as $\text{LCR}_{\text{approx}}(\gamma_{\text{th}})$, it can be written as
\begin{equation}
\label{eq:LCR_approx}
\text{LCR}_{\text{approx}}(\gamma_{\text{th}}) = \frac{2} {(L-1)!}  \sqrt{\frac{\pi^2 f_{\scriptscriptstyle \text{D}}^2 p_{\scriptscriptstyle D}^2 + \gamma_{\text{th}} p_{\scriptscriptstyle D} \dot{\sigma}_{\scriptscriptstyle \text{I}}^2} {2 \pi} }  \sum\limits_{n=1}^N \frac{ \mu_n e^{\frac{N_o}{p_n}} \Gamma_{\text{inc}}\left(L+\frac{1}{2}, \frac{\gamma_{\text{th}} N_o}{p_{\scriptscriptstyle \text{D}}}  \left( 1 + \frac{p_{\scriptscriptstyle \text{D}}} { \gamma_{\text{th}} p_n } \right) \right) } {\gamma_{\text{th}} p_n \left( 1 + \frac{p_{\scriptscriptstyle D}} {\gamma_{\text{th}} p_n} \right)^{L+\frac{1}{2}}}.
\end{equation}


As mentioned earlier, this approximate expression provides simplicity to get more insight on the behavior of the LCR w.r.t. different parameters easily. For example it is much easier to plot the behavior of the LCR w.r.t. $N$ and $L$ from the approximate expression \eqref{eq:LCR_approx} than from the exact one \eqref{eq:LCR_final}.

\section{Validation of Theoretical Results}
\label{sec:LCR_num_results}

In this section we demonstrate the validity and accuracy of both the exact and approximated LCR expressions derived in this paper, \eqref{eq:LCR_final} and \eqref{eq:LCR_approx} respectively, through comparing them with simulation results for different scenarios. We note that due to the complexity of the LCR expression and the fact that it is highly non-linear function in these parameters, and the abundance of parameters affecting the LCR and consequently the very wide variety of combinations of these parameters that need to be investigated, it takes a lot more examples and complex cross-analysis to investigate the interrelated effects of the different parameters of the system on the LCR's behavior. Since our main focus is to validate the novel theoretical LCR expressions presented in this paper and not the cross-analysis of the effect of the system's parameters on the LCR, we suffice by presenting a few examples showing some interesting observations on the effect of the powers and speeds of the desired user and interferers on the LCR behavior, while detailed analysis of the effect of different system's parameters on the LCR is left for future work. To show the validity of our LCR expressions, specially the exact one \eqref{eq:LCR_final}, we simulate three different systems with different number of Rx antennas $L$, interferers $N$, powers $p_{\scriptscriptstyle \text{D}}$ and $p_{\scriptscriptstyle \text{I}}$, Doppler frequencies $f_{\scriptscriptstyle \text{D}}$ and $f_{\scriptscriptstyle \text{I}}$, and SNR $p_{\scriptscriptstyle \text{D}} / N_o$. The values of the aforementioned parameters are chosen randomly across a range of possible values in practical cellular system.

For the simulation results, the Rayleigh fading channels are generated using the method described in \cite{507} which is a variation of Jake's method. The simulations are performed using Matlab R2014b, and the results are averaged over $400$ channel realizations each for time duration of $5$ sec with $1$ MHz sampling rate which is much larger than the maximum Doppler frequencies simulated, hence, capturing the channel's variations. It is also worth mentioning that since we are performing baseband simulations, then the sampling frequency's value is determined by considering the simulated Doppler frequencies' values only and not the carrier frequency's.

\subsection{Effect of Desired User and Interferers Powers}
\label{section:effect_pD_pI}

\begin{figure}[!htb]
        \centering
        \includegraphics[trim=1.5cm 6.5cm 2cm 6.5cm,clip=true,width=3.5in]{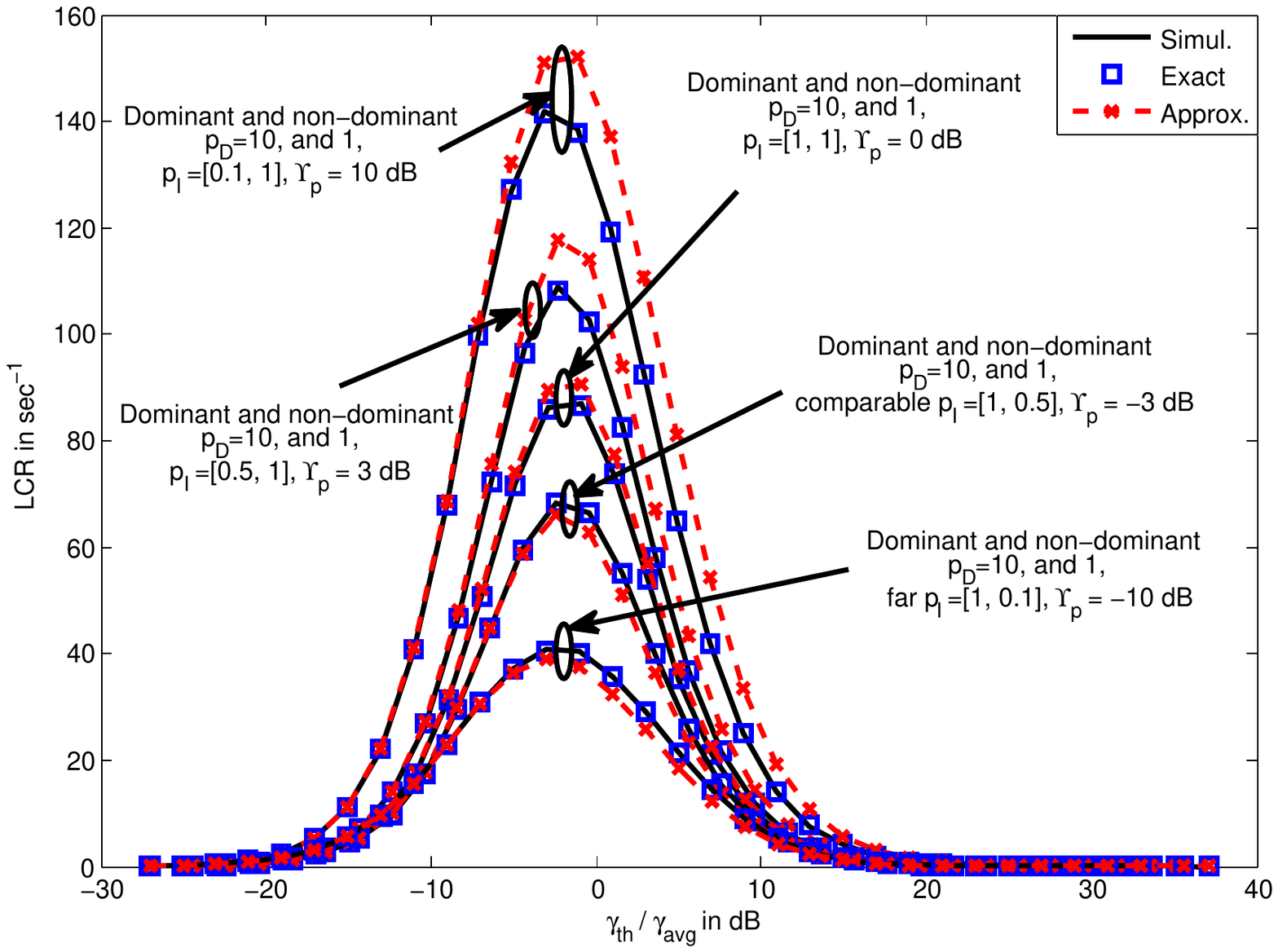}
        \caption{The effect of power values on the LCR where $L=2$, $N=2$, $f_{\scriptscriptstyle \text{D}}=32$ Hz, and $f_{\scriptscriptstyle \text{I}} = [32, 162]$ Hz.}
        \label{fig:pD_and_pn_effect_All}
\end{figure}

In Fig. \ref{fig:pD_and_pn_effect_All}, the LCR is plotted for an interference-limited system where $L=2, N=2$, for different values of the interferers' powers, $p_{\scriptscriptstyle \text{I}}$, and different values of the desired Tx power $p_{\scriptscriptstyle \text{D}}$. In particular, five different combinations of interferers' powers are plotted, namely when the interferers' powers ratios, $\Upsilon_{\text{p}} = p_{\scriptscriptstyle \text{I},2} / p_{\scriptscriptstyle \text{I},1}$ are equal to $-10$ dB, $-3$ dB, $0$ dB, $3$ dB, and $10$ dB. Each of these cases is simulated for two values of $p_{\scriptscriptstyle \text{D}}$, namely $10$, and $1$, corresponding to dominant and non-dominant desired Tx power cases, respectively, w.r.t. the maximum of the interferers powers. In all systems, the maximum Doppler frequencies of the interferers, $f_{\scriptscriptstyle \text{I},n}$, are $32$ and $162$ Hz, for $n=1,2$, respectively, while that of the desired user, $f_{\scriptscriptstyle \text{D}}$, is set to the minimum of both, i.e. $32$ Hz. These frequencies correspond to vehicular speeds of $10$ and $50$ km/hr, respectively, for a cellular system with carrier frequency $3.5$ GHz, which is band $22$ in the LTE FDD bands \cite[Table 5.5-1]{601}.

The LCR is plotted versus the SINR threshold $\gamma_{\text{th}}$ normalized w.r.t. the average SINR $\gamma_{\text{avg}}$ which is obtained through simulations. In the figure, we refer to the LCR obtained by equations \eqref{eq:LCR_final} and \eqref{eq:LCR_approx} as `Exact' and `Approx.', respectively. Fig. \ref{fig:pD_and_pn_effect_All} shows that the exact LCR matches perfectly the LCR values obtained from simulations. On the other hand, there is a deviation between the approximated LCR and the exact LCR.

We also note that as $p_{\scriptscriptstyle \text{D}}$ changes, the average SINR value changes, however, the LCR value does not change for a fixed $\gamma_{\text{th}} / \gamma_{\text{avg}}$. This can be seen in Fig. \ref{fig:pD_and_pn_effect_All} where the LCR curves are the same for both dominant and non-dominant desired user transmitting power cases, where $p_{\scriptscriptstyle \text{D}}=10$ and $1$, respectively.

Fig. \ref{fig:pD_and_pn_effect_All} indicates that as the power ratio, $\Upsilon_p$, increases, the LCR increases. This can be explained as follows. In general as the power of an interferer increases, its effect becomes more pronounce. Since the second interferer has a higher Doppler frequency, hence when its power increases, more fluctuation is induced in the resultant received signal leading to more frequent crossings around a given threshold. This shows in the cases $\Upsilon_{\text{p}} = -10$ dB, $-3$ dB, and $0$ dB, where $p_{\scriptscriptstyle \text{I},1}$ is fixed to $1$ and $p_{\scriptscriptstyle \text{I},2}$ increases. In the cases $\Upsilon_{\text{p}}=0$ dB, $3$ dB, and $10$ dB, still the LCR increases yet it is because the power $p_{\scriptscriptstyle \text{I},1}$ of the slower first interferer decreases while $p_{\scriptscriptstyle \text{I},1}$ is fixed to 1. This observation can be stated in other words: the LCR increases as the ratio of the faster interferer's power to the slower interferer's power increases, and vice versa.

In order to analyze the gap between the approximated and exact LCR and how it is affected by the system parameters, we define a metric named \textit{maximum relative gap}, MRG, as
\begin{equation}
\text{MRG} = \frac{\max\left| \text{LCR}_{\text{exact}} - \text{LCR}_{\text{approx}} \right|} {\text{LCR}_{\text{exact, MRG}}},
\end{equation}
where $\text{LCR}_{\text{exact, MRG}}$ is the exact LCR at which the difference between the exact and approximate LCR is maximum.

Table \ref{table_MRG_pD_effect} lists the values of MRG for the cases plotted in Fig. \ref{fig:pD_and_pn_effect_All}. It can be seen from the table that as the difference in the powers between the interferers increases, the MRG increases as well, indicating that the accuracy of the approximation declines. This is because the approximated LCR assumes that the time derivative of the interference signal is independent of the interference signal itself which is not accurate, \emph{unless} all of the interferers have the same transmitting powers and speeds. Thus, for given speeds of the interferers, as the interferers' powers get closer in values, the assumption of independence of the two signals, and consequently the approximation made within the derivation, become more accurate, and vice versa. This shows in the low MRG values that the equal-power case exhibits.

It is also noticed that for the same interferers' powers ratio, the MRG values are larger when the faster interferer has larger power compared to the opposite case. This is obvious when comparing the MRG value in cases of $\Upsilon_{\text{p}}=-10$ dB and $10$ dB together and comparing the cases where $\Upsilon_{\text{p}}=-3$ dB and $3$ dB together. Again this is attributed to the assumption made during the derivation of the approximated LCR, where as the speeds and powers of the interferers get closer in value, i.e. more homogeneous interferers, the more accurate the assumption mentioned in the previous paragraph is. One metric of the interferers homogeneity is the maximum product of the interferer's power and speed, $\max\limits_{n} p_{\scriptscriptstyle \text{I},n} f_{\scriptscriptstyle \text{I}, n}$. Hence it shows that as the power-frequency product increases like the case of $\Upsilon_{\text{p}} = 10$ dB, the more inhomogeneous the interferers are compared to the case of $\Upsilon_{\text{p}} = -10$ dB, and hence the greater the MRG and the less accurate the approximated LCR expression is. On another note, it is seen that still the approximated LCR provides reasonable accuracy over the relatively low normalized threshold range, $\gamma_{\text{th}} / \gamma_{\text{avg}} < 0$ dB,  as shown from Fig. \ref{fig:pD_and_pn_effect_All}, noting that the high values of MRG tabulated in Table \ref{table_MRG_pD_effect} usually occur at relatively high normalized thresholds.

\begin{table}[!htb]
\centering
\caption{MRG values for the system in Fig. \ref{fig:pD_and_pn_effect_All} }
\label{table_MRG_pD_effect}
\begin{tabular}{|c|c|c|c|c|c|}
  \hline
  $\Upsilon_{\text{p}}$ & -10 dB & -3 dB & 0 dB & 3 dB & 10 dB \\
  \hline
  MRG & 11.4\% & 7.3\% & 3.7\% & 11.5\% & 19.9\% \\
  \hline
\end{tabular}
\end{table}

\subsection{Effect of Desired User's Speed}
\label{sec:desired_speed}

\begin{figure}[!htb]
        \centering
        \includegraphics[trim=1.5cm 6.5cm 2cm 6.5cm,clip=true,width=3.5in]{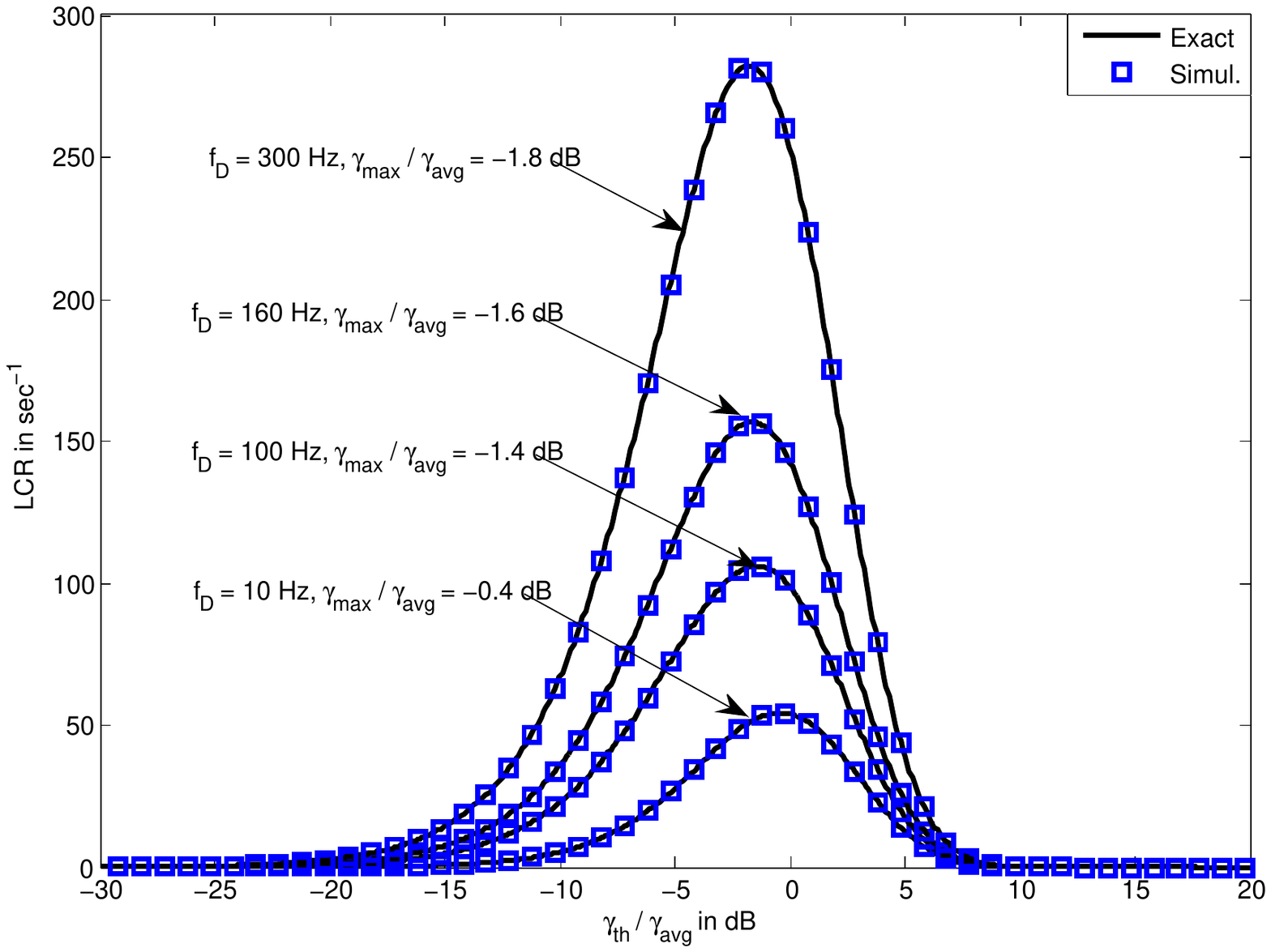}
        \caption{The effect of $f_{\scriptscriptstyle \text{D}}$ values on the LCR where $L=2$, $N=4$, and $f_{\scriptscriptstyle \text{I}} = [30, 160, 65, 100]$ Hz, $p_{\scriptscriptstyle \text{D}}=10$, $p_{\scriptscriptstyle \text{I}} = [0.07, 0.1, 0.05, 0.12]$ and $p_{\scriptscriptstyle \text{D}} / N_o = 20$ dB.}
        \label{fig:fD_effect_Near_pn}
\end{figure}

In this subsection, the effect of the desired user's speed of motion, i.e. its maximum Doppler frequency, is investigated. \mbox{Fig. \ref{fig:fD_effect_Near_pn}} plots the exact and the simulated LCR for different speeds of the desired user in an $L=2, N=4$ system where $p_{\scriptscriptstyle \text{D}} = 10$ while the interferers' powers are $p_{\scriptscriptstyle \text{I}}=[0.07, 0.1, 0.05, 0.12]$ and the SNR $p_{\scriptscriptstyle \text{D}} / N_o$. This setting is chosen such that the desired Tx power is dominant w.r.t. the interferers' to ensure that the effect of the desired user's speed is evident. On the other hand, the choice of the interferers' powers is irrelevant since they are fixed over the different values of $f_{\scriptscriptstyle \text{D}}$, and indeed the same behavior was observed when the system was simulated for different combinations of $p_{\scriptscriptstyle \text{I}}$. The corresponding noise power, $N_o = 0.1$, is chosen that way to be comparable to the interferers' powers to demonstrate the accuracy of the LCR expression \eqref{eq:LCR_final} when taking the noise power into consideration. The figure again shows that the LCR values obtained from equation \eqref{eq:LCR_final} match the simulations. The curves of the approximated LCR obtained from equation \eqref{eq:LCR_approx} are not plotted for the sake of clarity of the figure, however, it is worth mentioning that they are still close to the exact LCR curves.

The values of $f_{\scriptscriptstyle \text{D}}$ plotted correspond to the speed values of approximately $3, 30, 50, $ and $100$ km/hr, hence, ranging from a pedestrian Tx to a high-speed moving vehicle. The figure shows that for a given combination of interferers' speeds, as the desired user's speed increases, the LCR increases. This is because more fluctuation is introduced to the received signal due to the higher Doppler frequency of the desired Tx, causing faster crossings around a given threshold.

An interesting finding is that as the desired user's Doppler frequency increases, the normalized threshold at which the maximum LCR occurs, $\gamma_{\text{max}} / \gamma_{\text{th}}$, decreases, i.e. shifts to the left as Fig. \ref{fig:fD_effect_Near_pn} indicates. This is shown explicitly from the values of $\gamma_{\text{max}} / \gamma_{\text{th}}$ mentioned for each curve. This is in accordance with \eqref{eq:gamma_max} though this equation is derived for interference-limited systems with equal-power and equal-speed interferers. This indicates that although the analytical expression of $\gamma_{\text{max}}$ is different and more complicated in case of systems with significant noise power and interferers with unequal powers and speeds, a similar behavior is observed as in the simplified system of \eqref{eq:gamma_max}, hence it provides a useful insight on the general LCR behavior. It is worth mentioning that a high value of LCR indicates that the signal crosses the associated threshold frequently which implies that the received signal spends more time in the vicinity. Consequently, the figure shows that as the desired user's speed increases, the received signal spends more time at a lower SINR value. This indicates that the threshold should be lowered as $f_{\scriptscriptstyle \text{D}}$ increases to maintain same outage probability.

\subsection{Effect of Interferers Speeds}
\label{sec:interferer_speed}

In this subsection, the LCR is simulated for a system with $L=3$ and $N=2$ while fixing $p_{\scriptscriptstyle \text{D}}=1$, and $f_{\scriptscriptstyle \text{D}} = 100$ Hz. The interferers have close powers where $p_{\scriptscriptstyle \text{I}} = [0.1, 0.09]$, and equal Doppler frequencies $f_{\scriptscriptstyle \text{I},1} = f_{\scriptscriptstyle \text{I},2} = f_{\scriptscriptstyle \text{I}}$, while the SNR $p_{\scriptscriptstyle \text{D}} / N_o = 10$ dB. The Tx power, $p_{\scriptscriptstyle \text{D}}$, could have been chosen differently and still the same behavior would be observed as long as it is fixed over the different combinations of the interferers' speeds simulated. Again, the curves of the approximated LCR obtained from equation \eqref{eq:LCR_approx} are not plotted for the sake of clarity of the figure, however, it is worth mentioning that they are still close to the exact LCR curves.

\begin{figure}[!htb]
        \centering
        \includegraphics[trim=1.5cm 6.5cm 2cm 6.5cm,clip=true,width=3.5in]{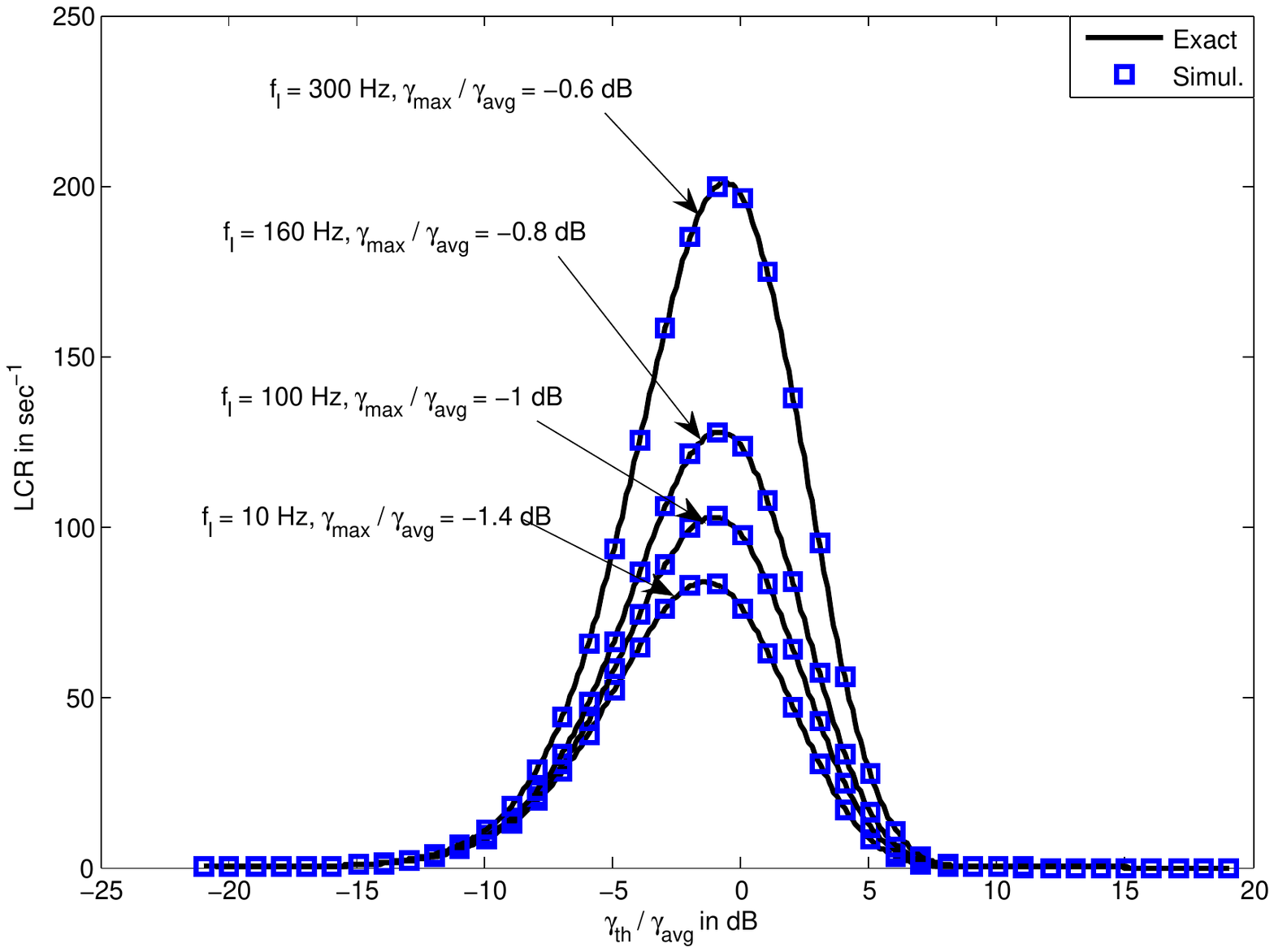}
        \caption{The effect of $f_{\scriptscriptstyle \text{I}}$ values on the LCR where $L=3$, $N=2$, $f_{\scriptscriptstyle \text{D}} = 100$ Hz, $f_{\scriptscriptstyle \text{I,1}} = f_{\scriptscriptstyle \text{I,2}} = f_{\scriptscriptstyle \text{I}}$, $p_{\scriptscriptstyle \text{D}}=1$, $p_{\scriptscriptstyle \text{I}} = [0.1, 0.09]$, and $p_{\scriptscriptstyle \text{D}} / N_o = 10$ dB.}
        \label{fig:fI_effect_veh_Vclose_pn}
\end{figure}

In Fig. \ref{fig:fI_effect_veh_Vclose_pn}, the LCR is plotted for fours systems where the speeds of the interferers are given by approximately $3, 30, 50,$ and $100$ km/hr. Again the figure shows that as the speeds of the interferers increase, the LCR increases, however, the normalized threshold at which the maximum LCR value occurs increases as opposed to the case in the last subsection. Again this is in accordance with \eqref{eq:gamma_max} which was derived for interference-limited equal-power and equal-speed interferers which confirms that this special case is insightful even for cases with medium to low SNR and not interference-limited as in this system ,where $N_o = 0.1$.

\section{AOD Derivation}
\label{sec:AOD_deriv}

In this section, the AOD of the system under study is derived. As AOD is defined in \eqref{eq:AOD_def}, the CDF $F_{\mathit{\Gamma}}(\gamma)$ needs to be obtained. In \cite{538}, the outage probability $\text{Prob}\left\{\mathit{\Gamma} < \gamma_{\text{th}}\right\}=F_{\mathit{\Gamma}}(\gamma_{\text{th}})$ was derived for Rayleigh distributed desired signal and Nakagami-m distributed interference signals with distinct unequal powers. By substituting $m=1$ in \cite{538}, where $m$ is the Nakagami parameter, and adapting the notations, the CDF of our system $F_{\mathit{\Gamma}}(\gamma_{\text{th}})$ is given as
\begin{align}
\label{eq:CDF}
F_{\mathit{\Gamma}}(\gamma_{\text{th}}) &= 1 - \sum\limits_{n=1}^N \sum\limits_{m=0}^{L-1} p_{\scriptscriptstyle D} c_n \gamma_{\text{th}}^m e^{-\gamma_{\text{th}} N_o / p_{\scriptscriptstyle D}} \sum\limits_{l=0}^m \frac{1}{(m-l)!} \left( \frac{N_o}{p_{\scriptscriptstyle D}} \right)^{m-l} \frac{p_n^l} {\left( \gamma_{\text{th}} p_n + p_{\scriptscriptstyle D} \right)^{l+1}} \nonumber\\
&= 1 - \sum\limits_{n=1}^N \sum\limits_{m=0}^{L-1} p_{\scriptscriptstyle D} c_n e^{ N_o / p_n} \frac{ \gamma_{\text{th}}^m \ p_n^m} {m! \ \left( \gamma_{\text{th}} p_n + p_{\scriptscriptstyle D} \right)^{m+1}} \Gamma_{\text{inc}} \left(m+1, \frac{\gamma_{\text{th}} N_o}{p_{\scriptscriptstyle D}} + \frac{N_o}{p_n} \right),
\end{align}
where the second equality follows from some algebraic manipulations, and from the following series representation of the upper incomplete gamma function $\Gamma_{\text{inc}}(m+1,x)$ \cite[Sec. 8.352, Eq. 2]{372}
\begin{equation}
\Gamma_{\text{inc}}\left(m+1,x\right) = m! e^{-x} \sum\limits_{l=0}^m \frac{x^l}{l!},
\end{equation}
and $c_n$ is defined as
\begin{equation}
c_n = \prod\limits_{\substack{k=1 \\ k \neq n}}^N \frac{p_n} {p_n - p_k}.
\end{equation}
By substituting \eqref{eq:CDF} and \eqref{eq:LCR_final} in \eqref{eq:AOD_def}, an exact closed-form expression for the AOD is achieved.

\begin{figure}[!htb]
        \centering
        \includegraphics[trim=1.5cm 6.5cm 2cm 6.5cm,clip=true,width=3.5in]{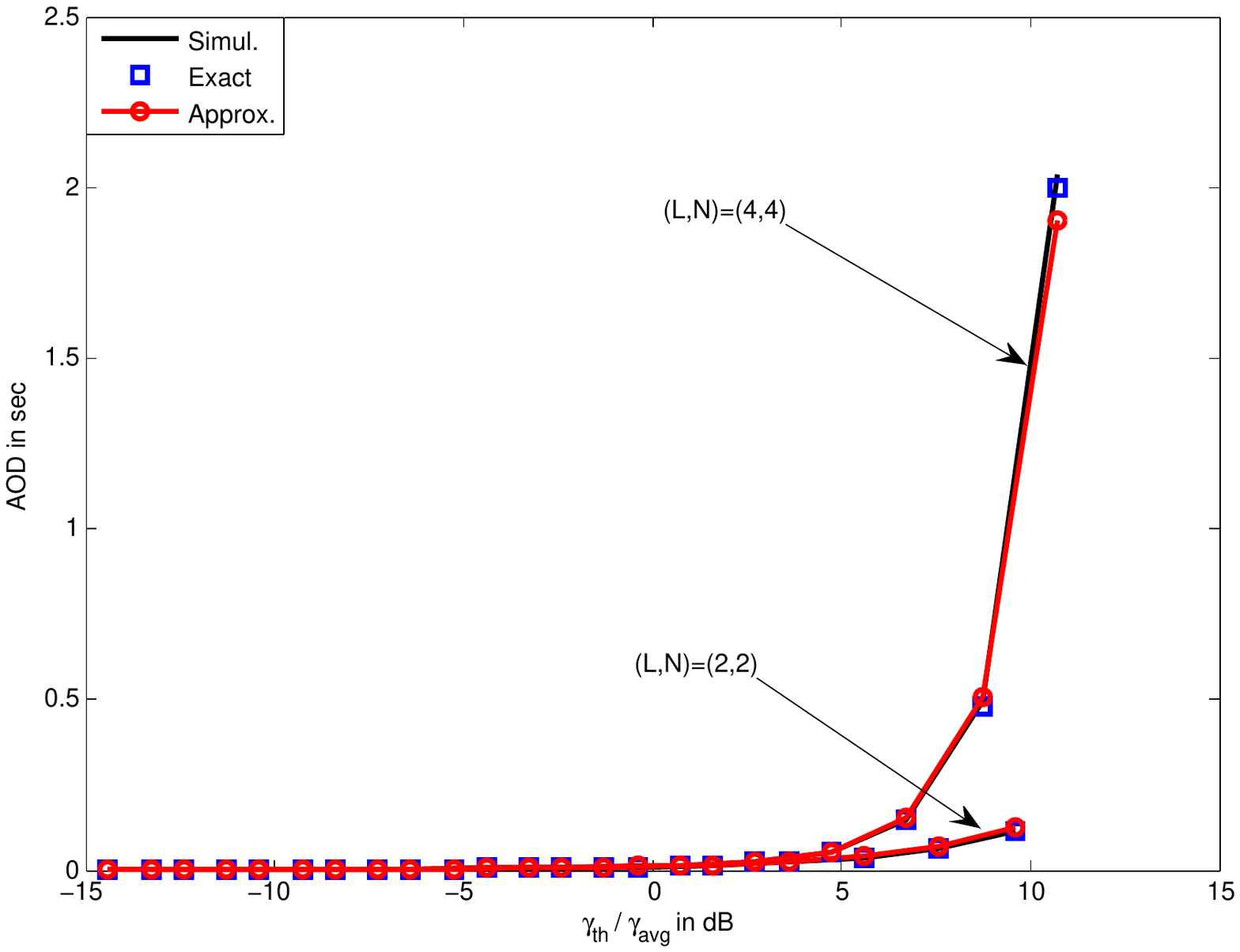}
        \caption{The AOD vs normalized threshold for interference-limited systems $(L,N)=(2,2)$, and $(L,N)=(4,4)$, $f_{\scriptscriptstyle \text{D}} = 32$ Hz, $f_{\scriptscriptstyle \text{I}} = [32, 162, 65, 97]$ Hz, $p_{\scriptscriptstyle \text{D}}=1$, $p_{\scriptscriptstyle \text{I}} = [1, 0.5, 0.9, 0.3]$, and $p_{\scriptscriptstyle \text{D}} / N_o = 50$ dB.}
        \label{fig:AOD_L2N2_and_L4N4_Nearpn}
\end{figure}

Fig. \ref{fig:AOD_L2N2_and_L4N4_Nearpn} plots the AOD vs the normalized threshold for the interference-limited systems with $(L,N)=(2,2)$ and $(L,N)=(4,4)$. The maximum Doppler frequencies of the interferers are $f_{\scriptscriptstyle \text{I}}=[32,162, 65, 97]$ Hz, selected across a range of possible Doppler shifts in practical cellular system. Their corresponding transmitting powers are $p_{\scriptscriptstyle \text{I}}=[1,0.5, 0.8, 0.3]$, while for the desired user $f_{\scriptscriptstyle \text{D}}=32$ Hz and $p_{\scriptscriptstyle \text{D}}=1$. The two-interferer system consists of the first two interferers only. The figure compares the AOD obtained from the simulation with the exact and approximate AOD expressions resulting from the exact and approximate LCR expressions \eqref{eq:LCR_final} and \eqref{eq:LCR_approx}, respectively. As expected, the AOD increases monotonically as the threshold increases since the probability of the received signal SINR being greater than the threshold decreases as the threshold increases. This results in spending more time in fade which in turn increases the AOD. We note that the exact expression matches almost perfectly the simulation results whereas the approximate solution is very close to the simulation as well. It is also observed that as $L$ and $N$ increase, the AOD increases as well as also observed in \cite{505}.

\section{Application 1: PER Calculation}
\label{sec:FSMC}

\begin{figure}[!htb]
        \centering
        \includegraphics[trim=1.5cm 6.5cm 2cm 6.5cm,clip=true,width=3.5in]{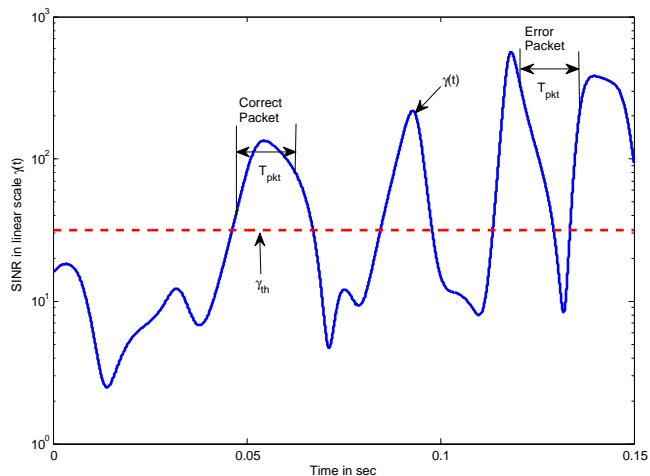}
        \caption{The received SINR over a time interval where $L=2,N=2, \gamma_{\text{avg}}=\gamma_{\text{th}}=15$ dB, and $f_{\scriptscriptstyle \text{D}}=f_{\text{I,max}}=35$ Hz.}
        \label{fig:SINR_env}
\end{figure}

In this section, the PER is derived for uncoded transmission. Ideally, the calculation of PER requires the demodulation of all bits in the packet to determine if all bits in the packet are received correctly, else the packet is considered erroneous. However, this mandates a large computational time in a conventional computer simulation. Hence for simplicity, the data packet is considered to be received correctly only if the received SINR is above a certain threshold $\gamma_{\text{th}}$ during the whole data packet duration $T_{\text{pkt}}$ \cite{505}, as shown in Fig. \ref{fig:SINR_env}. If at any instant, the SINR drops below the threshold, the received data packet is considered to be erroneous. This process can be modeled by using the two-state Markov model \cite{505}. The system is in good state $G$ if the SINR is greater than or equal to the threshold $\gamma_{\text{th}}$, which can identify a target PER. The system is in bad state $B$ otherwise. Hence, PER $P_e$ is the probability of being in state $B$. In other words, $P_e = P_b = 1 - P_g$, where $P_b$ and $P_g$ are the probabilities of being in states $B$ and $G$, respectively, also known as steady state probabilities. $P_g$ is given as \cite{505}
\begin{equation}
\label{eq:Pg}
P_g = P_{\scriptscriptstyle \text{CF}} e^{-N_c T_{\text{pkt}} / P_{\scriptscriptstyle \text{CF}}},
\end{equation}
where $P_{\scriptscriptstyle \text{CF}} = 1-F_{\mathit{\Gamma}}(\gamma_{\text{th}})$ is the complementary CDF. To simplify the notation, define $N_c = \text{LCR}(\gamma_{\text{th}})$. Substituting the LCR and the CDF derived in previous subsections, the PER of the uncoded system is obtained. It is worth mentioning that the PER considers the channel time variations since it is derived using the FSMC model which captures the channel time variations and correlation, second-order channel statistics, through LCR.

\begin{figure}[!htb]
        \centering
        \includegraphics[trim=1.5cm 6.5cm 2cm 6.5cm,clip=true,width=3.5in]{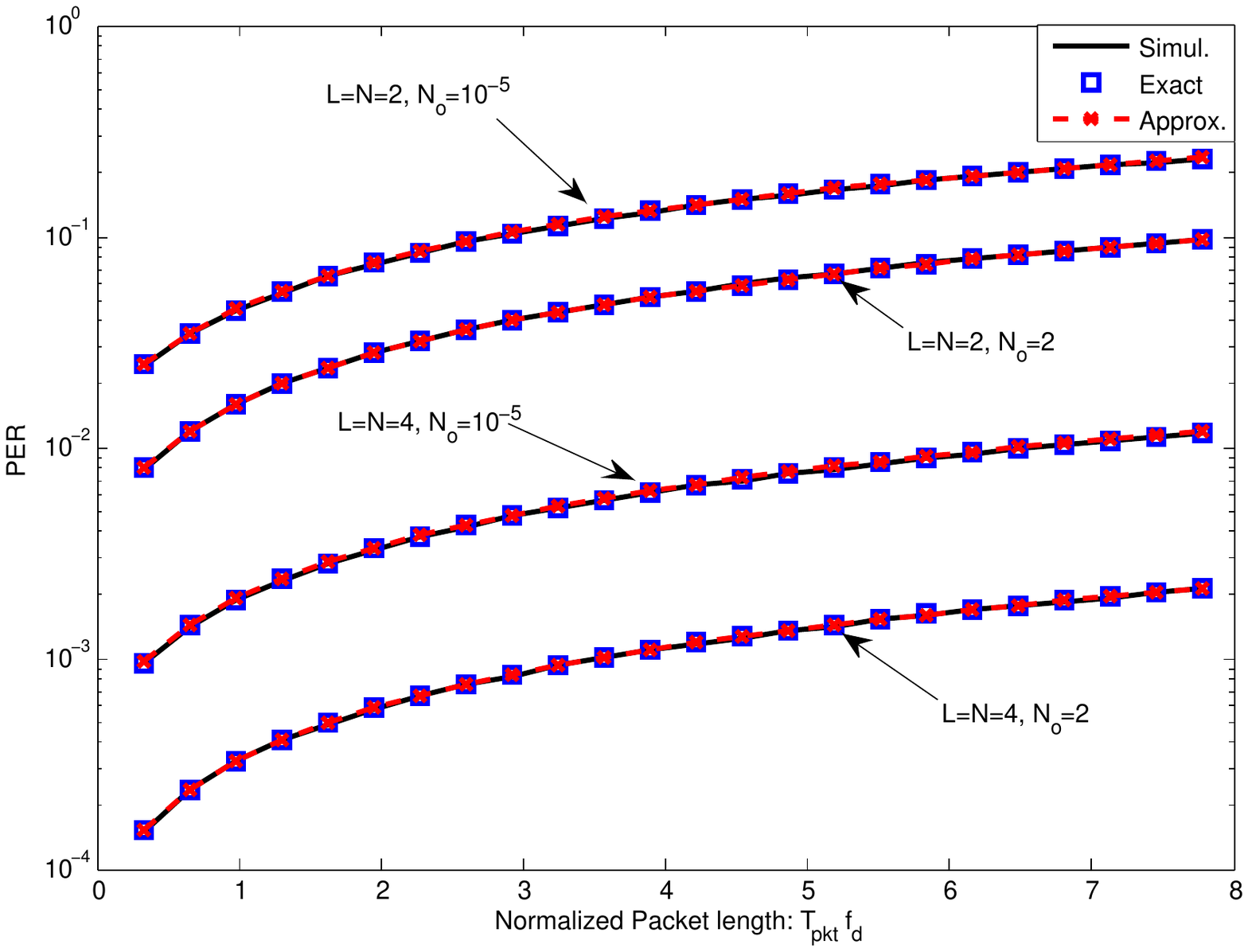}
        \caption{The PER vs normalized packet length, $T_{\text{pkt}}*f_{\text{d}}$, for $(L,N)=(2,2)$ and $(4,4)$, where $f_{\scriptscriptstyle \text{D}} = 32$ Hz, $f_{\scriptscriptstyle \text{I}} = [32, 162, 65, 97]$ Hz, $f_{\text{d}}=\max\{f_\text{D}, f_{\text{I,max}}\}$, $p_{\scriptscriptstyle \text{D}}=1$, $p_{\scriptscriptstyle \text{I}} = [1, 0.5, 0.8, 0.3]$, and $\gamma_{\text{th}} / \gamma_{\text{avg}} = -14$ dB.}
        \label{fig:PER_thresh_14dB_fs1e5_interferece_noise}
\end{figure}

 In Fig. \ref{fig:PER_thresh_14dB_fs1e5_interferece_noise}, the PER plots of two systems with $(L,N)=(2,2)$ and $(4,4)$ are compared at the same normalized threshold value of $-14$ dB as the packet length changes. The PER is plotted versus the packet length normalized w.r.t. the maximum Doppler frequency in the system. The desired user in both systems has a transmitting power of $p_{\scriptscriptstyle \text{D}} =1$, and a maximum Doppler frequency equal to $32$ Hz which is equal to the minimum Doppler frequency among the interferers in both systems. The powers of the interferers are in the same order, thus no single interferer is dominant over the others. In particular, the powers of the interferers are $1, 0.5, 0.8,$ and $0.3$ and their corresponding maximum Doppler frequencies are $32$ Hz, $162$ Hz, $65$ Hz and $97$ Hz. The two-interferer system consists of the first two interferers only. We also plot the two scenarios: interference-limited system ($p_{\scriptscriptstyle \text{D}} / N_o = 50$ dB), and noise-limited system, ($p_{\scriptscriptstyle \text{D}} / N_o = -3$ dB). For the interference-limited systems, Fig. \ref{fig:PER_thresh_14dB_fs1e5_interferece_noise} shows that the theoretical PER derived through using the exact LCR expression matches the simulation results very well. It also shows that the PER calculated via the approximated LCR expression matches very well, almost perfectly, the actual PER values obtained through simulation. This is because the approximate LCR is almost the same as the exact LCR for this system at $\gamma_{\text{th}} / \gamma_{\text{avg}} = -14$ dB, as shown in Fig. \ref{fig:pD_and_pn_effect_All}. In fact, it is noticed that the approximate LCR is almost identical to the exact LCR over relatively low values normalized threshold, which are the typical range of values used in practical systems. The discrepancy between the approximate and exact LCR appears most around $\gamma_{\text{th}} / \gamma_{\text{avg}} = 0$ dB, and then diminishes again for relatively high values of normalized threshold, i.e. where the LCR value itself decreases again. Thus, for practical systems, the approximate LCR expression can be used with high accuracy. The same behavior is experienced for the noise-limited systems.

Fig. \ref{fig:PER_thresh_14dB_fs1e5_interferece_noise} also shows that as the packet length increases the PER increases as well. This is intuitive since the probability of having at least one erroneous bit increases. It also shows that as $L$ increases, the PER decreases though the number of interferers increases. The PER improvement is attributed to the increased diversity rather than increase in the SINR. Moreover, the amount of improvement depends on the speeds, i.e., Doppler frequencies, of the desired user and the interferers as well as their relative powers.

It is worth noting that this model is very useful to system designers due to the accuracy of PER calculation compared with the simulation results. Hence, a large amount of simulation time can be saved, specially for large number of interferers, receive antennas, and long packets. We stress that this accurate result is due to the exact LCR expression proposed in this paper.

\section{Application 2: Throughput Maximization for an ARQ-Based System}
\label{sec:ARQ}

\subsection{ARQ Basics}
\label{sec:ARQ_basics}

Many practical wireless packet systems deploy ARQ schemes to improve the transmission reliability. In this section, the stop and wait ARQ (SW-ARQ) scheme with unlimited number of retransmissions is deployed in the system under study. In such a scheme, during a single ARQ round, the transmitter transmits a packet and waits for an acknowledgement (ACK) from the receiver. If the transmitter receives an ACK, then the transmitter transmits the next packet, otherwise it retransmits the last packet. The duration of a single ARQ round is given by $T_{\text{ARQ}} = T_s (m_{\text{t}} + m_\text{o})$, where $T_s$ is the transmitted symbol duration, and $m_{\text{t}}$ is the number of symbols per data packet, i.e., data packet length. $m_\text{o}$ is the equivalent number of symbols induced by the ARQ protocol overhead, e.g., ACK transmission duration and the guard interval for processing.

For an SW-ARQ scheme with unlimited number of retransmissions, assuming a PER $P_e$ and that the receiver detects any data packet error that occurs, then the average number of retransmissions $\bar{M}$ is given as \cite{506}
\begin{equation}
\bar{M} = (1-P_e) + 2 \ P_e(1-P_e) + 3 \ P_e^2(1-P_e) + \dots + k \ P_e^{k-1}(1-P_e) + \dots = (1-P_e) \sum\limits_{k=1}^{\infty} k(P_e)^{k-1} = \frac{1} {1-P_e}.
\end{equation}
Then the throughput $R$ of such a system is given by \cite{579}
\begin{equation}
\label{eq:throughput1}
R = \frac{m_\text{r}}{T_{\text{tot}}} = \frac{m_{\text{t}}\left(1-P_e\right)} {\left(m_{\text{t}} + m_\text{o}\right) T_s},
\end{equation}
where $m_\text{r}$ is the total number of user's information symbols to be transmitted, and $T_{\text{tot}}$ is the total transmission time given by
\begin{equation}
T_{\text{tot}} = \frac{m_\text{r}}{m_{\text{t}}} \ \bar{M} \ T_{\text{ARQ}}.
\end{equation}

\subsection{Optimal Packet Length}
\label{sec:optimal}

Since $P_e = 1 - P_g$, then from \eqref{eq:Pg} and \eqref{eq:throughput1}, the throughput can be rewritten as
\begin{equation}
\label{eq:throughput2}
R(m_{\text{t}}) = \frac{P_{\scriptscriptstyle \text{CF}} \ m_{\text{t}} \ e^{-\frac{N_c Ts}{P_{\scriptscriptstyle \text{CF}}} m_{\text{t}}}} {\left(m_{\text{t}} + m_\text{o}\right) T_s},
\end{equation}
where the data packet duration is $T_{\text{pkt}} = m_{\text{t}} T_s$. Thus the throughput of the ARQ-based uncoded \mbox{multi-user} system under correlated fading channel is a function of the data packet length $m_{\text{t}}$. In the following, the optimal data packet length that maximizes the throughput of such a system is obtained as a function of the system parameters by taking the derivative of $R(m_{\text{t}})$ w.r.t. $m_{\text{t}}$ and equating it to zero
\begin{equation}
\label{eq:quad}
\frac{\partial R(m_{\text{t}})} {\partial m_{\text{t}}} = \frac{e^{-N_c T_s / P_{\scriptscriptstyle \text{CF}}} }  {\left( m_{\text{t}} + m_\text{o} \right)^2} \left( \frac{m_\text{o} P_{\scriptscriptstyle \text{CF}}} {T_s} - N_c m_{\text{t}}^2 - N_c m_\text{o} m_{\text{t}} \right) = 0.
\end{equation}

By solving the quadratic equation \eqref{eq:quad}, the optimal data packet length $m_{\text{opt}}$ is given as
\begin{equation}
\label{eq:m_opt}
m_{\text{opt}} = \frac{m_\text{o}}{2} \left( \sqrt{1 + \frac{4 P_{\scriptscriptstyle \text{CF}}} { m_\text{o} N_c T_s } } - 1 \right),
\end{equation}
where $P_{\scriptscriptstyle \text{CF}}$ and $N_c$ are functions of $\gamma_{\text{th}}$ and the system parameters, i.e., the powers of the desired and interfering users, $p_{\scriptscriptstyle D}$ and $p_n$, respectively, the maximum Doppler frequencies of the desired and interfering users, $f_{\scriptscriptstyle \text{D}}$ and $\ f_{\scriptscriptstyle I,\text{max}}$, respectively, the number of interferers $N$, and the number of diversity paths $L$.

We note that while in this paper we are considering obtaining the optimum packet length for one receiver, the same framework can be used in broader networks such as interference networks, where each transmitter has only one intended receiver. This can be done by obtaining the optimum packet length for every receiver in the network through following the same approach presented here for one receiver. Note that the packet length will not be the same for each receiver, since for each receiver, there is a different desired transmitter and the transmitters in the network have unequal powers and speeds. Thus, for every receiver, the corresponding powers and speeds of the desired transmitter and interferers are different, hence the optimum packet length differs for every transmitter and receiver pair.

\subsection{Discussion}
\label{sec:discussion}

From \eqref{eq:throughput2}, it is seen that the data packet length $m_{\text{t}}$ has adverse effects on the throughput. To elaborate, the throughput is rewritten as \mbox{$R(m_{\text{t}}) = \frac{P_{\scriptscriptstyle \text{CF}}}{T_s} f_1(m_{\text{t}}) f_2(m_{\text{t}})$}, where
\begin{equation}
f_1(m_{\text{t}}) = \frac{m_{\text{t}}}{m_{\text{t}} + m_\text{o}}, \text{ and} \ \ \ \ \ \  f_2(m_{\text{t}}) = e^{-\frac{N_c Ts}{P_{\scriptscriptstyle \text{CF}}} m_{\text{t}}}.
\end{equation}

It can be seen that $f_1(m_{\text{t}})$ is monotonically increasing while $f_2(m_{\text{t}})$ is monotonically decreasing with $m_{\text{t}}$. However, the steepness of each, i.e. how fast the curve increases or decreases, depends on the values of $m_\text{o}$ and $N_c Ts/P_{\scriptscriptstyle \text{CF}}$, respectively. As $m_\text{o}$ increases, the slope (steepness) of $f_1(m_{\text{t}})$ decreases, thus taking longer time to reach the asymptotic \mbox{value $1$.} This is confirmed by \eqref{eq:m_opt}, where for a fixed $N_c T_s/P_{\scriptscriptstyle \text{CF}}$, as $m_\text{o}$ increases, the term inside the parenthesis decreases while the term outside the parenthesis increases. However, the effect of the latter is greater, thus resulting in increasing $m_{\text{opt}}$ as $m_\text{o}$ increases. This result agrees with the intuition to increase the data packet length as the overhead increases. Note that this result does not contradict with the fact that as the data packet length increases, the packet bears more channel changes thus increasing the probability of error. This demonstrates the importance of obtaining an optimum data packet length that is long enough to overcome the large overhead and small enough to reduce the PER. On the other hand, as $m_\text{o}$ decreases, $m_{\text{opt}}$ decreases as well, until the point where the throughput is monotonically decreasing, depending on the relative values of $m_\text{o}$ and $N_c T_s/P_{\scriptscriptstyle \text{CF}}$. In this case, the optimum data packet length would be the minimum data packet length allowed by the system.


With all other parameters fixed, as $\gamma_{\text{th}}$ increases, the complementary CDF $P_{\scriptscriptstyle \text{CF}}$ increases monotonically while $N_c$ increases until around the average SINR value of the received signal then starts decreasing as $\gamma_{\text{th}}$ further increases as shown in Fig. \ref{fig:pD_and_pn_effect_All}. The relation between $m_{\text{opt}}$ and $\gamma_{\text{th}}$ is not straightforward, which needs further investigation through \eqref{eq:m_opt}.

\begin{figure}[!htb]
        \centering
        \includegraphics[trim=1.5cm 6.5cm 2cm 6.5cm,clip=true,width=3.5in]{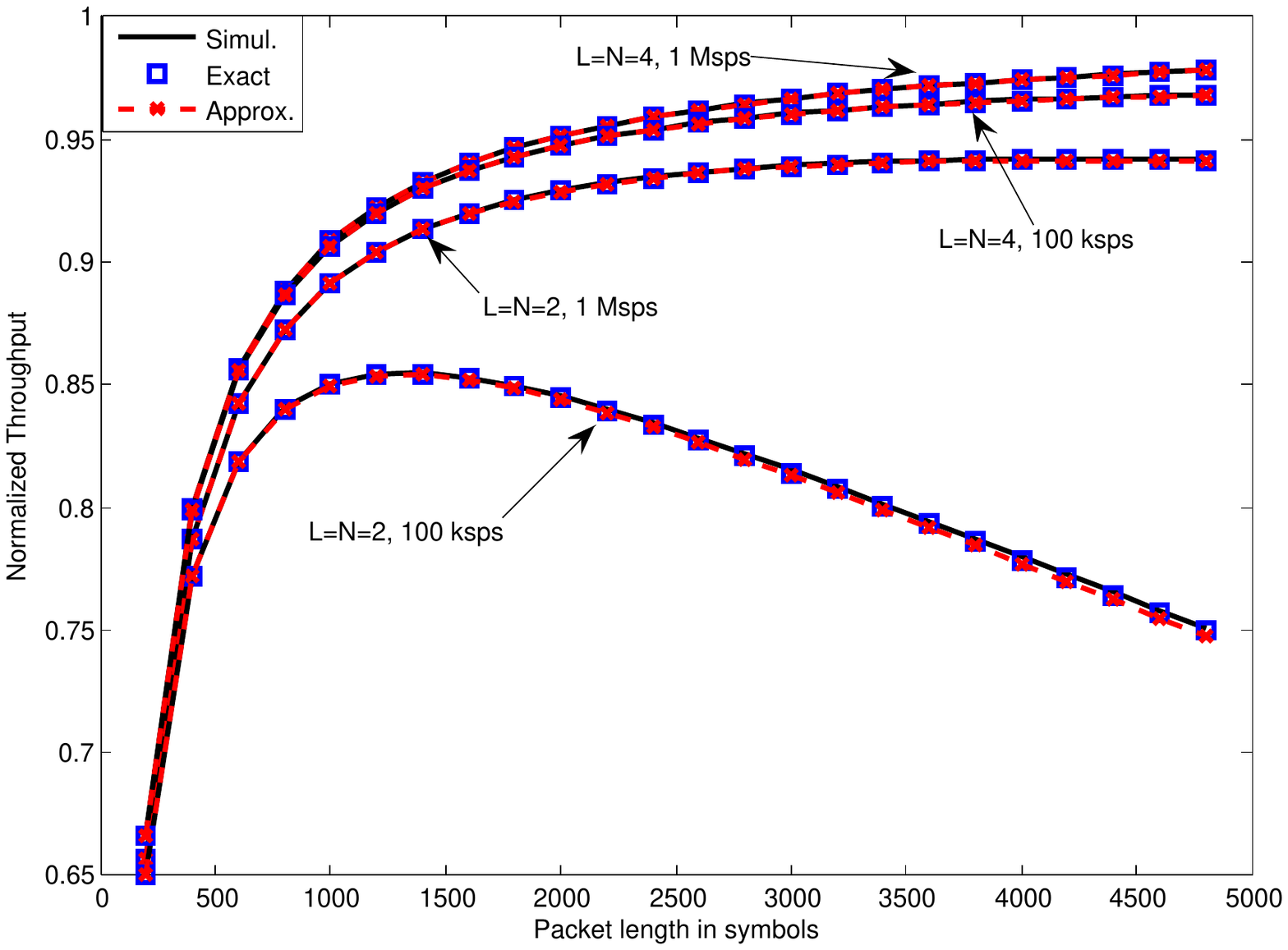}
        \caption{The normalized throughput vs packet length in symbols for $(L,N)=(2,2)$ and $(4,4)$ interference-limited systems where $p_{\scriptscriptstyle \text{D}} / N_o = 50$ dB, $f_{\scriptscriptstyle \text{D}} = 32$ Hz, $f_{\scriptscriptstyle \text{I}} = [32, 162, 65, 97]$ Hz, $p_{\scriptscriptstyle \text{D}}=1$, $p_{\scriptscriptstyle \text{I}} = [1, 0.5, 0.8, 0.3]$, and $\gamma_{\text{th}} / \gamma_{\text{avg}} = -14$ dB.}
        \label{fig:Thrput_thresh_14dB_fs1e6and1e5}
\end{figure}

Fig. \ref{fig:Thrput_thresh_14dB_fs1e6and1e5} plots the throughput normalized to the data rate ($R_s = 1/T_s$), i.e. $R(m)/R_s$, for two interference-limited systems, namely the two- and four-interferer systems described in Section \ref{sec:FSMC} for two data rates of values $10^5$ samples/sec ($100$ ksps) and $10^6$ samples/sec ($1$ Msps) at normalized threshold value of $-14$ dB. In these systems, the ARQ protocol overhead is set to $m_{\text{o}} = 100$ symbols. The figure shows that the exact and approximate theoretical throughputs, obtained via the exact and approximate LCR expressions and the FSMC model, match very well the simulation results.

We note that although the throughput is \textit{normalized} in the figures, as the data rate increases, the normalized throughput increases. As seen in \eqref{eq:throughput2}, the throughput is dependent on the data rate, $R_s=1/T_s$ in the exponential function that is a resultant term from the PER expression. Thus, as $R_s$ increases, $T_s$ decreases and the normalized throughput increases as obvious from \eqref{eq:throughput2} given a fixed number of symbols in the packet and a fixed SINR threshold.

As for the four-interferer system, it is noticed that the normalized throughput is very close for both data rates and the approximation is pretty close to the exact values. This is because as noticed from Fig. \ref{fig:PER_thresh_14dB_fs1e5_interferece_noise}, the PER for the $(4,4)$ system is small compared to that of the $(2,2)$ system. Hence the $1-P_{\text{e}}$ factor in the throughput in \eqref{eq:throughput1} approaches $1$ and the effect of the data rate, which is inherent in the $P_e$ expression, is diminished in the normalized throughput.

Fig. \ref{fig:Thrput_thresh_14dB_fs1e6and1e5} also shows that there is indeed an optimum packet length $m_{\text{opt}}$ which is more evident in the two-interferer system curves. In particular, $m_{\text{opt}}$ obtained from \eqref{eq:m_opt} for the two-interferer system with data rates $100$ ksps and $1$ Msps by using the exact LCR expression are $1,328$ and $4,304$ symbols, respectively. It is obvious from the figure that they match the simulation results very well. It is intuitive that as $R_s$ increases while fixing all other parameters, $m_{\text{opt}}$ increases as well which agrees with \eqref{eq:m_opt}. This is because fixed $N_c$ and $P_{\scriptscriptstyle \text{CF}}$ correspond to fixed Doppler frequencies in which case as the symbol time decreases, more symbols can be included within one data packet such that the data packet time duration $T_{\text{pkt}}$ and the normalized data packet length $T_{\text{pkt}} f_d$ are still same, where $f_d = \text{max}(f_{\scriptscriptstyle \text{D}},f_{\scriptscriptstyle I,max})$. Thus the packet experiences the same fading behavior as with larger $T_s$ and less number of symbols. It is worth mentioning that this argument implies linear increase of $m_{\text{opt}}$ w.r.t. $R_s$, however, this is true only if the ARQ overhead $m_{\text{o}}$ is increased by the same factor by which $R_s$ increased. Recall that $m_{\text{o}}$ incorporates the ACK transmission duration which is dependent on the number of symbols in an ACK packet which can be fixed and is independent of the data rate. Thus, $m_{\text{o}}$ cannot increase linearly with the increase of $R_s$ and hence, neither can $m_{\text{opt}}$. This is clear when examining the effects of $m_{\text{o}}$ and $T_s$ on $m_{\text{opt}}$ in \eqref{eq:m_opt}. It also shows in the $m_{\text{opt}}$ values obtained for the two-interferer system in Fig. \ref{fig:Thrput_thresh_14dB_fs1e6and1e5} where for a fixed $m_{\text{o}} = 100$, when $R_s$ increases by a factor of $10$ from $100$ ksps to $1$ Msps, $m_{\text{opt}}$ increases from $1,328$ to only $4,304$ and not to $13,280$.

For the $(4,4)$ system, the normalized throughput seems to keep increasing and not decreasing afterwards, however, this is only because of the short packet lengths over which the simulation was performed. Indeed, the theoretical optimum packet lengths for this system with data rates $100$ ksps and $1$ Msps are obtained from \eqref{eq:m_opt} to be $6,450$ and $20,503$ symbols, respectively, which are beyond the simulated range in the figure.


\section{Conclusions}
\label{sec:conclusion}

In this paper, the exact and approximate closed-form expressions of the LCR and AOD are derived for \mbox{multi-user} \mbox{multi-antenna} wireless communication systems under AWGN and fading channels. The effects of system parameters, i.e., the desired user's and interferers' powers, and their speeds, on the LCR of the system are analyzed. The derived expressions are used in two different applications via FSMC model, evaluating the PER, and maximizing the throughput for an \mbox{SW-ARW} based system by setting the optimum packet length as a function of the system parameters. The theoretical expressions are validated through comparisons with the simulation results. Extending the model presented in this work to coded transmission systems, and throughput maximization via joint optimization of the packet length and modulation size are deferred for future work.

\appendices

\section{Solving the integration \eqref{eq:Ia_def}}
\label{sec:LCR_deriv_appendix}


In this appendix, the detailed derivation steps to solve the integral $I_a$ in \eqref{eq:Ia_def} are provided. As mentioned in Section \ref{sec:LCR_derivation}, to solve the integral in \eqref{eq:Ia_def}, $U_n, \forall n \in \mathcal{N}$ are assumed to be i.i.d. standard exponential RVs, then their joint PDF is given by $f_{\scriptscriptstyle U_1,\dots,U_{\scriptscriptstyle N}}(u_1,\dots,u_{\scriptscriptstyle N}) = e^{-\sum_{n=1}^N u_n}$. Let $g(U_1,\dots,U_{\scriptscriptstyle N})$ be a function of these RVs, then its expected value is defined as $E\{ g(U_1,\dots,U_{\scriptscriptstyle N}) \} = \int\limits_{u_{\scriptscriptstyle N}} \dots \int\limits_{u_1} g(u_1,\dots,u_{\scriptscriptstyle N}) e^{-\sum_{n=1}^N u_n} du_1 \dots du_{\scriptscriptstyle N} $. Using this argument, $I_a$ can be regarded as in \eqref{eq:Ia_def2}
\begin{equation}
I_a = E\left\{ \left( 1 + \sum\limits_{n=1}^N a_n U_n \right)^{L-1} \sqrt{1 + \sum\limits_{n=1}^N b_n U_n } \right\}. \tag{\ref{eq:Ia_def2} revisited}
\end{equation}

In order to find the above expectation, the following RV transformations are introduced.
\begin{align}
Q_1 &= \sum\limits_{n=1}^N b_n U_n, \tag{\ref{eq:Q1} revisited} \\
Q_2 &= \sum\limits_{n=2}^N \left(a_n - \frac{a_1 b_n}{b_1} \right) U_n, \text{ and} \tag{\ref{eq:Q2} revisited} \\
Q_n &= a_n U_n, \ \ \ \ \  n = 3,\dots,N. \tag{\ref{eq:Qn} revisited}
\end{align}

The inverse transformation, i.e. $u_n$ in terms of $Q_n$, can be obtained as follows. From (\ref{eq:Qn}), by substituting for $U_n = Q_n/a_n, n=3,\dots,N$ in (\ref{eq:Q2}), $U_2$ can be  written as
\begin{equation}
\label{eq:u2}
U_2 = \frac{b_1} {a_2 b_1 - a_1 b_2} Q_2 - \sum\limits_{n=3}^N \frac{a_n b_1 - a_1 b_n} {a_n \left(  a_2 b_1 - a_1 b_2 \right)} Q_n.
\end{equation}

Using \eqref{eq:Q1}, \eqref{eq:Qn} and \eqref{eq:u2}, $U_1$ is given as
\begin{equation}
\label{eq:u1}
U_1 = \frac{1}{b_1} Q_1 - \frac{b_2}{a_2 b_1 - a_1 b_2} Q_2 - \sum\limits_{n=3}^N \left( \frac{b_n}{a_n b_1} - \frac{b_2 \left( a_n b_1 - a_1 b_n \right)} {a_n b_1 \left( a_2 b_1 - a_1 b_2 \right)} \right) Q_n.
\end{equation}

Note that although $U_n, \forall n \in \mathcal{N}$ are defined over the non-negative domain only, i.e. $U_n \geq 0$, it might seem straightforward that $Q_n \geq 0, \forall n$, as they are either just weighted sum of $U_n, n=1,2$ or a scaled version of $U_n, n=3,\dots,N$. However, this is not always the case for $Q_1$ and $Q_2$ as it turns out. To find the domain of $Q_1$, we substitute for $U_1$ by \eqref{eq:u1} in the inequality $U_1 \geq 0$, then it follows that $Q_1 \geq \sum\limits_{n=2}^N \beta_n Q_n$, where $\beta_n$ is defined as
\begin{equation}
\label{eq:beta}
\beta_n = \begin{cases}
\frac{b_1 b_2}{a_2 b_1 - a_1 b_2}, &n = 2\\
\frac{b_1 \left( a_2 b_n - a_n b_2 \right)} {a_n \left( a_2 b_1 - a_1 b_2 \right)}, &n = 3,\dots,N
\end{cases}
\end{equation}
and $\beta_1$ is not defined.

Similarly, from \eqref{eq:u2} it follows that $Q_2 \geq \sum\limits_{n=3}^N h_n Q_n$, where $h_n$ are defined as
\begin{equation}
\label{eq:hn}
h_n = \frac{a_n b_1 - a_1 b_n} {a_n b_1}, \ \ \ \ \ \ n = 3,\dots,N
\end{equation}
while $h_1$ and $h_2$ are undefined. One very important note here is that the above domain of $Q_2$ is correct only if $a_2 b_1 > a_1 b_2$ else the inequality is reversed. For the rest, of the paper this condition is assumed to be always satisfied and in Appendix \ref{sec:notes} we show the implication of this and how to satisfy it physically given a real cellular system. Finally, the domain of $Q_n, n=3,\dots,N$ is simply $Q_n \geq 0$ since they have one-to-one relation with $u_n$.

Thus, the joint PDF $f_{Q_1,\dots,Q_{\scriptscriptstyle N}}(q_1,\dots,q_{\scriptscriptstyle N})$ is given as
\begin{equation}
f_{Q_1,\dots,Q_{\scriptscriptstyle N}}(q_1,\dots,q_{\scriptscriptstyle N}) = \frac{1}{\begin{vmatrix} J \end{vmatrix}} f_{U_1,\dots,U_{\scriptscriptstyle N}}(u_1,\dots,u_{\scriptscriptstyle N}) = \frac{1}{\begin{vmatrix} J \end{vmatrix}} e^{-\sum\limits_{n=1}^N u_n} = \frac{1}{\begin{vmatrix} J \end{vmatrix}} e^{-\sum\limits_{n=1}^N \alpha_n q_n}, \tag{\ref{eq:joint_pdf_allQ} revisited}
\end{equation}
where $|J|$ is the Jacobian. $|J|$ and $\alpha_n$ can be obtained to be
\begin{equation}
|J| = \begin{vmatrix} \frac{\partial \left( Q_1,\dots,Q_{\scriptscriptstyle N} \right)} {\partial \left( U_1,\dots,U_{\scriptscriptstyle N} \right)} \end{vmatrix} = \left( a_2 b_1 - a_1 b_2 \right) \prod\limits_{n=3}^N a_n, \text{ and}  \tag{\ref{eq:jacob} revisited}
\end{equation}
\begin{equation}
\alpha_n = \begin{cases}
\frac{1} {b_1}, &n = 1 \\
\frac{b_1 - b_2} {a_2 b_1 - a_1 b_2}, &n = 2 \\
\frac{\left( b_2 - b_1 \right) \left( a_n b_1 - a_1 b_n \right)} {a_n b_1 \left( a_2 b_1 - a_1 b_2 \right)} - \frac{b_n} {a_n b_1} + \frac{1}{a_n},  &n = 3,\dots,N.
\end{cases}  \tag{\ref{eq:alpha} revisited}
\end{equation}

Using the characteristic function approach, detailed in Appendix \ref{sec:fQ1Q2_deriv_appendix}, the joint PDF $f_{Q_1,Q_2}(q_1,q_2)$ is given by
\begin{align}
f_{Q_1,Q_2}(q_1,q_2) &= \mathcal{L}_{s_2}^{-1} \left\{ \mathcal{L}_{s_1}^{-1} \left\{ \varphi_{\scriptscriptstyle Q_1,Q_2}\left(s_1,s_2\right) \right\} \right\} = \Lambda \sum\limits_{n=1}^N \delta_n \sum\limits_{\substack{t=1 \\ t \neq n}} \mathcal{L}_{s_2}^{-1} \left\{  \frac{\psi_{t,n} e^{-g_n q_1} e^{-f_n q_1 s_2}} {s_2 + \lambda_{t,n}} \right\} \nonumber\\
&= \Lambda \sum\limits_{n=1}^N \delta_n \sum\limits_{\substack{t=1 \\ t \neq n}} \psi_{t,n} e^{-g_n q_1} e^{-\lambda_{t,n} \left( q_2 - f_n q_1\right)}, \tag{\ref{eq:f_Q1Q2} revisited}
\end{align}
where $q_2 \geq f_n q_1$ for every $n \in \mathcal{N}$, and $g_n, f_n, \delta_n, \lambda_{t,n}, \psi_{t,n}$ and $\Lambda$ are defined in \eqref{eq:gn_simple}-\eqref{eq:psi_simp}.

Using the RVs transformation \eqref{eq:Q1}-\eqref{eq:Qn}, \eqref{eq:Ia_def2} can be rewritten as
\begin{align}
&I_a = E\left\{ \sqrt{1+q_1} \left( 1 + \frac{a_1}{b_1}q_1 + q_2 \right)^{L-1} \right\} \tag{\ref{eq:Ia1} revisited} \\
\label{eq:Ia2}
&= \left( \frac{a_1}{b_1}\right)^{L-1} E\left\{ \sqrt{1+q_1} \left( \left[1 + q_1\right] + \left[\frac{b_1}{a_1} - 1 + \frac{b_1}{a_1}q_2 \right] \right)^{L-1} \right\} \\
\label{eq:Ia3}
&= \left( \frac{a_1}{b_1}\right)^{L-1} E\left\{ \sqrt{1+q_1} \sum\limits_{k=0}^{L-1} \binom{L-1}{k} \left( 1 + q_1 \right)^{k} \left( \left[ \frac{b_1}{a_1} - 1 \right] + \left[ \frac{b_1}{a_1}q_2 \right] \right)^{L-1-k} \right\} \\
\label{eq:Ia4}
&= \left( \frac{a_1}{b_1}\right)^{L-1} E\left\{ \sum\limits_{k=0}^{L-1} \binom{L-1}{k} \left( 1 + q_1 \right)^{k+\frac{1}{2}} \sum\limits_{m=0}^{L-1-k} \binom{L-1-k}{m} \left( \frac{b_1}{a_1} - 1 \right)^{L-1-k-m} \left( \frac{b_1}{a_1} \right)^m q_2^m  \right\} \\
&= \sum\limits_{k=0}^{L-1} \binom{L-1}{k} \sum\limits_{m=0}^{L-1-k} \binom{L-1-k}{m} \left( \frac{a_1}{b_1}\right)^{L-1-m} \left( \frac{b_1}{a_1} - 1 \right)^{L-1-k-m} \underbrace{\int\limits_{q_1} \int\limits_{q_2} \left( 1 + q_1 \right)^{k+\frac{1}{2}} q_2^m f_{Q_1,Q_2}(q_1,q_2) dq_2 \ dq_1}_{I_b}, \tag{\ref{eq:Ia5} revisited}
\end{align}
where \eqref{eq:Ia3} follows from the binomial expansion of the two terms in square brackets in \eqref{eq:Ia2}. Similar argument leads to \eqref{eq:Ia4}.

Substituting \eqref{eq:f_Q1Q2} in \eqref{eq:Ia5}, $I_b$ can be rewritten as
\begin{equation}
I_b = \Lambda \left[ \delta_1 \sum\limits_{t=2}^N \psi_{t,1} I_1 + \sum\limits_{n=2}^N \delta_n \sum\limits_{\substack{t=1 \\ t \neq n}}^N \psi_{t,n} I_2 \right], \tag{\ref{eq:Ib} revisited}
\end{equation}
where
\begin{align}
\label{eq:I1_I2}
I_1 &= \underbrace{\int\limits_{q_1=0}^{\infty} \left( 1 + q_1 \right)^{k+\frac{1}{2}} e^{-g_1 q_1} \ dq_1}_{I_{11}} \ \underbrace{\int\limits_{q_2=0}^{\infty} q_2^m e^{-\lambda_{t,1} q_2} \ dq_2 }_{I_{12}}, \text{ and} \nonumber\\
I_2 &= \int\limits_{q_1=0}^{\infty} \left( \left( 1 + q_1 \right)^{k+\frac{1}{2}} e^{-\left( g_n - \lambda_{t,n} f_n \right) q_1} \ \underbrace{\int\limits_{q_2 = f_n q_1}^{\infty} q_2^m e^{-\lambda_{t,n} q_2} dq_2}_{I_{21}} \right) dq_1.
\end{align}

The integration limits are derived based on the conditions imposed by the double inverse Laplace transform as mentioned previously. Note that in \eqref{eq:Ib} the term corresponding to $n=1$ is separated from the rest of the terms because it is a special case where $f_1=0$ which affects the domain on which $q_2$ is integrated as shown in $I_{12}$ and $I_{21}$ in \eqref{eq:I1_I2}, leading to two different integration results as will be shown next.

It is obvious that $I_1$ can be easily solved since it is decoupled into the product of two integrals $I_{11}$ and $I_{12}$. Hence, from \cite[Sec. 3.382, Eq. 4]{372} and \cite[Sec. 3.381, Eq. 4]{372}, it can be shown that
\begin{equation}
\label{eq:I11_I12}
I_{11} = \frac{e^{g_1} \Gamma_{\text{inc}}\left(k+\frac{3}{2}, g_1\right)} {g_1^{k+\frac{3}{2}}}, \text{ and} \ \ \ \ \ \ \ \ \ \ \ \ \ \ \ \ \ \ I_{12} = \frac{m!} {\lambda_{t,1}^{m+1}},
\end{equation}
under condition that $g_1 > 0$ and $\lambda_{t,1} > 0$ which are discussed in detail in Appendix \ref{sec:notes}.

On the other hand, $I_2$ is nested, so we integrate over $q_2$ first, then over $q_1$ since the integration limits of $q_2$ are function of $q_1$ as imposed by the inverse Laplace transform conditions. From \cite[Sec. 3.381, Eq. 3]{372}, $I_{21}$ and $I_2$ can be obtained as
\begin{align}
\label{eq:I21}
I_{21} & = \frac{\Gamma_{\text{inc}}\left(m+1, \lambda_{t,n} f_n q_1 \right)} {\lambda_{t,n}^{m+1}} = \frac{m!}{\lambda_{t,n}^{m+1}} e^{-\lambda_{t,n} f_n q_1} \sum\limits_{r = 0}^m \frac{ \left(\lambda_{t,n} f_n \right)^r q_1^r } {r!},
\end{align}
where the second equality results from expanding the incomplete gamma function in the series form using \cite[Sec. 8.352, Eq. 2]{372}. Substituting by \eqref{eq:I21} in \eqref{eq:I1_I2}, $I_2$ can be solved as follows
\begin{align}
I_2 &= \frac{m!}{\lambda_{t,n}^{m+1}} \sum\limits_{r = 0}^m \frac{\left(f_n \lambda_{t,n}\right)^r} {r!} \int\limits_{q_1=0}^{\infty} q_1^r \left( 1 + q_1 \right)^{k+\frac{1}{2}} e^{-g_n q_1} \ dq_1
\end{align}
\begin{align}
\label{eq:second_equation}
&= \frac{m! e^{g_n} }{\lambda_{t,n}^{m+1}} \sum\limits_{r = 0}^m \frac{\left(f_n \lambda_{t,n}\right)^r} {r!} \int\limits_{y=1}^{\infty} \left( y-1 \right)^r y^{k+\frac{1}{2}} e^{-g_n y} \ dy \\
\label{eq:I2_integral}
&= \frac{m! e^{g_n} }{\lambda_{t,n}^{m+1}} \sum\limits_{r = 0}^m \frac{\left(f_n \lambda_{t,n}\right)^r} {r!} \sum\limits_{w=0}^r (-1)^{r-w} \binom{r}{w} \int\limits_{y=1}^{\infty} y^{k+w+\frac{1}{2}} e^{-g_n y} \ dy \\
&= \frac{m! e^{g_n} }{\lambda_{t,n}^{m+1}} \sum\limits_{r = 0}^m \frac{\left(f_n \lambda_{t,n}\right)^r} {r!} \sum\limits_{w=0}^r (-1)^{r-w} \binom{r}{w} \frac{\Gamma_{\text{inc}}\left( k+w+\frac{3}{2}, g_n \right)} {g_n^{k+w+\frac{3}{2}}}, \tag{\ref{eq:I2_LCR} revisited}
\end{align}
where the variable transformation $y=1+q_1$ is used in \eqref{eq:second_equation}, while \eqref{eq:I2_integral} follows from applying the binomial expansion on $(y-1)^r$. Equation \eqref{eq:I2_LCR} follows from solving the integration in \eqref{eq:I2_integral} using \cite[Sec. 3.381, Eq. 3]{372}, under condition that $g_n > 0$.

Finally, using \eqref{eq:LCR4}, \eqref{eq:Ia5}, \eqref{eq:Ib},\eqref{eq:I1_LCR} and \eqref{eq:I2_LCR}, the closed-form expression of the LCR can be written as
\begin{align}
\text{LCR} \left(\gamma_{\text{th}}\right) &= \frac{ \sqrt{2 \sigma_{\scriptscriptstyle \text{D}}^2 } \left( \frac{\gamma_{\text{th}} N_o}{p_{\scriptscriptstyle D}} \right)^{L-\frac{1}{2}} e^{-\frac{\gamma_{\text{th}} N_o}{p_{\scriptscriptstyle D}}} }   {\sqrt{\pi} \prod\limits_{n=1}^N \left(1 + \frac{\gamma_{\text{th}} p_n} {p_{\scriptscriptstyle D}}\right)} \ \Lambda \sum\limits_{k=0}^{L-1} \sum\limits_{m=0}^{L-1-k} \Xi_{k,m} \Bigg[ \sum\limits_{t=2}^N \delta_1 \psi_{t,1} \frac{e^{g_1} \Gamma_{\text{inc}}\left(k+\frac{3}{2}, g_1 \right) } {g_1^{k+\frac{3}{2}} \ \lambda_{t,1}^{m+1} } \nonumber\\
 & +  \sum\limits_{n=2}^N \sum\limits_{\substack{t=1 \\ t \neq n}}^N \sum\limits_{r=0}^m \sum\limits_{w=0}^r \delta_n \psi_{t,n} e^{g_n} \frac{(-1)^{r-w} \binom{r}{w} f_n^r \Gamma_{\text{inc}}\left(k+w+\frac{3}{2}, g_n \right)} {r! \ \lambda_{t,n}^{m-r+1} g_n^{k+w+\frac{3}{2}}}  \Bigg], \tag{\ref{eq:LCR_final} revisited}
\end{align}
where $\Xi_{k,m}$ is defined as
\begin{align}
\Xi_{k,m} &= \binom{L-1}{k} \ \binom{L-1-k}{m} \left( \frac{a_1}{b_1} \right)^{L-1-m} \left( \frac{b_1}{a_1} - 1 \right)^{L-1-k-m} \frac{m!}{\Gamma_{\text{s}}(L)} = \frac{a_1^k \left( b_1 - a_1 \right)^{L-1-m-k}} {k! \ \left( L-1-k-m \right)! b_1^{L-1-m} } \nonumber\\
&= \frac{1} {k! (L-1-k-m)!} \frac{1} { \varepsilon_1^k \left( 1+\frac{1}{\varepsilon_1} \right)^{L-1-m}}, \tag{\ref{eq:Xi_def} revisited}
\end{align}
since $\Gamma_{\text{s}}(L)=(L-1)!$.

\section{Derivation of the Joint PDF $f_{Q_1, Q_2}(q_1,q_2)$}
\label{sec:fQ1Q2_deriv_appendix}

First, the joint characteristic function of $Q_1$ and $Q_2$, $\varphi_{\scriptscriptstyle Q_1,Q_2}(s_1,s_2)$, is obtained, where $s_1$ and $s_2$ are the Laplace variables. To be able to use the Laplace Transform concept, we change slightly the definition of the characteristic function to be $\varphi_{\scriptscriptstyle Q_1,Q_2}(s_1,s_2) = E\{e^{-s_1 Q_1} e^{-s_2 Q_2}\}$ instead of the conventional definition $\varphi_{\scriptscriptstyle Q_1,Q_2}(s_1,s_2) = E\{e^{s_1 Q_1} e^{s_2 Q_2}\}$. Of course the effect of introducing the negative sign has been taken into account through the derivation by adjusting all the conditions and domains accordingly. Using the joint PDF $f_{Q_1,\dots,Q_{\scriptscriptstyle N}}(q_1,\dots,q_n)$ in \eqref{eq:joint_pdf_allQ}, $\varphi_{\scriptscriptstyle Q_1,Q_2}(s_1,s_2)$ can be obtained as follows

\begin{align}
&\varphi_{\scriptscriptstyle Q_1,Q_2}(s_1,s_2) = \int\limits_{Q_{\scriptscriptstyle N}} \dots \int\limits_{Q_1} e^{-s_1 q_1} e^{-s_2 q_2} f_{Q_1,\dots,Q_{\scriptscriptstyle N}}(q_1,\dots,q_{\scriptscriptstyle N}) dq_1 \dots dq_{\scriptscriptstyle N} \\
&= \frac{1}{|J|} \int\limits_{q_{\scriptscriptstyle N}=0}^{\infty} \dots \int\limits_{q_1=\sum\limits_{n=2}^N \beta_n q_n}^{\infty} \ e^{-(s_1+\alpha_1) q_1} \ e^{-(s_2+\alpha_2) q_2} \ e^{-\sum\limits_{n=3}^N \alpha_n q_n} \ dq_1 \dots dq_{\scriptscriptstyle N} \\
&= \frac{1}{\left(s_1 + \alpha_1 \right) |J|} \int\limits_{q_{\scriptscriptstyle N}=0}^{\infty} \dots \int\limits_{q_2=\sum\limits_{n=2}^N h_n q_n}^{\infty} \ e^{-\left[ \left(s_1+\alpha_1\right) \beta_2 + s_2 + \alpha_2 \right] q_2} \ e^{-\sum\limits_{n=3}^N \left[ \left( s_1 + \alpha_1 \right) \beta_n + \alpha_n \right] q_n} \ dq_2 \dots dq_{\scriptscriptstyle N} \\
&= \frac{1}{|J| \left(s_1 + \alpha_1 \right) \left( \left(s_1 + \alpha_1\right) \beta_2 + s_2 + \alpha_2 \right)} \int\limits_{q_{\scriptscriptstyle N}=0}^{\infty} \dots \int\limits_{q_3=0}^{\infty} \ e^{-\sum\limits_{n=3}^N \left( \left( h_n \beta_2 + \beta_n \right) \left( s_1 + \alpha_1 \right) + h_n \left( s_2 + \alpha_2 \right) + \alpha_n \right) q_n} \ dq_3 \dots dq_{\scriptscriptstyle N} \\
&= \frac{1}{|J| \left(s_1 + \alpha_1 \right) \left( \left(s_1 + \alpha_1\right) \beta_2 + s_2 + \alpha_2 \right) \prod\limits_{n=3}^N \left( \left( h_n \beta_2 + \beta_n \right) s_1 + \left( h_n \beta_2 + \beta_n \right) \alpha_1 + h_n s_2 + h_n \alpha_2 + \alpha_n \right)}.
\end{align}

After obtaining the joint characteristic function of $Q_1$ and $Q_2$, partial fraction is done over $s_1$. Thus it can be rewritten as
\begin{equation}
\label{eq:cc_fn1}
\varphi_{\scriptscriptstyle Q_1,Q_2}(s_1,s_2) = \frac{1}{|J| \beta_2 \prod\limits_{n=3}^N \left( h_n \beta_2 + \beta_n \right)} \left[ \sum_{n=1}^N \frac{\zeta_n(s_2)}{s_1 + (f_n s_2 + g_n)} \right], \\
\end{equation}
where
\begin{equation}
\label{eq:fn}
f_n = \begin{cases}
0, &n = 1 \\
\frac{1}{\beta_2}, &n = 2 \\
\frac{h_n}{h_n \beta_2 + \beta_n}, &n=3,\dots,N
\end{cases}
\end{equation}

\begin{equation}
\label{eq:gn}
g_n = \begin{cases}
\alpha_1, &n=1  \\
\frac{\alpha_1 \beta_2 + \alpha_2}{\beta_2}, &n=2  \\
\frac{\alpha_1 \left( h_n \beta_2 + \beta_n \right) + h_n \alpha_2 + \alpha_n } {h_n \beta_2 + \beta_n}, &n=3,\dots,N
\end{cases}
\end{equation}

\begin{equation}
\label{eq:zeta}
\zeta_n(s_2) = \frac{\delta_n} {\prod\limits_{\substack{t=1 \\ t\neq n}}^N \left( s_2 + \lambda_{t,n} \right)}, \ \ \ \ \ n=1,\dots,N
\end{equation}

\begin{equation}
\label{eq:delta}
\delta_n = \begin{cases}
\frac{\beta_2 \prod\limits_{t=3}^N \left( h_t \beta_2 + \beta_t \right)} {\prod\limits_{t=3}^N h_t}, &n=1 \\
\frac{(-\beta_2)^{N-1} \prod\limits_{t=3}^N \left( h_t \beta_2 + \beta_t \right) } {\prod\limits_{t=3}^N \beta_t},  &n=2 \\
\frac{-\beta_2 \left( h_n \beta_2 + \beta_n \right)^{N-1} \prod\limits_{\substack{t=3 \\ t\neq n}}^N \left( h_t \beta_2 + \beta_t \right) } {h_n \beta_n \prod\limits_{\substack{t=3 \\ t\neq n}}^N \left( h_t \beta_n - h_n \beta_t \right)}, &n=3,\dots,N
\end{cases}
\end{equation}

\begin{equation}
\label{eq:lambda}
\lambda_{t,n} = \begin{cases}
\alpha_2,                                               &(t,n) = (1,2) \mbox{ or } (2,1)\\
\frac{\alpha_2 h_t + \alpha_t}{h_t},                    &t=3,\dots,N, \ n = 1 \\
\frac{\alpha_2 \beta_t - \alpha_t \beta_2}{\beta_t},    &t = 3,\dots,N, \ n = 2 \\
\frac{h_n \alpha_2 + \alpha_n} {h_n},                   &t = 1, \ n = 3,\dots,N \\
\frac{\alpha_2 \beta_n - \alpha_n \beta_2} {\beta_n},   &t = 2, \  n=3,\dots,N \\
\frac{ \alpha_2 \left( h_t \beta_n - h_n \beta_t \right) + \beta_2 \left( \alpha_t h_n - \alpha_n h_t \right) + \alpha_t \beta_n - \alpha_n \beta_t } {h_t \beta_n - h_n \beta_t},   &t=3,\dots,N, \ n=3,\dots,N, \ t \neq n.
\end{cases}
\end{equation}

Note that $\lambda_{t,n}$ are undefined for $t=n$. Noting that $\zeta_n(s_2)$ is a function of $s_2$, then it is expressed in partial fraction form over $s_2$ as follows:
\begin{equation}
\label{eq:zeta_def}
\zeta_n(s_2) = \frac{\delta_n} {\prod\limits_{\substack{t=1 \\ t\neq n}}^N \left( s_2 + \lambda_{t,n} \right)} = \delta_n \sum\limits_{\substack{t=1 \\ t \neq n}}^N \frac{\psi_{t,n}} {s_2 + \lambda_{t,n}},
\end{equation}
where the partial fraction coefficients $\psi_{t,n}$ are given by
\begin{equation}
\label{eq:psi}
\psi_{t,n} = \frac{1} {\prod\limits_{\substack{q=1 \\ q \neq t \neq n}}^N \left( \lambda_{q,n} - \lambda_{t,n} \right)}, \ \ \ t \in \mathcal{N}, \ \ \ n \in \mathcal{N}, \ \ t \neq n,
\end{equation}
and $\psi_{t,n}$ are undefined for $t = n$.

By substituting \eqref{eq:zeta_def} into \eqref{eq:cc_fn1}, the characteristic function can be rewritten as
\begin{equation}
\varphi_{\scriptscriptstyle Q_1,Q_2}(s_1,s_2) = \Lambda \sum\limits_{n=1}^N \delta_n \sum\limits_{\substack{t=1\\t\neq n}}^N \frac{\psi_{t,n}}{\left( s_1 + f_n s_2 + g_n \right) \left(s_2 + \lambda_{t,n}\right)}, \tag{\ref{eq:cc_fn2} revisited}
\end{equation}
where $\Lambda$ is given by
\begin{equation}
\label{eq:Lambda}
\Lambda = \frac{1}{|J| \beta_2 \prod\limits_{n=3}^N \left( h_n \beta_2 + \beta_n \right)}.
\end{equation}

In order to find the joint PDF $f_{Q_1,Q_2}(q_1,q_2)$, the double inverse Laplace transform is performed over $s_1$ and $s_2$. The inverse Laplace transform is taken over $s_1$ first, followed by $s_2$ as follows:
\begin{align}
f_{Q_1,Q_2}(q_1,q_2) &= \mathcal{L}_{s_2}^{-1} \left\{ \mathcal{L}_{s_1}^{-1} \left\{ \varphi_{\scriptscriptstyle Q_1,Q_2}\left(s_1,s_2\right) \right\} \right\} = \Lambda \sum\limits_{n=1}^N \delta_n \sum\limits_{\substack{t=1 \\ t \neq n}} \mathcal{L}_{s_2}^{-1} \left\{  \frac{\psi_{t,n} e^{-g_n q_1} e^{-f_n q_1 s_2}} {s_2 + \lambda_{t,n}} \right\} \nonumber\\
&= \Lambda \sum\limits_{n=1}^N \delta_n \sum\limits_{\substack{t=1 \\ t \neq n}} \psi_{t,n} e^{-g_n q_1} e^{-\lambda_{t,n} \left( q_2 - f_n q_1\right)}, \tag{\ref{eq:f_Q1Q2} revisited}
\end{align}
where this is valid under condition $q_2 \geq f_n q_1$ for $\forall n \in \mathcal{N}$. This condition results from the inverse Laplace transformation and it affects the range upon which $q_2$ is integrated as was seen in Appendix \ref{sec:LCR_deriv_appendix}.

\section{Simplifications}
\label{sec:simplification_appendix}

After some tedious algebraic manipulations, the parameters of \eqref{eq:LCR_final} can be simplified in terms of $a_n$ and $\varepsilon_n$ as follows:
\begin{align}
\alpha_1 &= \frac{1}{b_1} = \frac{1}{a_1 (1+\varepsilon_1)}, \\
\label{eq:fn_simple}
f_n &= \begin{cases}
0, & n =1 \\
\frac{a_n b_1 - a_1 b_n} {b_1 b_n} = \frac{\varepsilon_1 - \varepsilon_n} {(1+\varepsilon_1)(1+\varepsilon_n)},  & n=2,\dots,N,
\end{cases}\\
\label{eq:gn_simple}
g_n &= \frac{1}{b_n} = \frac{1}{a_n (1+\varepsilon_n)}, \forall n \in \mathcal{N}, \\
\label{eq:delta_n_simp}
\delta_n &= \frac{ b_n^{N-2} \prod\limits_{t=1}^N b_t } {\prod\limits_{\substack{t=1 \\ t \neq n}}^N \left( a_t b_n - a_n b_t \right)} = \frac{ \left( 1 + \varepsilon_n \right)^{N-1} \prod\limits_{q=1\neq n}^N \left( 1 + \varepsilon_q \right) } { \prod\limits_{\substack{q=1 \\ q \neq n}} \left( \varepsilon_n - \varepsilon_q \right)}, \forall n \in \mathcal{N}, \\
\label{eq:lambda_simp}
\lambda_{t,n} &= \frac{ b_n - b_t } {a_t b_n - a_n b_t} = \frac{a_n \left( 1 + \varepsilon_n \right) - a_t \left( 1 + \varepsilon_t \right)} {a_t a_n \left( \varepsilon_n - \varepsilon_t \right) }, \ \ \ \forall \ (t,n), \ t \neq n, \\
\label{eq:Lambda_simp}
\Lambda &= \frac{1}{\prod\limits_{n=1}^N b_n} = \frac{1} { \left( \prod\limits_{n=1}^N a_n \right) \prod\limits_{n=1}^N \left( 1 + \varepsilon_n \right) },
\end{align}

\begin{align}
\psi_{t,n} &= \frac{1} {\prod\limits_{\substack{q=1 \\ q \neq t \neq n}}^N \left( \lambda_{q,n} - \lambda_{t,n} \right)} = \frac{\left( a_t b_n - a_n b_t \right)^{N-2} \prod\limits_{\substack{q=1 \\ q \neq n \neq t}}^N \left( a_q b_n - a_n b_q \right) } { \prod\limits_{\substack{q=1 \\ q \neq n \neq t}}^N \left[ b_n^2 \left( a_t - a_q \right) + b_t b_n \left( a_q - a_n \right) + b_n b_q \left( a_n - a_t \right) \right] } \nonumber\\
\label{eq:psi_simp}
&= \frac{ \left( a_t a_n \left( \varepsilon_n - \varepsilon_t \right) \right)^{N-2} \prod\limits_{\substack{q=1 \\ q \neq n \neq t }}^N \left[ a_q \left(  \varepsilon_n - \varepsilon_q \right) \right] } {\prod\limits_{\substack{q=1 \\ q \neq n \neq t}}^N \left[ a_q \left( \varepsilon_q \left( a_n - a_t \right) + \left( a_t \varepsilon_t - a_n \varepsilon_n \right) \right) + a_t a_n \left( \varepsilon_n - \varepsilon_t \right) \right] }, \ \ \ \ \forall \ (t,n), \ t\neq n.
\end{align}


\section{Notes on the LCR Derivation}
\label{sec:notes}

It is worth noting that the LCR expression in \eqref{eq:LCR_final} is only valid if the conditions mentioned within the derivation are satisfied. For the sake of completeness, these conditions are discussed here in detail. The conditions are given as follows:
\begin{align}
\label{eq:cond_a2b1}
&a_2 b_1 > a_1 b_2 \\
\label{eq:cond_gn}
&g_n > 0, \ \ \ \ \ \ \ \ \  \ \forall n \in \mathcal{N} \\
\label{eq:cond_lambda}
&\lambda_{t,1} > 0,
\end{align}
where the condition \eqref{eq:cond_gn} is essential to have the domain of $q_2 \geq \sum\limits_{n=3}^N \beta_n q_n$, else the inequality is reversed as mentioned earlier. In order to find this domain $U_2$ is substituted for by \eqref{eq:u2} in the inequality $U_2 > 0$ and then both sides are multiplied by $(a_2 b_1 - a_1 b_2)$ which implicitly means that this quantity is assumed positive. Using \eqref{eq:anbneps}, we can write $a_1 b_1 - a_1 b_2 = a_1 a_2 \left( \varepsilon_1 - \varepsilon_2 \right) > 0$. Since $a_n > 0, \forall n$, then this condition simplifies to
\begin{equation}
\varepsilon_1 > \varepsilon_2  \ \Rightarrow \ p_1 \sigma_1^2 > p_2 \sigma_2^2,
\end{equation}
where \eqref{eq:anbneps} is used to reach the equivalence on the R.H.S. We note that this condition does not impose any constraints on the practical systems that this LCR expression can be on. In fact, this condition can be easily satisfied, without loss of generality, by ordering the interferers such that the first interferer has a power-variance product ($p_n \sigma_n^2$) greater than that of the second interferer.

Conditions \eqref{eq:cond_a2b1} arises from \eqref{eq:I1_I2} and \eqref{eq:I2_integral}, since these solutions are only possible under these conditions \cite[Sec. 3.382, Eq. 4]{372}, \cite[Sec. 3.381, Eq. 3]{372}. It can be shown that these conditions will always be satisfied. As shown in Appendix \ref{sec:simplification_appendix}, $g_n$ is simplified as $g_n = 1/b_n, \forall n \in \mathcal{N}$ and since $b_n > 0, \forall n \in \mathcal{N}$, hence, $g_n$ is always positive and condition (\ref{eq:cond_a2b1}) is always satisfied.

Finally, condition \eqref{eq:cond_lambda} is necessary to solve \eqref{eq:I1_I2} \cite[Sec. 3.381, Eq. 4]{372}. Unfortunately this condition is not guaranteed to be satisfied for all values of the systems' parameters. Hence, in the following this condition is investigated along with the range of $\gamma_{\text{th}}$ over which this condition is satisfied and thus over which the LCR expression \eqref{eq:LCR_final} is valid. From \eqref{eq:lambda_simp}, $\lambda_{t,1}$ can be written as
\begin{equation}
\label{eq:lambda_t1}
\lambda_{t,1} = \frac{b_1 - b_t} {a_t b_1 - a_1 b_t} = \frac{a_1(1+\varepsilon_1) - a_t(1 + \varepsilon_t)} {a_1 a_t (\varepsilon_1 -\varepsilon_t)}, \ \ t=2,\dots,N.
\end{equation}

Thus it is obvious that to satisfy \eqref{eq:cond_lambda}, both the numerator and denominator of \eqref{eq:lambda_t1} have to be simultaneously positive or simultaneously negative. Since for $t=2$, the denominator has to be positive to satisfy \eqref{eq:cond_a2b1} as well, the interferers shall be ordered such that the first interferer has a power-variance product ($p_n \sigma_n^2$) greater than that of all of the other interferers or
\begin{equation}
\label{eq:order_epsilon}
p_1 \sigma_1^2 > p_t \sigma_t^2  \Rightarrow \ \varepsilon_1 > \varepsilon_t \Rightarrow \ a_t b_1 - a_1 b_t > 0, \ \ t=2,\dots,N.
\end{equation}
Accordingly, \eqref{eq:cond_lambda} is satisfied if the numerator is positive as well, i.e. if $ b_1 > b_t , t=2,\dots,N$. This inequality can be expanded in terms of the systems' parameters using \eqref{eq:anbneps} and some algebraic manipulation as follows
\begin{equation}
\label{eq:b1_great_bt}
b_1 > b_t \ \ \Rightarrow \ \ \frac{p_{\scriptscriptstyle \text{D}} \sigma_{\scriptscriptstyle \text{D}}^2 + \gamma_{\text{th}} p_1 \sigma_1^2 } {\frac{p_{\scriptscriptstyle \text{D}}}{p_1} + \gamma_{\text{th}} }    >    \frac{p_{\scriptscriptstyle \text{D}} \sigma_{\scriptscriptstyle \text{D}}^2 + \gamma_{\text{th}} p_t \sigma_t^2 } {\frac{p_{\scriptscriptstyle \text{D}}}{p_t} + \gamma_{\text{th}} }, \ \ \ \ t=2,\dots,N.
\end{equation}

From \eqref{eq:order_epsilon} and \eqref{eq:b1_great_bt}, it follows that \eqref{eq:cond_lambda} is satisfied for all values of $\gamma_{\text{th}}$ iff $p_1 > p_t, t=2,\dots,N$. However, there might be some interferers in the system where $p_1 < p_t$ for some value of $t \neq 1$. In this case, we derive the range of $\gamma_{\text{th}}$ over which condition \eqref{eq:cond_lambda}, and consequently the LCR expression \eqref{eq:LCR_final}, are valid as follows.

Condition \eqref{eq:b1_great_bt} can be rewritten as
\begin{equation}
\label{eq:quad_inequality}
\gamma_{\text{th}}^2 p_1 p_t \left( p_1 \sigma_1^2 - p_t \sigma_t^2 \right) + \gamma_{\text{th}} p_{\scriptscriptstyle \text{D}} \left( p_1^2 \sigma_1^2 - p_t^2 \sigma_t^2 \right) + p_{\scriptscriptstyle \text{D}} \sigma_{\scriptscriptstyle \text{D}}^2 \left( p_1 - p_t \right) > 0.
\end{equation}

Solving the quadratic equation on the left-hand side of \eqref{eq:quad_inequality}, the roots of $\gamma_{\text{th}}$ are given as
\begin{equation}
\label{eq:gamma_th_roots}
\gamma_{\text{th}} = \frac{ -p_{\scriptscriptstyle \text{D}} \left( p_1^2 \sigma_1^2 - p_t^2 \sigma_t^2 \right) \pm \sqrt{ p_{\scriptscriptstyle \text{D}}^2 \left( p_1^2 \sigma_1^2 - p_t^2 \sigma_t^2 \right)^2 - 4 p_1 p_t p_{\scriptscriptstyle \text{D}} \sigma_{\scriptscriptstyle \text{D}}^2 \left( p_1 \sigma_1^2 - p_t \sigma_t^2 \right) \left( p_1 - p_t \right) } }   { 2 p_1 p_t \left( p_1 \sigma_1^2 - p_t \sigma_t^2 \right) }.
\end{equation}

Given that the interferers are ordered such that \eqref{eq:order_epsilon} is satisfied, it can be seen from \eqref{eq:gamma_th_roots} that if $p_1 > p_t$, then the roots are always negative and since the roots represent SINR threshold which has to be a positive value (on linear scale), then this means that the inequality \eqref{eq:b1_great_bt} is satisfied for any positive value of $\gamma_{\text{th}}$. However, if $p_1 < p_t$, then the root with the positive square root is positive which means that to satisfy the inequality \eqref{eq:b1_great_bt}, $\gamma_{\text{th}}$ has to be greater than this root in value.

In summary, given that the interferers are ordered to satisfy \eqref{eq:order_epsilon}, the LCR expression \eqref{eq:LCR_final} is valid for all values of $\gamma_{\text{th}}$ if $p_1 > p_t, t=2,\dots,N$. If for $\forall t \in \mathcal{T}, \ p_1 < p_t$, the LCR expression \eqref{eq:LCR_final} is valid over the range of SINR threshold given by
\begin{equation}
\label{eq:gamma_range}
\gamma_{\text{th}} > \max_{t \in \mathcal{T}} \frac{ -p_{\scriptscriptstyle \text{D}} \left( p_1^2 \sigma_1^2 - p_t^2 \sigma_t^2 \right) + \sqrt{ p_{\scriptscriptstyle \text{D}}^2 \left( p_1^2 \sigma_1^2 - p_t^2 \sigma_t^2 \right)^2 - 4 p_1 p_t p_{\scriptscriptstyle \text{D}} \sigma_{\scriptscriptstyle \text{D}}^2 \left( p_1 \sigma_1^2 - p_t \sigma_t^2 \right) \left( p_1 - p_t \right) } }   { 2 p_1 p_t \left( p_1 \sigma_1^2 - p_t \sigma_t^2 \right) }.
\end{equation}

On another note, taking a look at the procedure of the derivation and the conditions under which the LCR expression is derived, it shows that the LCR expression in \eqref{eq:LCR_final} is in general valid for equal, unequal power co-channel interferer, with equal/unequal speeds, and any combination of these scenarios, except the case: interferers of equal powers with equal speeds, where this case is derived as a special case in Section \ref{sec:Equal-Power Equal-Speed Interferers}. 

It is worth noting that the derived LCR expression does not ignore the noise in the SINR definition unlike many other works in literature that are valid for high SNR regime or interference-limited scenarios only. In conclusion, the LCR expression given in \eqref{eq:LCR_final} is valid for any combination of interferers powers and speeds, in interference-limited, noise-limited systems and all SNR regimes.

\section{LCR Derivation for Interferers with Equal Powers and Equal Speeds}
\label{sec:eq_pow_eq_speed_LCR_deriv_appendix}

In this section, the LCR expression in \eqref{eq:LCR_equal_pow_equal_speed} is derived. Following the same steps as in the general case, the LCR is given by \eqref{eq:LCR4} and the integral in \eqref{eq:Ia_def2} can be rewritten as
\begin{equation}
\label{eq:Ia_def_eqpow_eqspeed}
I_a = E\left\{ \left( 1 + a \sum\limits_{n=1}^N u_n \right)^{L-1} \sqrt{1 + b \sum\limits_{n=1}^N u_n} \right\},
\end{equation}
where $a_n=a, \ b_n=b,$ and $\varepsilon_n=\varepsilon, \forall n \in \mathcal{N}$ and they are defined as in \eqref{eq:anbneps} and still $U_n$ are considered i.i.d. standard exponential RVs. The variable transformation $Q=\sum\limits_{n=1}^N U_n$ is introduced, which is an Erlang distributed variable whose PDF is given as
\begin{equation}
f_Q(q) = \frac{q^{N-1} e^{-q}} {\Gamma_{\text{s}}(N)}.
\end{equation}
Then the integration in \eqref{eq:Ia_def_eqpow_eqspeed} can be written as
\begin{align}
\label{eq:Ia_eqpow_eqspeed_1}
I_a &= E\left\{ \left( 1 + a q \right)^{L-1} \sqrt{1 + b q} \right\} \\
\label{eq:Ia_eqpow_eqspeed_2}
&= \int\limits_{q=0}^{\infty} \left( 1 + a q \right)^{L-1} \sqrt{1 + b q} \ \frac{q^{N-1} e^{-q}} {\Gamma_{\text{s}}(N)} \ dq \\
\label{eq:Ia_eqpow_eqspeed_3}
&= \frac{1} {\Gamma_{\text{s}}(N) (1 + \varepsilon)^{L-1}} \int\limits_{q=0}^{\infty} \left( \varepsilon + \left[1 + b q\right] \right)^{L-1} \sqrt{1 + b q} \ q^{N-1} e^{-q} \ dq \\
\label{eq:Ia_eqpow_eqspeed_4}
&= \frac{1} {\Gamma_{\text{s}}(N) (1 + \varepsilon)^{L-1}} \sum\limits_{l=0}^{L-1} \binom{L-1}{l} \varepsilon^{L-l-1} \int\limits_{q=0}^{\infty} q^{N-1} \left( 1 + b q \right)^{l+\frac{1}{2}} e^{-q} \ dq \\
\label{eq:Ia_eqpow_eqspeed_5}
&= \frac{e^{\frac{1}{b}}} {\Gamma_{\text{s}}(N) (1 + \varepsilon)^{L-1} \ b^N} \sum\limits_{l=0}^{L-1} \binom{L-1}{l} \varepsilon^{L-l-1} \int\limits_{y=1}^{\infty} \left( y-1 \right)^{N-1} y^{l+\frac{1}{2}} e^{-\frac{1}{b}y} \ dy
\end{align}
\begin{align}
\label{eq:Ia_eqpow_eqspeed_6}
&= \frac{e^{\frac{1}{b}}} {\Gamma_{\text{s}}(N) (1 + \varepsilon)^{L-1} \ b^N} \sum\limits_{l=0}^{L-1} \binom{L-1}{l} \sum\limits_{m=0}^{N-1} \binom{N-1}{m} \left( -1 \right)^{N-m-1} \varepsilon^{L-l-1} \int\limits_{y=1}^{\infty} y^{m+l+\frac{1}{2}} e^{-\frac{1}{b}y} \ dy \\
\label{eq:Ia_eqpow_eqspeed_7}
&= \frac{e^{\frac{1}{b}}} {\Gamma_{\text{s}}(N) (1 + \varepsilon)^{L-1}} \sum\limits_{l=0}^{L-1} \sum\limits_{m=0}^{N-1} \binom{L-1}{l} \binom{N-1}{m} \left( -1 \right)^{N-m-1} \varepsilon^{L-l-1} \frac{\Gamma_{\text{inc}} \left( m+l+\frac{3}{2}, \frac{1}{b} \right)} {b^{N-m-l-\frac{3}{2}}},
\end{align}
where \eqref{eq:Ia_eqpow_eqspeed_3} follows from \eqref{eq:Ia_eqpow_eqspeed_2} by substituting for $a = b / (1 + \varepsilon)$, and \eqref{eq:Ia_eqpow_eqspeed_4} follows from the binomial expansion of $\left( \varepsilon + \left[1 + b q\right] \right)^{L-1}$ in \eqref{eq:Ia_eqpow_eqspeed_3} and interchanging the order of the summation and integration. The variable transformation $y=1+bq$ is used to get \eqref{eq:Ia_eqpow_eqspeed_5} from which \eqref{eq:Ia_eqpow_eqspeed_6} follows by another binomial expansion of $\left( y-1 \right)^{N-1}$. Finally, the integration in \eqref{eq:Ia_eqpow_eqspeed_6} is given by \cite[Sec. 3.381, Eq. 3]{372} resulting in \eqref{eq:Ia_eqpow_eqspeed_7}.

By substituting \eqref{eq:Ia_eqpow_eqspeed_7} in \eqref{eq:LCR4}, the LCR for the equal-power equal-speed interferers case is given by
\begin{equation}
\text{LCR}( \gamma_{\text{th}} ) = \frac{ \sqrt{2 \sigma_{\scriptscriptstyle \text{D}}^2 } \left( \frac{\gamma_{\text{th}} N_o}{p_{\scriptscriptstyle D}} \right)^{L-\frac{1}{2}} e^{-\frac{\gamma_{\text{th}} N_o}{p_{\scriptscriptstyle D}}} e^{\frac{1}{b}}}   {\sqrt{\pi} (1 + \varepsilon)^{L-1} \left(1 + \frac{\gamma_{\text{th}} p_{\scriptscriptstyle \text{I}}} {p_{\scriptscriptstyle \text{D}}}\right)^N} \sum\limits_{l=0}^{L-1} \sum\limits_{m=0}^{N-1} \frac{ \left( -1 \right)^{N-m-1} \varepsilon^{L-l-1} } {l! \ m! \ \left(L-l-1\right)! \ \left(N-m-1\right)! }  \frac{\Gamma_{\text{inc}} \left( m+l+\frac{3}{2}, \frac{1}{b} \right)} {b^{N-m-l-\frac{3}{2}}}, \tag{\ref{eq:LCR_equal_pow_equal_speed} revisited}
\end{equation}
where the identity $\Gamma_{\text{s}}(x)=(x-1)!$ for an integer $x$ is used. It is worth mentioning that this expression has never been reported in literature up to our knowledge.

\section{LCR Derivation for Interferers with Equal Powers and Equal Speeds in an Interference-Limited System}
\label{sec:int_lim_eq_pow_eq_speed_LCR_deriv_appendix}

To find the LCR in an interference-limited system where all interferers have the same transmitting power and move with the same speed, the limit of \eqref{eq:LCR_equal_pow_equal_speed} as $N_o \rightarrow 0$ can be easily taken. First, \eqref{eq:LCR_equal_pow_equal_speed} is rewritten in terms of $N_o$. From \eqref{eq:anbneps}, $b$ is rewritten as
\begin{equation}
\label{eq:b_No_explicit}
b = \frac{1} {N_o} \underbrace{\frac{p_{\scriptscriptstyle \text{I}}} { 1 + \frac{\gamma_{\text{th}} p_{\scriptscriptstyle \text{I}}} {p_{\scriptscriptstyle \text{D}}} } }_{\tilde{a}} \left( 1 + \frac{ \gamma_{\text{th}} p_{\scriptscriptstyle \text{I}} \sigma_{\text{I}}^2 } { p_{\scriptscriptstyle \text{D}} \sigma_{\scriptscriptstyle \text{D}}^2 } \right) = \frac{\tilde{a}} {N_o} \left( 1 + \varepsilon \right),
\end{equation}
where the dependence of $b$ on $N_o$ is explicitly shown. We assert that neither $\tilde{a},$ nor $\varepsilon$ is a function of $N_o$ per their definitions in \eqref{eq:b_No_explicit} and \eqref{eq:anbneps}, respectively. Then, the LCR in \eqref{eq:LCR_equal_pow_equal_speed} can be rewritten as an explicit function of $N_o$ as follows
\begin{equation}
\label{eq:LCR_No_explicit}
\text{LCR}( \gamma_{\text{th}} ) = \frac{ \sqrt{2 \sigma_{\scriptscriptstyle \text{D}}^2 } \left( \frac{\gamma_{\text{th}} }{p_{\scriptscriptstyle D}} \right)^{L-\frac{1}{2}} e^{-\frac{\gamma_{\text{th}} N_o}{p_{\scriptscriptstyle D}}} e^{\frac{N_o}{\tilde{a} \left( 1+\varepsilon \right) }} }   {\sqrt{\pi} (1 + \varepsilon)^{L-1} \left(1 + \frac{\gamma_{\text{th}} p_{\scriptscriptstyle \text{I}}} {p_{\scriptscriptstyle \text{D}}}\right)^N} \sum\limits_{l=0}^{L-1} \sum\limits_{m=0}^{N-1} \frac{ \left( -1 \right)^{N-m-1} \varepsilon^{L-l-1} } {l! \ m! \ \left(L-l-1\right)! \ \left(N-m-1\right)! }  \frac{\Gamma_{\text{inc}} \left( m+l+\frac{3}{2}, \frac{N_o}{\tilde{a} \left( 1 + \varepsilon \right)} \right) \ (N_o)^{N+L-m-l-2} } {\left( \tilde{a} \left( 1 + \varepsilon \right) \right)^{N-m-l-\frac{3}{2}}}.
\end{equation}

From \eqref{eq:LCR_No_explicit}, it is noticed that the dependence on $N_o$ lies within the term $\Delta = e^{ \left( -\frac{\gamma_{\text{th}}} {p_{\scriptscriptstyle D}} + \frac{1} { \tilde{a} \left( 1 + \varepsilon \right)} \right) N_o } (N_o)^z \ \Gamma_{\text{inc}} \left( m+l+\frac{3}{2}, \frac{N_o}{\tilde{a} \left( 1 + \varepsilon \right)} \right)$, where $z=N+L-m-l-2$, $l=0,\dots,L-1$ and $m=0,\dots,N-1$. It can be seen that $z=0$, when $(l,m)=(L-,N-1)$ and $z>0$ for all other combinations of $l$ and $m$, which leads to
\begin{equation}
\lim_{N_o \to 0} \Delta = \begin{cases}
0, &(l,m) \neq (L-1,N-1)\\
\Gamma_{\text{s}} \left( m+l+\frac{3}{2} \right), &(l,m) = (L-1,N-1),
\end{cases}
\end{equation}
where the identity $\Gamma_{\text{inc}}(x,0)=\Gamma_{\text{s}}(x)$ is used. Then, the LCR of the interference-limited system under consideration can be obtained by taking the limit of \eqref{eq:LCR_No_explicit} as $N_o$ tends to zero as follows
\begin{align}
\label{eq:LCR_int1}
\text{LCR} &= \lim_{N_o \to 0} \text{LCR} ( \gamma_{\text{th}} ) = \frac{ \sqrt{2 \sigma_{\scriptscriptstyle \text{D}}^2 } \left( \frac{\gamma_{\text{th}} }{p_{\scriptscriptstyle D}} \right)^{L-\frac{1}{2}} }   {\sqrt{\pi} (1 + \varepsilon)^{L-1} \left(1 + \frac{\gamma_{\text{th}} p_{\scriptscriptstyle \text{I}}} {p_{\scriptscriptstyle \text{D}}}\right)^N} \frac{ \Gamma_{\text{s}} \left( N+L-\frac{1}{2} \right) } { (L-1)! \ (N-1)! } \frac{1} {\tilde{a}^{-L+\frac{1}{2}} \left( 1 + \varepsilon \right)^{-L+\frac{1}{2}}} \\
\label{eq:LCR_int2}
&= \sqrt{ \frac{2 \sigma_{\scriptscriptstyle \text{D}}^2} {\pi} } \ \left( 1 + \varepsilon \right)^{\frac{1}{2}} \ \frac{\Gamma_{\text{s}} \left( N+L-\frac{1}{2} \right)} {\Gamma_{\text{s}}\left( N \right) \Gamma_{\text{s}}\left( L \right)} \frac{ \left(\gamma_{\text{th}} p_{\scriptscriptstyle \text{I}} / p_{\scriptscriptstyle \text{D}} \right)^{L-\frac{1}{2}} } { \left( 1 + \gamma_{\text{th}} p_{\scriptscriptstyle \text{I}} / p_{\scriptscriptstyle \text{D}} \right)^{N+L-\frac{1}{2}} } \\
&= \sqrt{2 \pi} \frac{ \Gamma_{\text{s}}(N+L-\frac{1}{2}) } { \Gamma_{\text{s}}(N) \Gamma_{\text{s}}(L) } \left( f_{\scriptscriptstyle \text{D}}^2 + \frac{f_{\scriptscriptstyle \text{I}}^2} {\Omega_{\text{th}}} \right)^{\frac{1}{2}} \frac{ \Omega_{\text{th}}^N } {\left( 1 + \Omega_{\text{th}} \right)^{N+L-\frac{1}{2}}}, \tag{\ref{eq:int_lim_eqpow_eqspeed} revisited}
\end{align}
where $\Omega_{\text{th}} = p_{\scriptscriptstyle \text{D}} / \gamma_{\text{th}} p_{\scriptscriptstyle \text{I}}$. \eqref{eq:LCR_int2} follows from direct substitution for the variables $\tilde{a}$, and $\epsilon$ by their definitions in \eqref{eq:b_No_explicit} and \eqref{eq:anbneps}, respectively. To get to the final expression in \eqref{eq:int_lim_eqpow_eqspeed}, we also used $\sigma_{\scriptscriptstyle \text{D}}^2=\pi^2 f_{\scriptscriptstyle \text{D}}^2$ and $\sigma_{\text{I}}^2=\pi^2 f_{\scriptscriptstyle \text{I}}^2$ which are defined in Section \ref{sec:joint_pdf_SINR_time_der}.

\section{LCR Derivation for Single-Antenna Receiver}
\label{sec:appendix_MERL}

By directly substituting for $L=1$ in \eqref{eq:LCR_final}, the summations over $k,m,r$ and $w$ disappears and we substitute for $k=m=r=w=0$ and $\Xi_{k,m}=1$, then \eqref{eq:LCR_final} is given as
\begin{equation}
\text{LCR} (\gamma_{\text{th}}) = \sqrt{ \frac{ 2 \sigma_{\scriptscriptstyle \text{D}}^2 \gamma_{\text{th}} N_o } { \pi p_{\scriptscriptstyle \text{D}} } } \frac{ e^{-\frac{\gamma_{\text{th}} N_o} {p_{\scriptscriptstyle \text{D}}} } \Lambda } { \prod\limits_{n=1}^N \left( 1 + \frac{ \gamma_{\text{th}} p_n } { p_{\scriptscriptstyle \text{D}} } \right) } \sum\limits_{n=1}^N \delta_n \sum\limits_{\substack{t=1 \\ t \neq n}} \psi_{t,n} \frac{ e^{g_n} \Gamma_{\text{inc}}\left( \frac{3}{2}, g_n \right) } { g_n^{\frac{3}{2}} \lambda_{t,n} }. \tag{ \ref{eq:MERL_ours1} revisited}
\end{equation}

By re-ordering the terms, the LCR can be expressed as
\begin{equation}
\text{LCR} (\gamma_{\text{th}}) = \sqrt{ \frac{ 2 \sigma_{\scriptscriptstyle \text{D}}^2 \gamma_{\text{th}} N_o } { \pi p_{\scriptscriptstyle \text{D}} } } \frac{ e^{-\frac{\gamma_{\text{th}} N_o} {p_{\scriptscriptstyle \text{D}}`} } } { \prod\limits_{n=1}^N \left( 1 + \frac{ \gamma_{\text{th}} p_n } { p_{\scriptscriptstyle \text{D}} } \right) } \sum\limits_{n=1}^N \underbrace{\sum\limits_{\substack{t=1 \\ t \neq n}} \frac{\Lambda \delta_n \psi_{t,n}} {g_n \lambda_{t,n}}}_{\delta_n^{\text{MERL}}} \frac{ e^{g_n} } { \sqrt{g_n} } \Gamma_{\text{inc}}\left( \frac{3}{2}, g_n \right), \tag{ \ref{eq:MERL_ours2} revisited}
\end{equation}
where this equation is the same as equation (26) reported in \cite{370} where $\delta_n^{\text{MERL}}$ denotes $\delta_n$ in \cite{370} to avoid confusion, and $g_n$ is equivalent to $\mathcal{W}_n$ in \cite{370} which is proven as follows. First, $a_n$ and $\varepsilon_n$ are substituted for by \eqref{eq:anbneps} in \eqref{eq:gn_simple} resulting in
\begin{equation}
g_n = \frac{ 1 } {a_n \left(1 + \varepsilon_n \right) } = \frac{1 + \frac{ \gamma_{\text{th}} p_n} {p_{\scriptscriptstyle \text{D}}} } { \frac{p_n}{N_o} + \frac{ \gamma_{\text{th}} \sigma_n^2 p_n^2 } { N_o \sigma_{\scriptscriptstyle \text{D}}^2 p_{\scriptscriptstyle \text{D}} } } = \mathcal{W}_n,
\end{equation}
where the last equality follows from the definition of $\mathcal{W}_n$ in equation (24) in \cite{370}. Thus, it is obvious that $g_n = \mathcal{W}_n$.

Second we prove that $\delta_n^{\text{MERL}}$ is equal to the summation in \eqref{eq:MERL_ours2} as follows. Using equation (27) in \cite{370}, and the fact that $g_n = \mathcal{W}_n$, $\delta_n^{\text{MERL}}$ is rewritten as
\begin{align}
\delta_n^{\text{MERL}} &= \prod\limits_{\substack{t=1 \\ t \neq n}}^N \frac{\mathcal{W}_t} {\mathcal{W}_t - \mathcal{W}_n} = \prod\limits_{\substack{t=1 \\ t \neq n}}^N \frac{g_t} {g_t - g_n} = \frac{1} {\prod\limits_{\substack{t=1 \\ t \neq n}}^N b_t } \frac{1} {\prod\limits_{\substack{t=1 \\ t \neq n}}^N \left( \frac{1} {b_t} - \frac{1} {b_n} \right)} = \frac{b_n} {\prod\limits_{t=1}^N b_t} \frac{1} {\prod\limits_{\substack{t=1 \\ t \neq n}}^N \left( \frac{1} {b_t} - \frac{1} {b_n} \right)} \nonumber\\
&= \frac{\Lambda} {g_n} \frac{b_n^{N-1} \prod\limits_{t=1, t \neq n}^N b_t } { \prod\limits_{t=1, t \neq n}^N \left( b_n - b_t \right) } = \frac{\Lambda} {g_n} \frac{b_n^{N-2} \prod\limits_{t=1}^N b_t } { \prod\limits_{t=1, t\neq n}^N \left( a_t b_n - a_n b_t \right) }   \frac{ \prod\limits_{t=1, t\neq n}^N \left( a_t b_n - a_n b_t \right) } { \prod\limits_{t=1, t \neq n}^N \left( b_n - b_t \right) } \nonumber\\
\label{eq:delta_MERL}
&= \frac{\Lambda} {g_n}  \delta_n \prod\limits_{\substack{t=1 \\ t \neq n}} \frac{1}{\lambda_{t,n}}.
\end{align}

From the definition of the partial fraction coefficient $\zeta_n(s_2)$ in \eqref{eq:zeta_def}, by substituting for $s_2=0$, we find
\begin{equation}
\label{eq:zeta_n0}
\zeta_n(0) = \frac{\delta_n} { \prod\limits_{\substack{t=1 \\ t \neq n}}^N \lambda_{t,n} } = \delta_n \sum\limits_{\substack{t=1 \\ t \neq n}}^N \frac{\psi_{t,n}} {\lambda_{t,n}}.
\end{equation}

Combining \eqref{eq:delta_MERL} and \eqref{eq:zeta_n0}, it can be proved that
\begin{equation}
\delta_n^{\text{MERL}} = \sum\limits_{\substack{t=1 \\ t \neq n}} \frac{\Lambda \delta_n \psi_{t,n}} {g_n \lambda_{t,n}},
\end{equation}
and hence, \eqref{eq:MERL_ours2} is the same as equation (26) in \cite{370} which validates our general LCR expression in \eqref{eq:LCR_final}.

\section{Derivations of \eqref{eq:rD_pdf} - \eqref{eq:rdot_pdf}}
\label{sec:deriv_of_rdot_pdf_appendix}

The joint PDF $f_{\dot{R}_{\scriptscriptstyle \text{I}} | R_{\text{I}}}(r_{\scriptscriptstyle \text{I}},\dot{r}_{\scriptscriptstyle \text{I}})$ can be written as $f_{R_{\text{I}},\dot{R}_{\scriptscriptstyle \text{I}}}(r_{\scriptscriptstyle \text{I}},\dot{r}_{\scriptscriptstyle \text{I}}) = f_{\dot{R}_{\scriptscriptstyle \text{I}} | R_{\text{I}}}(\dot{r}_{\scriptscriptstyle \text{I}} | r_{\scriptscriptstyle \text{I}}) f_{R_{\text{I}}}(r_{\scriptscriptstyle \text{I}})$, where $f_{R_{\text{I}}}(r_{\scriptscriptstyle \text{I}})$ and $f_{\dot{R}_{\scriptscriptstyle \text{I}} | R_{\text{I}}}(\dot{r}_{\scriptscriptstyle \text{I}} | r_{\scriptscriptstyle \text{I}})$ are derived as follows.

$Derivation \ of \ f_{R_{\text{I}}}(r_{\scriptscriptstyle \text{I}})$: Let $H$ denote the interference component of the signal, then $H$ is defined as
\begin{equation}
\label{eq:eta_def}
H = \sum\limits_{n=1}^N p_n \ A_{\scriptscriptstyle \text{I},n}^2(t).
\end{equation}

The PDF of $H$ is given in \cite[Eq. 68]{505} for the UAP case, which is the case under consideration in this paper,  as
\begin{equation}
\label{eq:eta_pdf}
f_H(\eta) = \sum\limits_{n=1}^N \frac{\mu_n}{p_n} e^{-\frac{\eta} {p_n}},  \ \ \ \ \ \ \ \ \ \eta \geq 0
\end{equation}
where $\mu_n$ are constants given by \eqref{eq:mu}. Since $R_{\scriptscriptstyle \text{I}} = \sqrt{N_o + H}$, then the PDF $f_{R_{\text{I}}}(r_{\scriptscriptstyle \text{I}})$ can be given as
\begin{equation}
f_{R_{\text{I}}}(r_{\scriptscriptstyle \text{I}}) = f_H(\eta) \left| \frac{d H} {d R_{\scriptscriptstyle \text{I}}} \right| = \sum\limits_{n=1}^N \frac{ 2 \mu_n e^{\frac{N_o}{p_n}} } {p_n} r_{\scriptscriptstyle \text{I}} e^{-\frac{r_{\scriptscriptstyle \text{I}}^2} {p_n}}, \ \ \ \ \ \ \ r_{\scriptscriptstyle \text{I}} \geq \sqrt{N_o}.  \tag{\ref{eq:rI_pdf} revisited}
\end{equation}

$Derivation \ of \ f_{\dot{R}_{\scriptscriptstyle \text{I}}}(\dot{r}_{\scriptscriptstyle \text{I}})$: Since $R_{\scriptscriptstyle \text{I}} = \sqrt{N_o + \sum\limits_{n=1}^N p_n \ A_{\scriptscriptstyle \text{I},n}^2(t)}$, then its time derivative yields
\begin{equation}
\label{eq:rIdot}
\dot{R}_{\scriptscriptstyle \text{I}} = \frac{ 2 \sum\limits_{n=1}^N p_n A_{\scriptscriptstyle \text{I},n} \dot{A}_{\text{I},n} } { 2 \sqrt{N_o + \sum\limits_{n=1}^N p_n \ A_{\scriptscriptstyle \text{I},n}^2} } = \frac{ \sum\limits_{n=1}^N p_n A_{\scriptscriptstyle \text{I},n} \dot{A}_{\text{I},n} } { R_{\scriptscriptstyle \text{I}} }.
\end{equation}

Since, as mentioned in Section \ref{sec:joint_pdf_SINR_time_der}, $\dot{A}_{\text{I},n}$ is a zero mean Gaussian random process with variance $\sigma_{\text{I},n}^2 = \pi^2 f_{\scriptscriptstyle \text{I},n}^2$ \cite{370}, then it follows from \eqref{eq:rIdot} that $\dot{r}_{\scriptscriptstyle \text{I}}$ is a linear combination of Gaussian RVs $\dot{\alpha}_{\text{I},n}$ given $\alpha_{\scriptscriptstyle \text{I},n}$. Hence $r_{\scriptscriptstyle \text{I}}$ is conditionally Gaussian distributed with zero mean and variance given by
\begin{equation}
E\{\dot{R}_{\scriptscriptstyle \text{I}}^2\} = \frac{ \sum\limits_{n=1}^N p_n^2 A_{\scriptscriptstyle \text{I},n}^2 \sigma_{\text{I},n}^2 } { N_o + \sum\limits_{n=1}^N p_n \ A_{\scriptscriptstyle \text{I},n}^2 }.
\end{equation}

It can be seen from the above equation that the variance $E\{\dot{R}_{\scriptscriptstyle \text{I}}^2\}$ depends on $A_{\scriptscriptstyle \text{I},n}^2$ and consequently $R_{\scriptscriptstyle \text{I}}$ which makes finding $f_{\dot{R}_{\scriptscriptstyle \text{I}} | R_{\text{I}}}(\dot{r}_{\scriptscriptstyle \text{I}} | r_{\scriptscriptstyle \text{I}})$ a complicated process. Instead, the variance is approximated as follows
\begin{equation}
E\{\dot{R}_{\scriptscriptstyle \text{I}}^2\} \approx \frac{ \sum\limits_{n=1}^N p_n^2 E\{A_{\scriptscriptstyle \text{I},n}^2\} \sigma_{\text{I},n}^2 } { N_o + \sum\limits_{n=1}^N p_n \ E\{A_{\scriptscriptstyle \text{I},n}^2\} } = \frac{ \sum\limits_{n=1}^N p_n^2 \sigma_{\text{I},n}^2 } { N_o + \sum\limits_{n=1}^N p_n } = \frac{ \pi^2 \sum\limits_{n=1}^N f_{\scriptscriptstyle \text{I},n}^2 \ p_n^2  } { N_o + \sum\limits_{n=1}^N p_n }  = \dot{\sigma}_{\scriptscriptstyle \text{I}}^2,
\end{equation}
where the fact that $A_{\scriptscriptstyle \text{I},n}^2$ is a standard exponential RV with unit variance is used. Using this approximation, $E\{\dot{R}_{\scriptscriptstyle \text{I}}^2\}$ becomes independent of $R_{\scriptscriptstyle \text{I}}$, it can be seen that $\dot{R}_{\scriptscriptstyle \text{I}}$ is independent of $R_{\scriptscriptstyle \text{I}}$, and hence, the conditional PDF $f_{\dot{R}_{\scriptscriptstyle \text{I}} | R_{\text{I}}}(\dot{r}_{\scriptscriptstyle \text{I}} | r_{\scriptscriptstyle \text{I}})$ can be written as
\begin{equation}
f_{\dot{R}_{\scriptscriptstyle \text{I}} | R_{\text{I}}}(\dot{r}_{\scriptscriptstyle \text{I}} | r_{\scriptscriptstyle \text{I}}) = f_{\dot{R}_{\scriptscriptstyle \text{I}}}(\dot{r}_{\scriptscriptstyle \text{I}}) = \frac{1} {\sqrt{2 \pi \dot{\sigma}_{\scriptscriptstyle \text{I}}^2}} e^{-\frac{\dot{r}_{\scriptscriptstyle \text{I}}^2}{2 \dot{\sigma}_{\scriptscriptstyle \text{I}}^2}}   \tag{\ref{eq:rdot_pdf} revisited}.
\end{equation}

\textit{\textbf{Remark}} For $f_{R_{\scriptscriptstyle \text{D}},\dot{R}_{\scriptscriptstyle \text{D}}}(r_{\scriptscriptstyle \text{D}},\dot{r}_{\scriptscriptstyle \text{D}})$, it can be easily shown that $\dot{R}_{\scriptscriptstyle \text{D}}$ is independent of $R_{\scriptscriptstyle \text{D}}$ via following the same steps but without the need of doing any approximations. This is because the powers of all of the components of the desired signal are equal. Thus, the PDFs $f_{R_{\scriptscriptstyle \text{D}}}(r_{\scriptscriptstyle \text{D}})$ and $f_{\dot{R}_{\scriptscriptstyle \text{D}}}(\dot{r}_{\scriptscriptstyle \text{D}})$ can be easily obtained to be as in the EAP case in equations (23) and (24) in \cite{505}
\begin{align}
f_{R_{\scriptscriptstyle \text{D}}}(r_{\scriptscriptstyle \text{D}}) &= \frac{2}{(L-1)!} \frac{ r_{\scriptscriptstyle \text{D}}^{2L-1} } {p_{\scriptscriptstyle D}^L} e^{-\frac{r_{\scriptscriptstyle \text{D}}^2} {p_{\scriptscriptstyle D}}} \tag{\ref{eq:rD_pdf} revisited}, \text{ and} \\
f_{\dot{R}_{\scriptscriptstyle \text{D}}}(\dot{r}_{\scriptscriptstyle \text{D}}) &= \frac{1} {\sqrt{2 \pi p_{\scriptscriptstyle D} \sigma_{\scriptscriptstyle \text{D}}^2 }} e^{-\frac{\dot{r}_{\scriptscriptstyle \text{D}}^2} {2 p_{\scriptscriptstyle D} \sigma_{\scriptscriptstyle \text{D}}^2}},  \tag{\ref{eq:rD_dot_pdf} revisited}
\end{align}
where $\sigma_{\scriptscriptstyle \text{D}}^2=2 \pi^2 f_{\scriptscriptstyle \text{D}}^2$ is the variance of $A_{\scriptscriptstyle \text{D},l}^2$ for any $l \in \mathcal{L}$ as previously mentioned in Section \ref{sec:joint_pdf_SINR_time_der}.

\begin{figure}[!htb]
        \centering
        \includegraphics[trim=4cm 8.5cm 4cm 9cm,clip=true,width=3.5in]{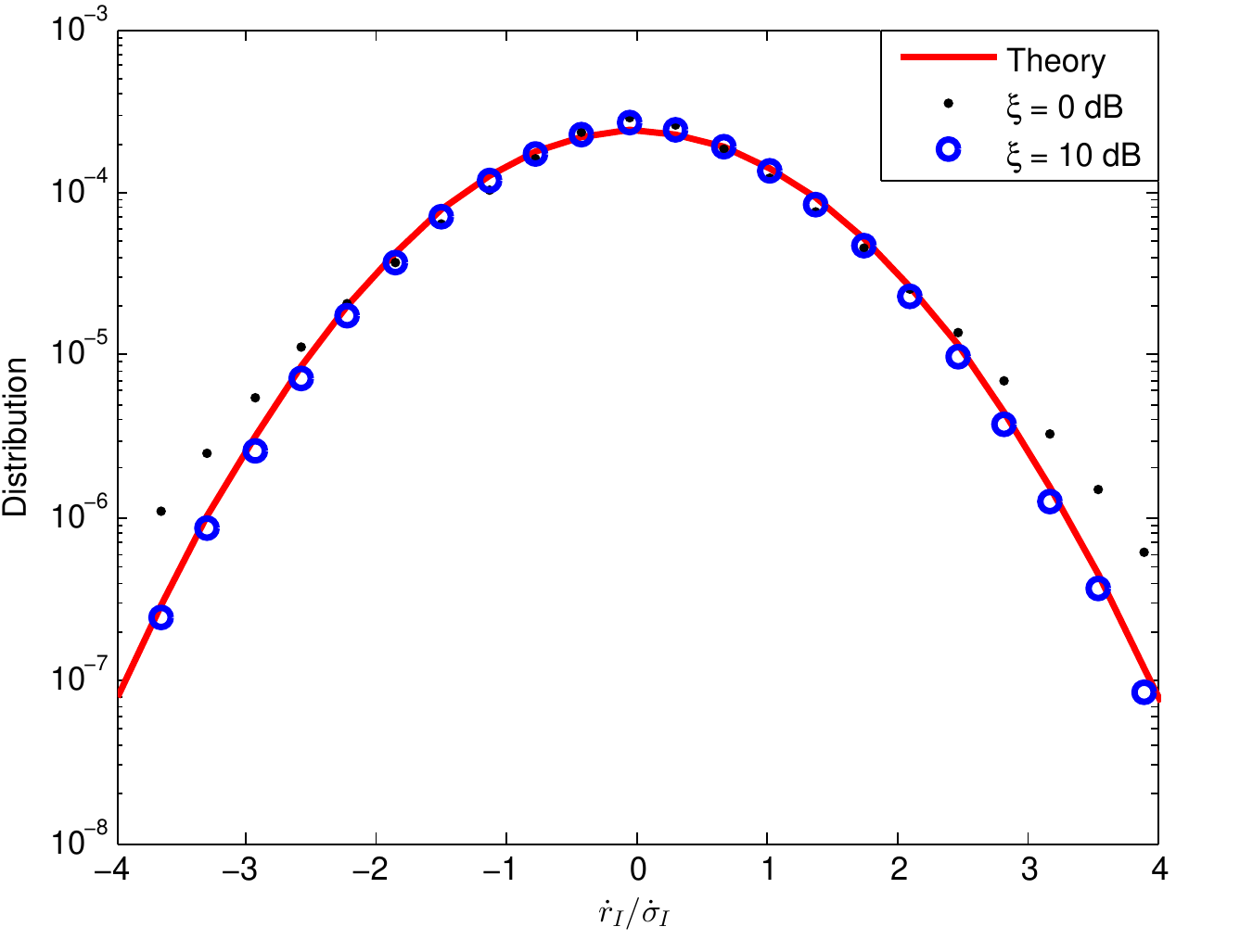}
        \caption{Comparing approximated and actual PDF of $r_{\scriptscriptstyle \text{I}}$ in the interference-limited $L=2$, $N=2$ system for the cases of equal and unequal powers where $N_o = 10^{-5}$, $p_{\scriptscriptstyle \text{I},2} = 1$, $p_{\scriptscriptstyle \text{D}}=1$, $f_{\scriptscriptstyle \text{I}} = [108, 540]$ Hz, and $f_{\scriptscriptstyle \text{D}}=540$ Hz.}
        \label{fig:pdf_No1e-5_step6e-4}
\end{figure}

\begin{figure}[!htb]
        \centering
        \includegraphics[trim=4cm 8.5cm 4cm 9cm,clip=true,width=3.5in]{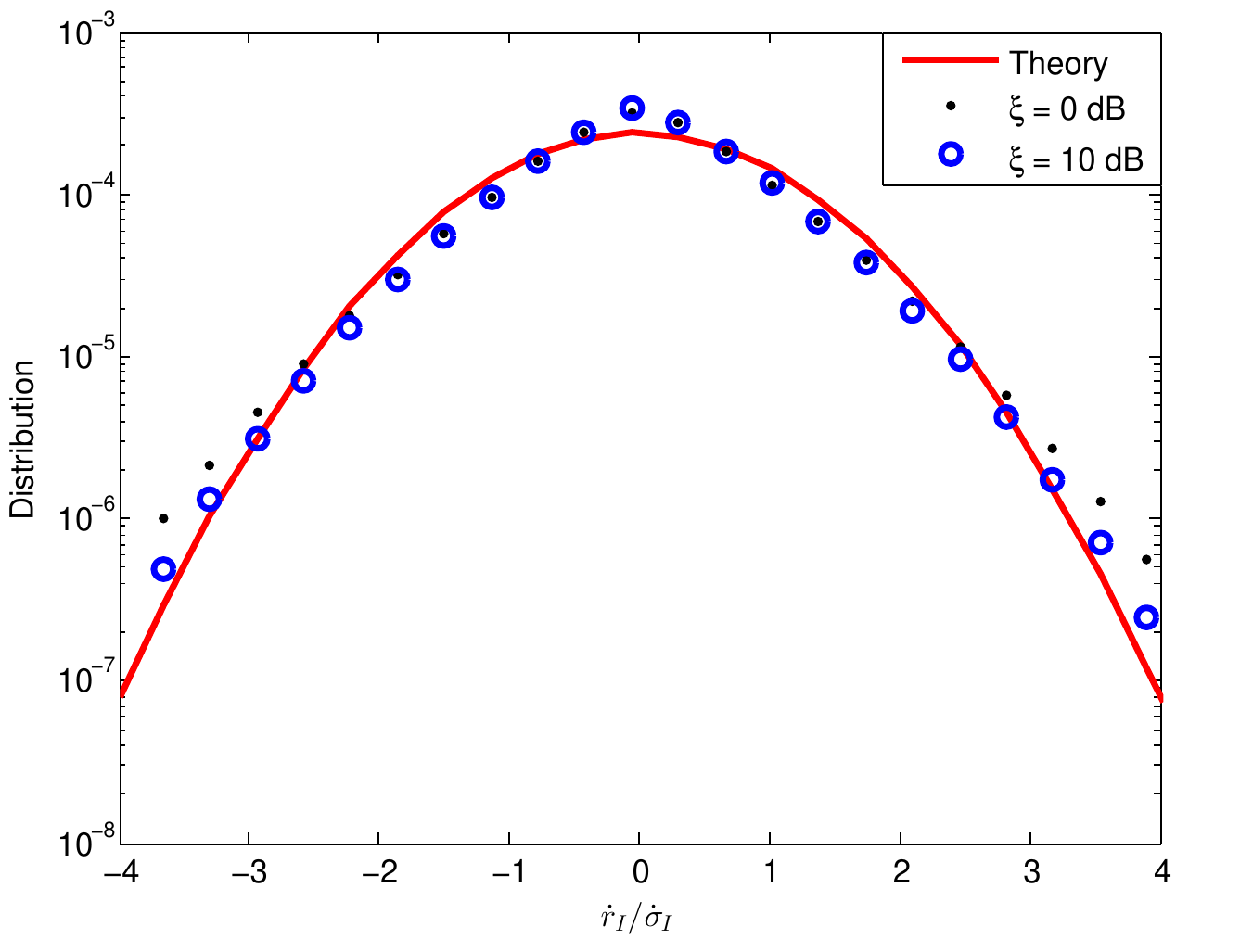}
        \caption{Comparing approximated and actual PDF of $r_{\scriptscriptstyle \text{I}}$ in the noise-limited $L=2$, $N=2$ system for the cases of equal and unequal powers where $N_o = 1$, $p_{\scriptscriptstyle \text{I},2} = 1$, $p_{\scriptscriptstyle \text{D}}=1$, $f_{\scriptscriptstyle \text{I}} = [108, 540]$ Hz, and $f_{\scriptscriptstyle \text{D}}=540$ Hz.}
        \label{fig:pdf_No1_step6e-4}
\end{figure}

The accuracy of approximation in \eqref{eq:rdot_pdf} is evaluated in Fig. \ref{fig:pdf_No1e-5_step6e-4}-\ref{fig:pdf_No1_step6e-4}. Fig. \ref{fig:pdf_No1e-5_step6e-4}-\ref{fig:pdf_No1_step6e-4} plot the approximated PDF and the distribution obtained from simulation for a system with $L=2$ and $N=2$ where the two interferers posses maximum Doppler frequencies of values $108$ Hz and $540$ Hz. Defining the ratio between the transmitting powers of the two interferers as $\xi=10 \log (p_2/p_1)$, and fixing the power of the second interferer $p_{2}=1$ for the sake of comparison, the simulation was done for the cases when $\xi=0$ dB, i.e. equal powers, and $\xi=10$ dB for two different values of noise power $N_o$, namely $N_o=10^{-5},$ and $1$, representing interference-limited and noise-limited systems. The transmitting power of the desired user is fixed to $p_{\scriptscriptstyle \text{D}}=1$ and consider that it moves with the maximum speed of the interferers, i.e. $f_{\scriptscriptstyle \text{D}}=540$ Hz. When the theoretical approximated PDFs \eqref{eq:rdot_pdf} for both cases $\xi=0$ dB and $\xi=10$ were calculated, it was found that they have very close values, almost the same, so we suffice by plotting only one of them and denote it by \textit{Theory} in the legend to avoid crowding the figure with many curves with different markers. When both PDFs were plotted, they were almost perfectly on top of each other, thus it does not matter which one is chosen to plot. Fig. \ref{fig:pdf_No1e-5_step6e-4} shows that the difference between the approximated PDF and the actual distribution is negligible in case of $\xi=10$ dB. Surprisingly, when $\xi=0$ dB the approximated PDF is less accurate. It is worth mentioning here that even when the interferers powers are equal the PDF \eqref{eq:rdot_pdf} is still an approximation to the exact $f_{\dot{R}_{\scriptscriptstyle \text{I}}}(\dot{r}_{\scriptscriptstyle \text{I}})$ unlike what is reported in \cite{505} because the interferers have different Doppler frequencies. Moreover, even if the interferers move with equal speeds, the PDF is still an approximation because we consider the SINR while \cite{505} considers the SIR only, hence ignoring the effect of noise.

As the noise power increases, the approximate PDF deviates noticeably from the exact PDF whether the interferers' powers are equal or not. This is obvious in Fig. \ref{fig:pdf_No1_step6e-4} where $N_o=1$ which is equal to the dominant interferer's power. Notice that although the deviation is noticeable in the figure, the approximation is still good as was shown in the numerical results.

\bibliographystyle{IEEEtran}
\bibliography{IEEEabrv,IEEEfull}

\end{document}